# Non-perturbative Studies of Non-conformal Field Theories

## Navdeep Singh Dhindsa

*A thesis submitted for the partial fulfillment of*
*the degree of Doctor of Philosophy*

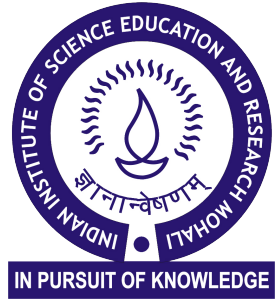

*This thesis is dedicated to my mom.*

*For your endless love, encouragement, and sacrifices.*

# Abstract

Many of the exciting features of the Standard Model of the elementary particles are inherently non-perturbative. A theoretical understanding of many physics aspects beyond the Standard Model of elementary particles also requires a non-perturbative framework. One such framework involves discretizing quantum field theories on a spacetime lattice. We can use this lattice regularization method to study supersymmetric versions of physics beyond the Standard Model. This method can also investigate the non-perturbative physics of quantum field theories that are holographically dual to theories containing gravity.

In this thesis, we discuss the spacetime lattice setup, and with the examples of different models, we will see the numerical capability of this tool in exploring field theory regimes that are not accessible through perturbation theory. We use an efficient version of the Monte Carlo algorithm to update the field configurations in the path integral and eventually reach the equilibrium configurations. The thesis reports the investigation of the possibilities of non-perturbative supersymmetry breaking in quantum mechanics models with different superpotentials to test the algorithms and numerical setup.

The gauge/gravity duality conjecture states that certain classes of field theories that do not contain gravity are equivalent to specific types of quantum gravitational theories. A version of the conjecture connects weakly coupled gravitational theories to strongly coupled field theories. Though there has been excellent progress in understanding and verifying the gauge/gravity duality conjecture by studying the maximally supersymmetric Yang-Mills theories in four spacetime dimensions, we will mainly focus on the non-conformal analogs of the conjecture in lower dimensions.

This thesis mainly discusses the numerical simulation results of two lower-dimensional models. One is the bosonic version of the BMN (Berenstein, Maldacena, and Nastase) matrix quantum mechanics and the other is a two-dimensional Yang-Mills theory containing four supersymmetries. We focus on the transition between different phases of the theories as the deconfinement phase transitions are dual to quantum black hole solutions.

The bosonic BMN matrix model is obtained by removing the fermionic degrees of freedom from the full BMN matrix model. Though the bosonic model does not admit a holographic dual, we find that it still undergoes a deconfinement phase transition. Our numerical results suggest that the phase diagram smoothly interpolates between the bosonic BFSS (Banks, Fischler, Susskind, and Shenker) and the gauged Gaussian model, with first-order deconfinement phase transition at all couplings. After this, we will focus on the thermal phase structure of a two-dimensional Yang-Mills theory that contains four supersymmetries. Our simulation results show that this model admits a deconfinement phase transition in the limit of a large number of colors. We also show that the nature of the transition looks similar to its maximally supersymmetric cousin in the weak coupling regime. The thesis is concluded with a discussion of the models investigated and their numerical results, along with possible future directions.

vii

# List of Publications/Proceedings/Preprints

# List of Figures

# List of Tables

# Contents

# 1
# Introduction

## 1.1 Motivation

Quantum field theory is a suitable framework for describing all the non-gravitational interactions between particles of the Standard Model. However, some features of the interactions in the Standard Model of elementary particles cannot be accessed by perturbation theory, for example, the physics of the strong force, which does not have a possible expansion using a small parameter. Also, many aspects of physics beyond the Standard Model (BSM) require a non-perturbative set up to formulate a theoretical understanding. Along with various non-perturbative methods used to tackle regimes of theories where perturbative tools fail, one that is highly used is *Lattice Field Theory*. It is based on the path integral formulation introduced by Feynman in the mid-20th century [6]. For an introduction to lattice field theory, the reader can refer to [7–9].

In lattice field theory, we discretize continuous fields on a spacetime lattice. With the help of Monte Carlo [10–12], we sample the fields of the theory with an associated probability distribution given by the action and let them evolve subject to a transition probability using the Metropolis algorithm. We can estimate the observables that we are interested in measuring by making use of the fields that evolved over a large Monte Carlo time. We successfully employ this method to investigate the thermodynamics of various supersymmetric systems in dimensions less than three in this thesis. Supersymmetry (SUSY) is an idea that predicts a partner particle for each particle in the Standard Model such that the spins of the two superpartners differ by half an integer. In simpler terms, it is an idea of connecting bosons with fermions and vice versa. We will discuss various supersymmetric gauge theories and their importance from a point of view of holographic duality in later chapters. Supersymmetric Yang-Mills (SYM) theories and their non-perturbative studies play an important role in BSM physics and string theory. Understanding these SYM theories can help us probe the thermodynamics of dual gravity theory. The gravity dual is expected to be only for field theories that are maximally supersymmetric and undergo a phase transition. However, only a limited study exists in probing the thermodynamics of the theories that are not maximally supersymmetric. This thesis attempts to probe the thermodynamics of such theories and motivate the possibility for these theories to be another set of candidates, which can help us understand the nature of the quantum gravitational theory, such as string theory.

In this chapter, we will look at the formalism related to the lattice approach and briefly touch on various problems that one can expect to face while working with this approach. Chapter 2 will focus on probing supersymmetry breaking in quantum mechanics with various superpotentials using the lattice approach. Chapter 3 will introduce the reader to various supersymmetric gauge theories and how these theories at strong coupling and finite temperature can help us to understand its dual gravity counterpart. It will also include the problems that appear in this non-perturbative approach, specifically to matrix models. Chapters 4 and 5 will discuss the study of supersymmetric matrix models in one and two dimensions, respectively. We provide a summary and future directions in Chapter 6.

## 1.2 Euclidean Path Integral

For a theory with a scalar field $\phi(x)$, the dynamics can be described by minimizing the action ($S[\phi(x)]$). According to Feynman [6], the contribution to the dynamics of the theory is not only from a classical path but rather from all possible paths. Every path contributes to the dynamics with equal amplitude but with a different phase, $\exp(iS[\phi(x)]/\hbar)$. Using the path integral approach, the partition function can be written as:

$$\mathcal{Z} = \int \mathcal{D}\phi \, e^{iS[\phi(x)]/\hbar}. \tag{1.1}$$

The expectation value of an observable $\mathcal{O}$ takes the form

$$\langle \mathcal{O} \rangle = \mathcal{Z}^{-1} \int \mathcal{D}\phi \, \mathcal{O}[\phi(x)] \, e^{iS[\phi(x)]/\hbar}. \tag{1.2}$$

The phase factors in the above equations give rise to oscillations that are difficult to deal with in the Monte Carlo method. We can avoid this difficulty by Wick rotating $t \to -i\tau$ and studying the thermal partition function instead of real-time processes. The thermal partition function is sufficient to study the phase structure of a theory, which is the focus of our work. Then, the contribution to the dynamics of the system will be coming from weight $\exp(-S_E[\phi(x)]/\hbar)$. We see that the contributions from trajectories that are away from the classical trajectory are exponentially small. The expectation value of an observable, given by Eq. (1.2) in Euclidean signature becomes [1]:

$$\langle \mathcal{O} \rangle = \mathcal{Z}^{-1} \int \mathcal{D}\phi \, \mathcal{O}[\phi] \, e^{-S[\phi]}, \tag{1.3}$$

where the partition function takes the form:

$$\mathcal{Z} = \int \mathcal{D}\phi \, e^{-S[\phi]}. \tag{1.4}$$

When we discretize the theory on a lattice to simulate using numerical methods, we impose certain boundary conditions on the fields. The different boundary conditions for fermions and bosons can be understood in various ways. Let us look at the implementation of boundary conditions using the thermal Green's function $G_B$ for a bosonic scalar field $\hat{\phi}$ [13]:

$$G_B(\mathbf{x}, \mathbf{y}; t_1, t_2) = \mathcal{Z}^{-1} \operatorname{Tr} \left( e^{-\beta K} T \left[ \hat{\phi}(\mathbf{x}, t_1) \hat{\phi}(\mathbf{y}, t_2) \right] \right), \tag{1.5}$$

---

[1] From now on, we will work with Euclidean signature; hence we will not use any subscript $E$ for simplification purposes. Also, from now on, we will work with units $\hbar = 1$.

where $T$ stands for time ordering, $\beta$ is the inverse temperature, $K$ is the Hamiltonian of the system, and the fields are taken at two different spacetime points $(\mathbf{x}, t_1)$ and $(\mathbf{y}, t_2)$.

Now take $t_1 = \tau$ and $t_2 = 0$, where the time ordering operator for the fields with $t_1 > t_2$ gives $\hat{\phi}(\mathbf{x}, \tau)\hat{\phi}(\mathbf{y}, 0)$. Then Eq. (1.5) becomes

$$G_B(\mathbf{x}, \mathbf{y}; \tau, 0) = \mathcal{Z}^{-1} \, \text{Tr}\left[e^{-\beta K}\hat{\phi}(\mathbf{x}, \tau)\hat{\phi}(\mathbf{y}, 0)\right]. \tag{1.6}$$

Making use of the cyclic property of the trace, we get

$$G_B(\mathbf{x}, \mathbf{y}; \tau, 0) = \mathcal{Z}^{-1} \, \text{Tr}\left[\hat{\phi}(\mathbf{y}, 0) \, e^{-\beta K} \, \hat{\phi}(\mathbf{x}, \tau)\right]. \tag{1.7}$$

Introducing identity as $\mathbb{I} = e^{-\beta K} e^{\beta K}$ the above expression becomes

$$G_B(\mathbf{x}, \mathbf{y}; \tau, 0) = \mathcal{Z}^{-1} \, \text{Tr}\left[e^{-\beta K} e^{\beta K} \hat{\phi}(\mathbf{y}, 0) \, e^{-\beta K} \, \hat{\phi}(\mathbf{x}, \tau)\right]. \tag{1.8}$$

Following on the same lines as that of the Heisenberg time evolution, we get $e^{\beta K}\hat{\phi}(\mathbf{y}, 0)\, e^{-\beta K} = \hat{\phi}(\mathbf{y}, \beta)$. Hence, the thermal Green's function takes the form

$$G_B(\mathbf{x}, \mathbf{y}; \tau, 0) = \mathcal{Z}^{-1} \, \text{Tr}\left[e^{-\beta K}\hat{\phi}(\mathbf{y}, \beta) \, \hat{\phi}(\mathbf{x}, \tau)\right]. \tag{1.9}$$

Since $\hat{\phi}$ are bosonic fields, we can interchange them in Eq. (1.9). On comparing Eqs. (1.6) and (1.9) we see that $\hat{\phi}(\mathbf{y}, \beta) = \hat{\phi}(\mathbf{y}, 0)$.

Instead of bosons, if we had fermions, then while performing the same procedure as above, we see that there can be an additional negative sign coming from interchanging the fermions in Eq. (1.9). Thus, for fermions $\hat{\psi}$ the relation becomes $\hat{\psi}(\mathbf{y}, \beta) = -\hat{\psi}(\mathbf{y}, 0)$; whenever we put fermions on a lattice with closed boundary conditions at finite temperature, we need to impose anti-periodic boundary conditions. On the other hand, if we put bosons on a lattice, we need to impose periodic boundary conditions.

## 1.3 Fields on a Lattice and Monte Carlo

As discussed in the previous section, with the help of the Euclidean path integral, we can understand the dynamics of the theory by regularising it on a spacetime lattice with appropriate boundary conditions for the fields. The circumference of the temporal circle, $\beta$, which is given by the product of lattice spacing ($\mathrm{a}$) and the total number of lattice sites $N_\tau$, gives us the information of the system at temperature $T = \beta^{-1}$. The fields on the lattice no longer stay continuous; rather, they live on discrete spacetime points. The derivative operators change to finite difference operators, and integration over continuous fields changes to summation over discrete lattice fields. To understand this in a quantum mechanical setup, we see that[2]

$$\phi(\tau) \to \phi_\tau, \qquad \frac{\partial \phi}{\partial \tau} \to \frac{\phi_{\tau+1} - \phi_\tau}{\mathrm{a}}, \qquad \int_0^\beta \to \mathrm{a} \sum_0^{N_\tau - 1}.$$

The fields can be simulated on a lattice using various numerical methods. The one which we have used throughout this thesis is the Monte Carlo method. We note that implementing fermions on a lattice is slightly non-trivial because of the *fermion doubling problem*. Instead of one fermion on the lattice, we get extra copies of fermions, which takes us away from our original setup. By naively discretizing the fermions, we get $2^d$ flavors of fermions, where $d$ is the number of spacetime dimensions of our theory. We need to take appropriate steps to remove the fermion doubling problem. We have used Wilson fermions in our supersymmetric quantum mechanics setup to counter fermion doubling, which is discussed in detail in the next chapter. Simulating gauge fields on a lattice is introduced in chapter 3.

Let us now move our focus toward simulating a field theory on a lattice. To simulate the fields on a lattice, we take the help of the hybrid Monte Carlo (HMC) method. For this, we deal with the Hamiltonian ($H$) of the theory, which requires us to introduce conjugate momenta ($p$) to the field $\phi$. The HMC method is described as follows:

- We can set the initial configurations of the field ($\phi$) either to be zero (*cold start*) or to be any random value (*hot start*).

- Draw momenta ($p$) from a Gaussian distribution.

---

[2]For a difference operator, we can take the forward, backward, or symmetric differences, depending on the problem.

- Save the initial values of fields $(\phi, p)$ so that, if required, we can access these values later.

- Update the fields using the leapfrog method for a specific trajectory length using Hamilton's equation of motion.

- The new configuration is either accepted or rejected with the help of the *Metropolis test*, which can be understood as follows:

    - If the Hamiltonian of the new configuration is less than the Hamiltonian of the old configuration, then the new configuration is accepted.
    - If the Hamiltonian of the new configuration is more than the Hamiltonian of the old configuration, then the new configuration is conditionally accepted. The condition can be understood as follows:
        * Generate a random number $u$ between 0 and 1.
        * Compute $\Delta H = H_{\text{old}} - H_{\text{new}}$, which is the change in action between the old and new configuration.
        * If $u < \exp(\Delta H)$, accept the new configuration.
        * If $u > \exp(\Delta H)$, reject the new configuration.

- If the step is rejected, load the old configuration again, which was initially saved.

- Repeat the steps of the HMC update for this new configuration.

Performing this procedure for a large number of steps, we get a Markov chain, which is a sequence of random field configurations

$$\phi^1 \rightarrow \phi^2 \rightarrow \phi^3 \rightarrow \phi^4 \rightarrow \cdots$$

Any member of this generated Markovian chain has memory only of its previous member. Hence, this chain with a very large length will approach the equilibrium configuration, after which the values of $\phi$'s will just fluctuate around the equilibrium value. The initial stage, when the members of the chain follow a certain trend towards the equilibrium value, is called *thermalization*. This data is discarded for any calculation as the configuration has not reached its equilibrium distribution. When the members of the chain start fluctuating around a particular value, that stage is called *thermalized* stage. Only the thermalized data is used for analyzing the data. The expectation value of the observable given in Eq. (1.3) becomes a statistical average over the thermalized data set as:

$$\langle \mathcal{O} \rangle = \frac{1}{N} \sum_{i=1}^{N} \mathcal{O}(\phi^i). \tag{1.10}$$

A larger sample size leads to a lesser statistical error on the estimated value of the observable as the size of the error goes like $\propto 1/\sqrt{N}$, where $N$ is the thermalized sample size. We should also take into account that while averaging, we should only consider the uncorrelated data from the generated chain. This can be done by calculating the *autocorrelation length* of the generated configurations and performing analysis after filtering out correlated data. The autocorrelation times can be reduced by using Hybrid Monte Carlo, in which the role of action in the transition probability gets replaced by the Hamiltonian. The conjugate momentum values, in that case, are sampled from a Gaussian distribution.

## 1.4 Supersymmetric Yang-Mills

Another important aspect of this thesis, apart from the non-perturbative setup, is dealing with supersymmetric theories. As discussed in the first section of this chapter, a non-perturbative study of supersymmetric Yang-Mills (SYM) theories is expected to play an important role in understanding physics beyond the Standard Model and quantum theories of gravity, such as string theory. Understanding these SYM theories can help us probe the dual gravity side using the *gauge/gravity* correspondence, based on original work by Maldacena, which connected $\mathcal{N} = 4$ SYM and Type-IIB supergravity on AdS$_5 \times$ S$_5$ [14]. Supersymmetric theories are based on the idea that every particle has a superpartner, which differs by half-integer spin from each other and is related via a supersymmetry transformation. The generator of supersymmetry transformation is called *supercharge*, $\mathcal{Q}$, which can be understood as

$$\mathcal{Q}|\text{Boson}\rangle = |\text{Fermion}\rangle, \qquad \mathcal{Q}|\text{Fermion}\rangle = |\text{Boson}\rangle. \tag{1.11}$$

SUSY is not an internal symmetry; rather, it is a spacetime symmetry. The two supercharges, $\mathcal{Q}$ and $\bar{\mathcal{Q}}$, anti-commute as follows

$$\{\mathcal{Q}, \bar{\mathcal{Q}}\} \propto P_\mu, \tag{1.12}$$

where $P_\mu$ is the generator for infinitesimal translations. We see that SUSY is naively broken on a lattice since we cannot realize infinitesimal translations on the lattice. However, we can still preserve a subgroup of the SUSY transformations on a lattice; we will discuss this in detail in the coming chapters. Hence, using the lattice setup, we can explore the strong coupling nature of the SYM theories and get information about the thermodynamics of the gravitational dual theories using gauge/gravity correspondence. The idea of AdS/CFT can be generalized to field theories without gravity in $d$ dimensions and theories with gravity in $d+1$ dimensions. We look to explore the non-conformal analog of this correspondence by studying theories in $d < 4$. One such example is discussed in chapter 5, which is $\mathcal{N} = (2, 2)$ SYM in two dimensions.

Another supersymmetric theory we discuss in chapter 4 is based on the BMN matrix model. Since a non-perturbative formalism of string theory is based on the Matrix theory, the BFSS matrix model (and its mass-deformed cousin, the BMN matrix model) are the ideal candidates to understand M theory. The BFSS model describes M theory in the light-cone gauge, in the large-$N$ limit. The BMN model describes a certain limit of the Type II string theory on a pp-wave background rather than the flat spacetime. Apart from broken SUSY on a lattice, one can encounter the issue of flat directions, which is discussed in detail in chapter 3. Another problem one encounter during lattice simulations, which can appear independent of whether fields are matrices or not, is the notorious *sign problem*. It is discussed in detail in the next section. In the next chapter, we will work with supersymmetric quantum mechanics on a lattice to test our simulation setup.

## 1.5 Witten Index and Sign Problem

The Witten index can be treated as a twisted partition function and can be defined as [15]

$$\tilde{\mathcal{Z}} = \mathcal{W} = \text{Tr}[(-1)^F e^{-\beta H}], \quad (1.13)$$

where $(-1)^F$ includes a minus sign for fermionic states of Hamiltonian $H$. The supersymmetry algebra in this Hamiltonian formalism is completed by nilpotent supercharges $\mathcal{Q}$ and $\bar{\mathcal{Q}}$ along with their commutation with Hamiltonian ($H$) i.e. $[\mathcal{Q}, H] = 0$. The Witten index corresponds to the sum of all differences between fermionic and bosonic energy states. If the Witten index is zero, then there is the same number of negative and positive contributions coming from configurations in the path integral, and there is no zero energy state, $|0\rangle$, which means SUSY is broken. If there are different numbers of normalizable zero energy states, then the Witten index is non-zero, indicating SUSY is intact in the theory. For a zero energy state, $\mathcal{Q}|0\rangle = \bar{\mathcal{Q}}|0\rangle = 0$. In Hamiltonian terms, we write it as $H_F|0\rangle_F = H_B|0\rangle_B = 0$. The multiplets of different bosonic and fermionic states are discussed in detail in the next chapter.

In our theory, when we integrate out fermions, the bosonic action picks contributions from the fermionic Pfaffian, which could have a fluctuating phase $e^{i\alpha}$. This can be understood by considering the following partition function for the action containing a boson $\phi$ and fermions $\psi$ and $\bar{\psi}$

$$\mathcal{Z} = \int \mathcal{D}\phi \mathcal{D}\bar{\psi}\mathcal{D}\psi \, e^{-S[\phi,\psi,\bar{\psi}]}. \tag{1.14}$$

On integrating out fermions, the partition function becomes

$$\mathcal{Z} = \int \mathcal{D}\phi \, \det(M(\phi)) \, e^{-S_B[\phi]}, \tag{1.15}$$

where $\det(M(\phi))$ is the fermionic determinant. If the fermionic determinant is not positive definite, then we encounter a sign problem in the numerical runs. The more this determinant switches its sign, the more dominant the sign problem will be. Here, in Eq. (1.15), we have a determinant instead of a Pfaffian because we have two fermionic fields instead of one. In SUSY broken case, the Witten index and hence the partition function is zero. Therefore, the observables defined in Eq. (1.3) becomes ill-defined. To counter this, we impose anti-periodic boundary conditions for fermions for which the partition function is not zero even in the SUSY broken case, and the simulation results can be treated as reliable. However, one still needs to be aware that this problem can still show up in certain regimes of the coupling.

Another drawback in Monte Carlo simulations is that they cannot be applied to systems with complex actions since the probability weight itself can be complex. This can be dealt with using other numerical methods, such as the complex Langevin method, but we will not discuss that as it is beyond the scope of this thesis.

# 2

# Probing Non-perturbative Supersymmetry Breaking through Lattice Path Integrals

*Contents of this chapter is partially based on:*

- Pub. [1]: N. S. Dhindsa and A. Joseph, "Probing Non-perturbative Supersymmetry Breaking through Lattice Path Integrals", Eur. Phys. J. Plus 137, **1155** (2022), arXiv:2011.08109 [hep-lat].

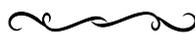

As discussed in the previous chapter, we can use supersymmetric quantum mechanics as a test bed to illustrate several properties of systems containing bosons and fermions. This work can also help us with a better understanding of the non-perturbative lattice setup and the numerical Monte Carlo algorithm before we shift our focus toward matrix models.

## 2.1 Supersymmetric Quantum Mechanics

Since Witten's seminal work [16], the idea of non-perturbative supersymmetry (SUSY) breaking has been investigated extensively in the literature. These investigations range from studying the properties of supersymmetric quantum mechanics to supersymmetric gauge theories in various spacetime dimensions [17–26]. This chapter investigates non-perturbative SUSY breaking in various quantum mechanics models by regularizing them on a Euclidean lattice. Supersymmetric quantum mechanics models have been the subject of thorough investigations in the context of various physical systems (see Ref. [27] for a review). For example, the supersymmetric anharmonic oscillator has been simulated on the lattice, by several groups, with great success (see Refs. [28–34]). In this chapter, we explore supersymmetric quantum mechanics with various types of superpotentials, including the shape-invariant potential and the interesting class of $PT$-symmetric potentials. After verifying the existing simulation results in the literature on the supersymmetric anharmonic oscillator, with the help of a lattice regularized action, and an efficient simulation algorithm, we use the same setup to probe SUSY breaking in models with three different types of potentials. They include a degree-five potential, a shape-invariant potential of Scarf I type, and a $PT$-invariant potential. The models discussed in this work were simulated using lattice regularized versions of the Euclidean path integrals. The Hybrid Monte Carlo (HMC) algorithm was used to update the field configurations in simulation time. We note that for general $PT$-symmetric models with complex actions, we can use the complex Langevin method to explore non-perturbative SUSY breaking. (See [35–37] for some recent results.)

The plan of this chapter is as follows. In this section, we continue the discussion of supersymmetric quantum mechanics in the continuum and on a lattice. In Sec. 2.2, we cross-check our simulation results with the ones existing in the literature for the supersymmetric anharmonic oscillator. In Sec. 2.3, we study the model with degree-five superpotential. Our simulation results indicate that SUSY is dynamically broken in this model, which agrees with the analytical result in Ref. [16]. In Sec. 2.4, we study the model with a particular type of shape-invariant potential known as the Scarf I potential. Our simulations indicate that this model also has an intact SUSY and thus confirms the numerical results given in Ref. [38]. In Sec. 2.5, we study the possibility of SUSY breaking in certain models with $PT$-invariant superpotentials. Our simulations point to intact SUSY in these models as well. The conclusions on $PT$-invariant models also agree with the recent simulation results obtained using the complex

Langevin method [36]. We provide conclusions and the possible future directions for this work in Sec. 2.6.

### 2.1.1 Continuum Theory

The Euclidean action of supersymmetric quantum mechanics has the off-shell form

$$S = \int_0^\beta d\tau \left( -\frac{1}{2}\phi \partial_\tau^2 \phi + \overline{\psi} \partial_\tau \psi - \frac{1}{2}B^2 + \overline{\psi} W''(\phi)\psi - BW'(\phi) \right). \quad (2.1)$$

Here, the integral is over a compactified time circle of circumference $\beta$ in Euclidean time. The fields $\phi$ and $B$ are bosonic, while $\overline{\psi}$ and $\psi$ are fermionic. They depend only on the Euclidean time variable $\tau$. The derivative with respect to $\tau$ is denoted as $\partial_\tau$. The superpotential $W(\phi)$ ultimately determines the interactions in the theory. The primes denote the derivatives of superpotential with respect to $\phi(\tau)$ at a given time $\tau$. It can be understood as $W'(\phi) = \frac{\partial W(\phi)}{\partial \phi(\tau)}\Big|_\tau$ and $W''(\phi) = \frac{\partial W'(\phi)}{\partial \phi(\tau)}\Big|_\tau$.

The above action is invariant under two supercharges, $Q$ and $\overline{Q}$. They obey the following algebra

$$\{Q, Q\} = 0, \quad \{\overline{Q}, \overline{Q}\} = 0, \quad \{Q, \overline{Q}\} = 2\partial_\tau. \quad (2.2)$$

The auxiliary field $B$ appearing in action can be integrated out using its equation of motion $B = -W'(\phi)$. Then the on-shell action takes the form

$$S = \int d\tau \left( -\frac{1}{2}\phi \partial_\tau^2 \phi + \overline{\psi} \partial_\tau \psi + \overline{\psi} W''(\phi)\psi + \frac{1}{2} \left[ W'(\phi) \right]^2 \right). \quad (2.3)$$

The supercharges $Q$ and $\overline{Q}$ act on the fields in the following way

$$Q\phi = \overline{\psi}, \quad Q\psi = -\partial_\tau \phi + W', \quad Q\overline{\psi} = 0, \quad (2.4)$$

and

$$\overline{Q}\phi = -\psi, \quad \overline{Q}\psi = 0, \quad \overline{Q}\overline{\psi} = \partial_\tau \phi + W'. \quad (2.5)$$

We can verify the algebra given in Eq. (2.2) by using the supersymmetric transformations provided in Eqs. (2.4) and (2.5). Two of the verification examples can be seen as such

$$\begin{aligned} \{Q,Q\}\phi &= (QQ + QQ)\phi, \\ &= 2QQ\phi, \\ &= 2Q\overline{\psi}, \\ &= 0, \end{aligned} \quad (2.6)$$

and

$$\begin{aligned} \{Q,\overline{Q}\}\phi &= (Q\overline{Q} + \overline{Q}Q)\phi, \\ &= -Q\psi + \overline{Q\psi}, \\ &= -(-\partial_\tau \phi + W') + (\partial_\tau \phi + W'), \\ &= 2\partial_\tau \phi. \end{aligned} \quad (2.7)$$

Similarly, this algebra can be tested for other fields in the theory; some identities will require the help of equations of motion for the proof.

To check whether the theory we are dealing with is inherently supersymmetric or not can be tested by a non-perturbative setup. SUSY breaking can be accounted for for many reasons. We can have a SUSY broken softly by explicitly adding SUSY breaking term. We will see an example of this in the coming chapters when we will add SUSY-breaking terms to counter flat directions. But in this chapter, we will check for whether the SUSY is broken spontaneously or not. Spontaneous supersymmetric breaking (SSB) happens when the system is supersymmetric, but its ground state is not supersymmetric. If the ground state is a singlet and the model is invariant under supersymmetric transformations, then SUSY is preserved in the model. If the ground state is degenerate and the model is invariant under supersymmetric transformations, that is the case of broken SUSY. In this chapter, we will refer to SSB as SUSY broken case.

Since SUSY is not broken explicitly in our models, we are interested in detecting the presence or absence of SUSY breaking that can arise from non-perturbative effects. A lattice formulation of quantum theory is inherently non-perturbative, and thus, we can detect SUSY breaking by simulating the model with the help of a suitable Monte Carlo algorithm.

The partition function, as discussed in the previous chapter, of the model is

$$\mathcal{Z} = \int \mathcal{D}\phi \mathcal{D}\overline{\psi}\mathcal{D}\psi \, e^{-S[\phi,\psi,\overline{\psi}]}, \tag{2.8}$$

with periodic temporal boundary conditions for all the fields.

In the Hamiltonian formalism, the supersymmetry algebra can be realized as $\{Q, \bar{Q}\} = 2H$. The Hamiltonian operator takes the form [39]

$$H = \frac{1}{2}\begin{pmatrix} H_B & 0 \\ 0 & H_F \end{pmatrix} = \frac{1}{2}\begin{pmatrix} -\partial_x^2 + W'^2 - W'' & 0 \\ 0 & -\partial_x^2 + W'^2 + W'' \end{pmatrix}, \tag{2.9}$$

where the upper component is for the bosonic sector, and the lower component is for the fermionic sector.

Consider the case when SUSY is preserved in the system. Then the Hamiltonian $H$, corresponding to the Lagrangian in Eq. (2.1), with energy levels $E_n$, $n = 0, 1, 2, \ldots$, will have the ground state energy $E_0 = 0$ which is a bosonic state $|b_0\rangle$ i.e. $|b_0\rangle = 0$. The normalized bosonic and fermionic excited states

$$|b_{n+1}\rangle = \frac{1}{\sqrt{2E_{n+1}}}\bar{Q}|f_n\rangle, \quad |f_n\rangle = \frac{1}{\sqrt{2E_{n+1}}}Q|b_{n+1}\rangle \tag{2.10}$$

form a SUSY multiplet, with $n = 0, 1, 2, \ldots$, satisfying the algebra Eq. (2.2). In the above, $|b_0\rangle$ is the ground state of the system. Assuming that the states $|b_n\rangle$ and $|f_n\rangle$ have the fermion number charges $F = 0$ and $F = 1$, respectively, when periodic temporal boundary conditions are imposed for both the bosonic and fermionic fields, the partition function Eq. (2.8) is equivalent to the Witten index $\mathcal{W}$ [15]. We can see that

$$\begin{aligned}\tilde{\mathcal{Z}} &\equiv \mathcal{W} = \text{Tr}\left[(-1)^F e^{-\beta H}\right], \\ &= \langle b_0|b_0\rangle + \sum_{n=0}^{\infty}\left[(\langle b_{n+1}|b_{n+1}\rangle - \langle f_n|f_n\rangle) e^{-\beta E_{n+1}}\right]\end{aligned} \tag{2.11}$$

does not vanish due to the existence of a normalizable ground state. This, in turn, makes the normalized expectation values of observables well-defined.

In the SUSY broken case, we end up in a not-so-trivial situation. The Hamiltonian $H$ corresponding to the Lagrangian in Eq. (2.1) has a positive ground state energy ($0 < E_0 < E_1 <$

$E_2 < \cdots$). The SUSY multiplet with $n = 0, 1, 2, \ldots$ is defined as

$$|b_n\rangle = \frac{1}{\sqrt{2E_n}} \bar{Q} |f_n\rangle, \quad |f_n\rangle = \frac{1}{\sqrt{2E_n}} Q |b_n\rangle. \tag{2.12}$$

Differently from the unbroken SUSY case, when SUSY is broken, the supersymmetric partition function

$$\begin{aligned}
\tilde{\mathcal{Z}} &\equiv \mathcal{W} = \text{Tr}\left[ (-1)^F e^{-\beta H} \right] \\
&= \sum_{n=0}^{\infty} \left[ (\langle b_n | b_n \rangle - \langle f_n | f_n \rangle) e^{-\beta E_n} \right]
\end{aligned} \tag{2.13}$$

vanishes due to the cancellation between bosonic and fermionic states. Now if we define the expectation value of an observable as

$$\langle \mathcal{O} \rangle = \frac{1}{\mathcal{Z}} \int \mathcal{D}\phi \, \mathcal{O}(\phi) e^{-S(\phi)}, \tag{2.14}$$

and in SUSY broken case partition function vanishes; as a consequence, the expectation values of observable as defined in Eq. (2.14) are ill-defined.

In order to avoid this difficulty, we can consider the system at non-zero temperature to examine the question of SUSY breaking. Thermal boundary conditions explicitly break SUSY. However, we can take the zero-temperature limit, and in this limit, if $E_0 > 0$, we infer that non-perturbative SUSY breaking occurs in the model.

We can make use of three different observables to probe SUSY breaking. These observables are the Ward identities, the expectation value of the action, and the expectation value of the first derivative of the superpotential. In a model with unbroken SUSY, the Ward identities should fluctuate around zero in the middle region of the lattice. The expectation value of the action should behave in a certain way (which we will see later) when SUSY is preserved. The expectation value of the first derivative of the superpotential should be zero in a model with intact SUSY and non-zero otherwise.

We can derive a Ward identity by rewriting the expectation value of an observable, say, $\mathcal{O}(\phi)$ by considering the infinitesimal transformations $\phi \to \phi' = \phi + \delta\phi$ with $\mathcal{D}\phi' = \mathcal{D}\phi$. Under these transformations $\langle \mathcal{O} \rangle$ becomes

$$\begin{aligned}\langle \mathcal{O} \rangle &= \frac{1}{\mathcal{Z}} \int \mathcal{D}\phi' \mathcal{O}(\phi') e^{-S(\phi')}, \\ &= \frac{1}{\mathcal{Z}} \int \mathcal{D}\phi \mathcal{O}(\phi) e^{-S(\phi)}[1 - \delta S(\phi)] + \frac{1}{\mathcal{Z}} \int \mathcal{D}\phi \delta \mathcal{O}(\phi) e^{-S(\phi)}[1 - \delta S(\phi)]. \end{aligned} \quad (2.15)$$

In the above, we have expanded the exponential up to the first order. Upon neglecting the $\delta S \delta \mathcal{O}$ term we get

$$\langle \mathcal{O} \rangle = \langle \mathcal{O} \rangle - \frac{1}{\mathcal{Z}} \int \mathcal{D}\phi \mathcal{O}(\phi) \delta S(\phi) e^{-S(\phi)} + \frac{1}{\mathcal{Z}} \int \mathcal{D}\phi \delta \mathcal{O}(\phi) e^{-S(\phi)}. \quad (2.16)$$

Thus we are left with $\langle \mathcal{O} \delta S \rangle = \langle \delta \mathcal{O} \rangle$. Since the action is invariant under the infinitesimal transformation $\delta$, we must have $\langle \mathcal{O} \delta S \rangle = 0$. This results in the Ward identity

$$\langle \delta \mathcal{O} \rangle = 0. \quad (2.17)$$

If this relation is respected in our numerical simulations, then we have a SUSY-preserving theory.

### 2.1.2 Lattice Theory

Let us consider the path integral quantized version of the model, discretized on a one-dimensional Euclidean lattice. The different variables in the above action are made dimensionless in terms of lattice spacing $\mathrm{a}$. The mass dimensions of different variables are given as: $[\partial_\tau] = 1, [\phi] = -1/2, [W'] = 1/2, [M] = 1, [\psi] = 0$. Hence the fields in discretized format are made dimensionless as $\mathrm{a}^{-1/2}\phi \to \phi$, $\psi \to \psi$, $\mathrm{a}^{1/2} W' \to W'$. The lattice has $N_\tau$ number of equally spaced sites with the lattice spacing $\mathrm{a} = \beta N_\tau^{-1}$, where $\beta$ is inverse temperature. The lattice action has the form

$$S = S_B + \sum_{ij} \overline{\psi}_i M_{ij} \psi_j, \quad (2.18)$$

where the bosonic part of the action is

$$S_B = \sum_{ij} \left(-\frac{1}{2}\phi_i D_{ij}^2 \phi_j\right) + \frac{1}{2}\sum_i W'_i W'_i. \quad (2.19)$$

The indices $i$ and $j$ represent the lattice sites, and they run from $0$ to $N_\tau - 1$. The fermion operator is denoted as $M$, and the elements of this matrix, $M_{ij}$, connect the sites $i$ and $j$. $W_i$ denotes the superpotential at site $i$.

Let us look at the Ward identities on the lattice. Consider the following two supersymmetry transformations

$$\begin{aligned}
\delta_1 \phi_i &= \overline{\psi}_i \epsilon, & \delta_2 \phi_i &= \bar{\epsilon} \psi_i, \\
\delta_1 \psi_i &= (D_{ik}\phi_k - W'_i)\epsilon, & \delta_2 \psi_i &= 0, \\
\delta_1 \overline{\psi}_i &= 0, & \delta_2 \overline{\psi}_i &= -\bar{\epsilon}(D_{ik}\phi_k + W'_i),
\end{aligned} \quad (2.20)$$

where $\epsilon$ and $\bar{\epsilon}$ denote the infinitesimal Grassmann odd variables that generate supersymmetry.

Let us take $\mathcal{O}_{ij} = \phi_i \overline{\psi}_j$. Upon using the transformation $\delta_2$ in the expression for the Ward identity, we get

$$\langle \delta_2 \mathcal{O}_{ij} \rangle = 0. \qquad (2.21)$$

This gives

$$\langle (\phi_i \cdot \delta_2 \overline{\psi}_j + \delta_2 \phi_i \cdot \overline{\psi}_j) \rangle = 0. \qquad (2.22)$$

Applying the supersymmetry transformation $\delta_2$ it becomes

$$\langle \phi_i(D_{jk}\phi_k + W'_j) + \overline{\psi}_j \psi_i \rangle = 0. \qquad (2.23)$$

We note that when we apply the transformation $\delta_1$ on the same observable $\phi_i \overline{\psi}_j$, our result vanishes trivially. Taking $\widetilde{\mathcal{O}}_{ij} = \phi_i \psi_j$ we have

$$\langle \delta_1 \widetilde{\mathcal{O}}_{ij} \rangle = 0. \qquad (2.24)$$

Expanding the terms

$$\langle (\phi_i \cdot \delta_1 \psi_j + \delta_1 \phi_i \cdot \psi_j) \rangle = 0. \qquad (2.25)$$

Upon simplification, this becomes

$$\langle \phi_i(D_{jk}\phi_k - W'_j) + \overline{\psi}_i \psi_j \rangle = 0. \qquad (2.26)$$

Equations. (2.23) and (2.26) are our Ward identities. Note that the Ward identities are functions of sites $i$ and $j$. In the simulations, we fix $i = 0$ and $j = n$ to monitor the Ward identities. We have

$$w_1(n) \equiv \langle \phi_0(D_{nk}\phi_k + W'_n) \rangle + \langle \overline{\psi}_n \psi_0 \rangle, \tag{2.27}$$

$$w_2(n) \equiv \langle \phi_0(D_{nk}\phi_k - W'_n) \rangle + \langle \overline{\psi}_0 \psi_n \rangle. \tag{2.28}$$

If SUSY is preserved in the model, these quantities should fluctuate around 0, at least in the middle region of the lattice, where the effects from higher excited states and other lattice artifacts are the lowest.

Let us try to show an example of how these supersymmetries keep the action invariant. For simplification purposes, we will start from continuum action as

$$\delta_2 S = \int d\tau \, \delta_2 \left( -\frac{1}{2}\phi \partial_\tau^2 \phi + \overline{\psi} \partial_\tau \psi + \overline{\psi} W''(\phi)\psi + \frac{1}{2}\left[W'(\phi)\right]^2 \right). \tag{2.29}$$

The dynamic $\phi$ term in action can be integrated and takes the form $\frac{1}{2}(\partial_\tau \phi)^2$. Now operating supersymmetry transformation on the resultant equation

$$\begin{aligned}
\delta_2 S &= \int d\tau \, \delta_2 \left( \frac{1}{2}(\partial_\tau \phi)^2 + \overline{\psi} \partial_\tau \psi + \overline{\psi} W''(\phi)\psi + \frac{1}{2}\left[W'(\phi)\right]^2 \right), \\
&= \int d\tau \, \left( (\partial_\tau \phi) \delta_2 (\partial_\tau \phi) + \delta_2 (\overline{\psi} \partial_\tau \psi + \overline{\psi} W''(\phi)\psi) + W'(\phi) \delta_2 (W'(\phi)) \right), \\
&= \int d\tau \, \left( \partial_\tau \phi \, \overline{\epsilon} \, \partial_\tau \psi - \overline{\epsilon}(\partial_\tau \phi + W')(\partial_\tau \psi + W''(\phi)\psi) + W'(\phi)W''(\phi)\overline{\epsilon}\psi \right). \tag{2.30}
\end{aligned}$$

One term in the above equation is not listed, which is operating transformation on the second derivative of superpotential, as it gives $\psi$ after operating supersymmetry and $\psi^2 = 0$. Now let us expand the above equation

$$\begin{aligned}
\delta_2 S &= \int d\tau \, \overline{\epsilon} \left( \partial_\tau \phi \, \partial_\tau \psi - (\partial_\tau \phi + W')(\partial_\tau \psi + W''(\phi)\psi) + W'(\phi)W''(\phi)\psi \right), \\
&= \int d\tau \, -\overline{\epsilon} \left( (\partial_\tau \phi W''(\phi)\psi + W' \partial_\tau \psi \right), \\
&= \int d\tau \, \partial_\tau \left( -\overline{\epsilon} W'(\phi)\psi \right). \tag{2.31}
\end{aligned}$$

Now, this is a total derivative term in the above equation. Hence when we discretize and impose periodic boundary conditions, the term vanishes, keeping action invariant under supersymmetry transformation, $\delta_2 S = 0$.. The same can be checked with other transformations. If we impose thermal boundary conditions, then this equation does not vanish, and we have to take zero temperature limit and continuum limit to preserve supersymmetry.

For supersymmetric quantum mechanics on the lattice, we can use a simple scaling argument (see Ref. [40]) to show that the expectation value of the total action

$$\langle S \rangle_{\text{exact}} = \frac{1}{2} N_{\text{d.o.f.}}, \tag{2.32}$$

where $N_{\text{d.o.f.}}$ is the total number of degrees of freedom. For a lattice with $N_\tau$ sites, there are two degrees of freedom per site, and thus there is a total of $2N_\tau$ degrees of freedom in the lattice theory. Thus we have

$$\langle S \rangle_{\text{exact}} = \frac{1}{2} N_{\text{d.o.f.}} = \frac{1}{2}(2N_\tau) = N_\tau. \tag{2.33}$$

In our simulations, we will use

$$\Delta S \equiv \langle S \rangle_{\text{exact}} - \langle S \rangle = N_\tau - \langle S \rangle \tag{2.34}$$

as an indicator of SUSY breaking.

In order to simulate the fermionic sector of the theory, we replace the fermions with pseudo-fermions $\chi$ [41]. Then the action in Eq. (2.18) takes the form

$$S = S_B + S_F. \tag{2.35}$$

Expressing the bosonic action $S_B$, given in Eq. (2.19), with the help of the symmetric difference operator defined in Eq. (2.44), we get

$$S_B = \sum_i \left[ -\frac{1}{8} \left( \phi_i \phi_{i+2} + \phi_i \phi_{i-2} - 2\phi_i^2 \right) + \frac{1}{2} W_i' W_i' \right]. \tag{2.36}$$

Simulating fermions on the lattice can be a very computationally challenging task. If we look at the partition function given in Eq. (2.8) for the action given in Eq. (2.3). The form of the partition function is given by

$$\mathcal{Z} = \int \mathcal{D}\phi \mathcal{D}\overline{\psi} \mathcal{D}\psi \, e^{-S_B - S_F}, \quad (2.37)$$

$$= \int \mathcal{D}\phi \mathcal{D}\overline{\psi} \mathcal{D}\psi \, e^{-S_B - (\overline{\psi}\partial_\tau \psi + \overline{\psi} W''(\phi)\psi)}, \quad (2.38)$$

$$= \int \mathcal{D}\phi \mathcal{D}\overline{\psi} \mathcal{D}\psi \, e^{-S_B - \overline{\psi} M \psi}. \quad (2.39)$$

The fermionic action is bilinear in fermionic fields, so the integration over the fermions can be carried out, and the partition function becomes

$$\mathcal{Z} = \int \mathcal{D}\phi \, \det(M) \, e^{-S_B}. \quad (2.40)$$

Now if we consider the effect of fermions on the numerical runs, we will need to compute this fermionic determinant for every field configuration at every lattice site. This computation, even for smaller lattices, is very expensive. To reduce the computational cost, quenched approximations have been used by fixing the fermionic determinant as constant. But as the fermionic determinant itself is a function of the bosonic fields, keeping it constant by updating the bosonic fields with a standard algorithm is highly non-trivial. Hence we have to work with dynamical fermions. The standard way of dynamically updating the fermionic determinant is by introducing a Grassmann even field known as the pseudo-fermion (we denote the pseudo-fermion as $\chi$). We can write $\det(M)$ as $\sqrt{\det(M^T M)}$ since the determinant of the fermionic matrix is real. Then we have

$$\sqrt{\det(M^T M)} = \int \mathcal{D}\chi \, e^{-\chi^T (M^T M)^{-1} \chi}. \quad (2.41)$$

On the lattice, the fermionic part of the action takes the following form in terms of the pseudo-fermion field

$$S_F = \sum_{ij} \left( \chi_i (M^T M)^{-1}_{ij} \chi_j \right). \quad (2.42)$$

The inverse of the matrix $M^T M$ is computed using the conjugate gradient method when it acts on a vector '$s$' and iteratively solves $(M^T M)s = \chi$.

To probe SUSY breaking, we can also study the expectation value of the first derivative of the superpotential. This quantity is related to the auxiliary field $B$ through the relation $B = -W'$. We have $\langle W' \rangle = 0$ when SUSY is preserved and non-zero otherwise. (See the work by Kuroki and Sugino [42] for more details.)

We can express the elements of the fermionic matrix ($M = \partial_\tau + W''$) as

$$M_{ij} = D_{ij} + W''_{ij}, \qquad (2.43)$$

where $D_{ij}$ is taken as symmetric difference operator

$$D_{ij} \equiv \frac{1}{2} \left( \delta_{j,i+1} - \delta_{j,i-1} \right). \qquad (2.44)$$

It is well known that for supersymmetric theories on a lattice, the presence of fermion doublers breaks SUSY. We could include a Wilson-mass term in the model to remove the doublers [43] and get the expected target theory in the continuum. The Wilson-mass matrix $K_{ij}$ has the following form

$$K_{ij} = m\delta_{ij} - \frac{1}{2} \left( \delta_{i,j+1} + \delta_{i,j-1} - 2\delta_{ij} \right). \qquad (2.45)$$

For a particular superpotential derivative of the form $W'(\phi) = m\phi + g\phi^n$, we replace mass $m$ by Wilson mass matrix on the lattice, and the superpotential derivative becomes

$$W'_i = \sum_j K_{ij}\phi_j + g\phi_i^n. \qquad (2.46)$$

The second derivative from this equation takes the form

$$\begin{aligned} W''_{ij} &= K_{ij} + ng\phi_i^{n-1}\delta_{ij}, \\ W''_{ij} &= m\delta_{ij} - \frac{1}{2} \left( \delta_{i,j+1} + \delta_{i,j-1} - 2\delta_{ij} \right) + ng\phi_i^{n-1}\delta_{ij}. \end{aligned} \qquad (2.47)$$

Put Eq. (2.47) and Eq. (2.44) in fermionic matrix equation (2.43)

$$\begin{aligned} M_{ij} &= \frac{1}{2} \left( \delta_{j,i+1} - \delta_{j,i-1} \right) + m\delta_{ij} - \frac{1}{2} \left( \delta_{i,j+1} + \delta_{i,j-1} - 2\delta_{ij} \right) + ng\phi_i^{n-1}\delta_{ij}, \\ M_{ij} &= \left( 1 + m + ng\phi_i^{n-1} \right) \delta_{ij} - \delta_{i,j+1}. \end{aligned} \qquad (2.48)$$

For the simulations, we will use a relatively more efficient algorithm, the Hybrid Monte Carlo (HMC) algorithm. As part of this algorithm, we take $p$ and $\pi$ as the momenta conjugate to the fields $\phi$ and $\chi$, respectively. Then the Hamiltonian of the system takes the form

$$H = \frac{1}{2} \sum_i (p_i^2 + \pi_i^2) + S_B + S_F. \tag{2.49}$$

This Hamiltonian is then evolved using a discretized version of Hamilton's equations in a fictitious time $\tau$, using a small step size $\epsilon$. At the end of the evolution, we get a new Hamiltonian $H'$. We then accept or reject $H'$ using the Metropolis test. (See Refs. [12, 44] for more details on the algorithm.)

## 2.2 Supersymmetric Anharmonic Oscillator

Let us consider the superpotential

$$W(\phi) = \frac{1}{2} m \phi^2 + \frac{1}{4} g \phi^4, \tag{2.50}$$

where $m$ is the mass, and $g$ is the coupling constant. In Ref. [28], it was concluded, using Monte Carlo simulations, that SUSY was preserved in this model. As a cross-check of our simulation code, we will reproduce the results.

The superpotential at a lattice site $i$ takes the following form

$$\begin{aligned} W'_i &= \sum_{j=0}^{N_\tau - 1} K_{ij} \phi_j + g \phi_i^3 \\ &= m \phi_i + \phi_i - \frac{1}{2} (\phi_{i-1} + \phi_{i+1}) + g \phi_i^3. \end{aligned} \tag{2.51}$$

The fermionic matrix, introduced in Eq. (2.43), becomes

$$M_{ij} = \left(1 + m + 3 g \phi_i^2\right) \delta_{ij} - \delta_{i,j+1}. \tag{2.52}$$

To understand how the boundary condition affects this fermionic matrix, let us take an example of the following $3 \times 3$ matrix.

$$M = \begin{bmatrix} 1+m+3gx_0^2 & 0 & -s \\ -1 & 1+m+3gx_1^2 & 0 \\ 0 & -1 & 1+m+3gx_2^2 \end{bmatrix}, \quad (2.53)$$

where $'s'$ is used as a factor for twisting the boundary values. In periodic boundary conditions, take $s = 1$, and in anti-periodic boundary conditions, take $s = -1$. The determinant of this matrix is given as

$$|M| = \prod_{i=1}^{3} \left(1 + m + 3gx_i^2\right) - s. \quad (2.54)$$

Hence with anti-periodic boundary conditions, this determinant from Eq. (2.54) is given by $\prod_{i=1}^{3} \left(1 + m + 3gx_i^2\right) + 1$, which makes sure that it stays positive in the simulations, which is not the case with periodic boundary conditions.

In the simulations we used the following dimensionless variables: $m = m_{\text{phys}} a$, $g = g_{\text{phys}} a^2$, and $\phi = \phi_{\text{phys}} a^{-1/2}$. We performed the simulations for $N_\tau = 16, 32,$ and $48$ with inverse temperature $\beta = 0.5, 1.0, 1.5,$ and $2.0$. We keep $m_{\text{phys}} = 10, g_{\text{phys}} = 100$ constant in the runs by changing $N_\tau$ and hence moving towards continuum.

Let us look at the time series plot for various observables for one set of simulations. It can be clearly seen in this plot (Fig. 2.1) that all observables behave nicely over the Monte Carlo time. The plot shown is for a subset of total runs.

In Fig. 2.2 (left) $\langle W' \rangle$ per site against $N_\tau$ is shown for various $\beta$ values. The data fluctuate around zero within error bars indicating that the model has intact SUSY. A linear $\beta \to \infty$ fit to the $N_\tau \to \infty$, extrapolated data gives us $-0.0036(27)$. In Fig. 2.2 (right) we show $\Delta S$ per site against $N_\tau$ for various $\beta$ values. A linear $\beta \to \infty$ fit to the $N_\tau \to \infty$, extrapolated data gives us $0.0032(26)$. Thus we conclude that SUSY is preserved in the model.

In Fig. 2.3, we show the simulation results for Ward identities. They show small fluctuations around zero in the middle regions of the lattice. Note that the simulation data are contaminated due to excited states and other lattice artifacts around the edge regions of the lattice. Thus we do not include them in the fits. To be on the conservative side, we discarded data in the first and last $N_\tau/4$ regions. We note that the strength of the fluctuations tends to reduce as $\beta$ is increased (temperature is reduced). A linear $\beta \to \infty$ fit for the averaged value of Ward

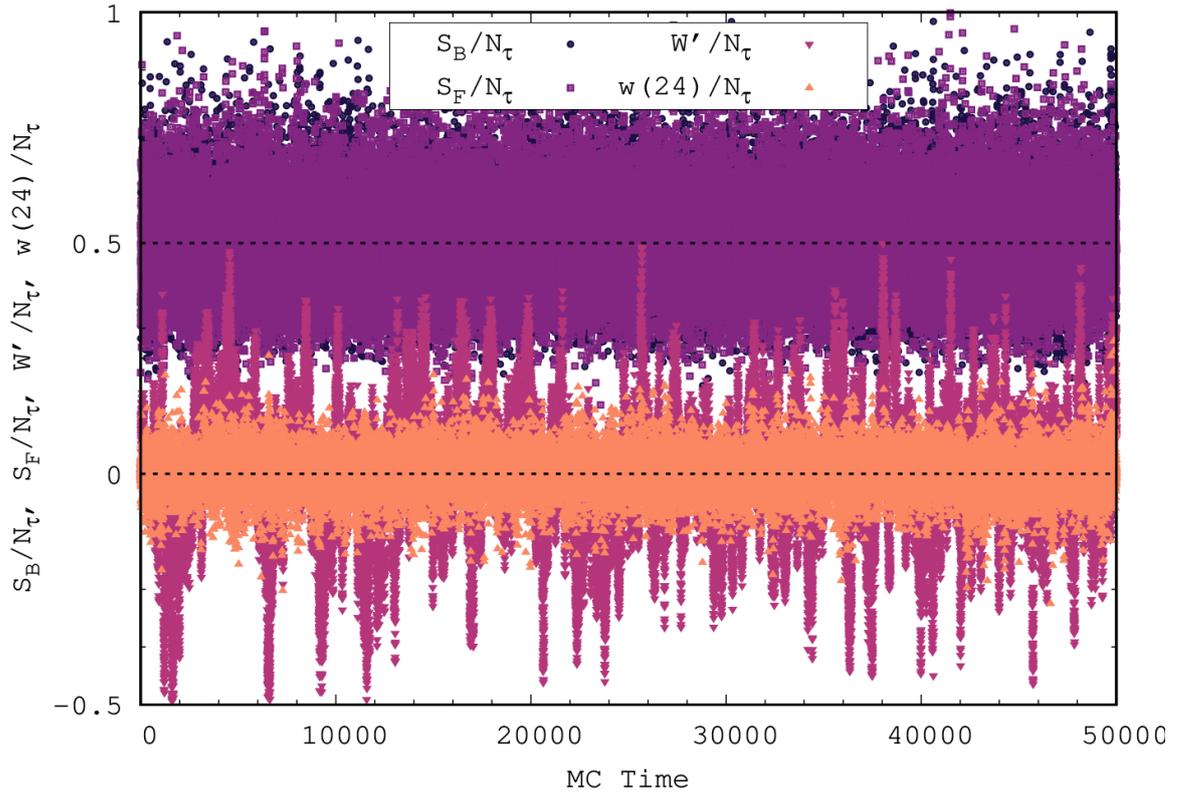

FIGURE 2.1: Time series plot for normalized bosonic action, fermionic action, the derivative of superpotential, and one of the ward identities at the 24th lattice site for degree four superpotential with $N_\tau = 48, \beta = 0.5, m_{\text{phys}} = 10,$ and $g_{phys} = 100$. It can be clearly seen that fermionic and bosonic action values normalized by $N_\tau$ fluctuate around $0.5$, which represents intact SUSY. Ward identity at a lattice site at the center of the lattice and the derivative of superpotential fluctuates around $0$, which confirms that SUSY is preserved in this model.

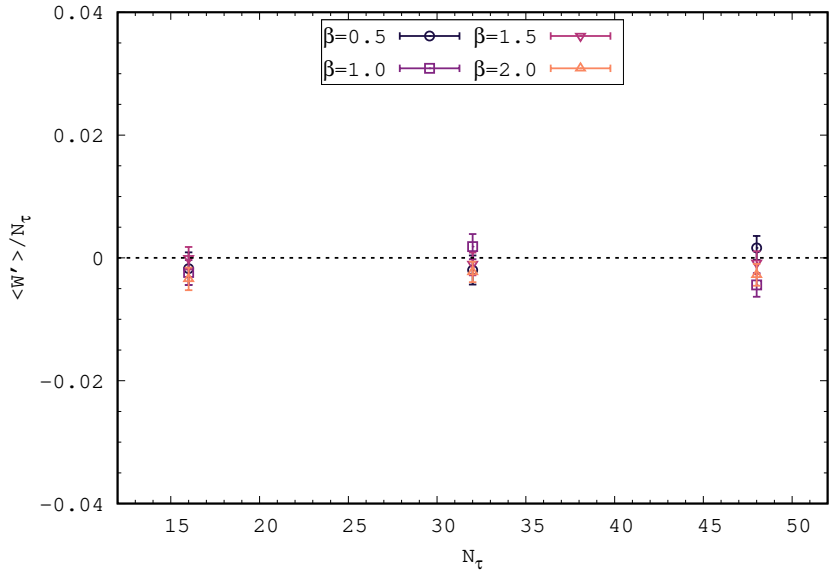

(a) $\langle W' \rangle$ per site against $N_\tau$.

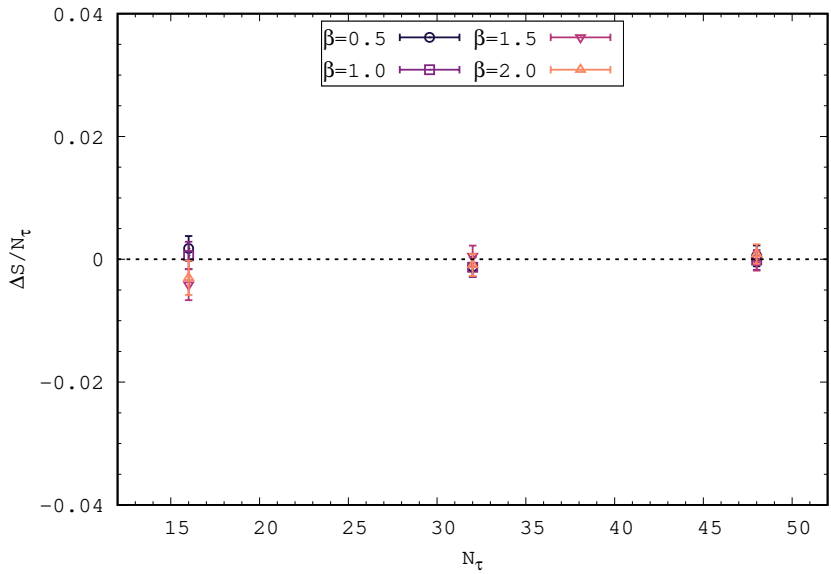

(b) $\Delta S$ per site against $N_\tau$.

FIGURE 2.2: Supersymmetric anharmonic oscillator. (Top) The expectation value of $\langle W' \rangle$ per site against $N_\tau$. (Bottom) $\Delta S$ per site against $N_\tau$. We used $\beta = 0.5, 1.0, 1.5,$ and $2.0$ with $m_{\text{phys}} = 10$ and $g_{\text{phys}} = 100$.

identity in the middle region after discarding first and last $N_\tau/4$ regions gives us $-0.019(2)$ for Fig. 2.3(a) and $0.008(2)$ for Fig. 2.3(b). Here also, the data suggest that SUSY is preserved in the model.

Taking into account the combined results shown in Figs. 2.2 and 2.3, we conclude that SUSY is preserved in the supersymmetric anharmonic oscillator.

## 2.3 Model with Odd-Degree Superpotential

In this section, we consider a model with degree-five superpotential. According to Ref. [20], SUSY is dynamically broken in this model. Our goal is to cross-check this result numerically.

Our lattice prescription gives the following form for the derivative of the superpotential

$$\begin{aligned} W'_i &= \sum_{j=0}^{N_\tau-1} K_{ij}\phi_j + g\phi_i^4 \\ &= m\phi_i + \phi_i - \frac{1}{2}\left(\phi_{i-1} + \phi_{i+1}\right) + g\phi_i^4. \end{aligned} \qquad (2.55)$$

The dimensionless parameters are $m = m_{\text{phys}}a$, $g = g_{\text{phys}}a^{5/2}$ and $\phi = \phi_{\text{phys}}a^{-1/2}$. We keep $m_{\text{phys}} = 10, g_{\text{phys}} = 100$ constant in the runs by changing $N_\tau$ and hence moving towards continuum.

In Fig. 2.4 (left) $\langle W' \rangle$ per site against $N_\tau$ is shown for various $\beta$ values. It is non-vanishing for all $\beta$ values and deviates away from zero as $\beta$ (temperature) is increased (decreased). In Fig. 2.4 (right), we show the plot of $\Delta S$ per site against $N_\tau$ for various $\beta$ values. The data do not fluctuate around zero. As $\beta$ is increased $\Delta S$ increases. Thus we conclude that SUSY is broken in this model.

The simulation results for Ward identities are shown in Fig. 2.5. We show the Ward identity Eq. (2.27) on the left panel. We see that the fluctuations are reducing in the middle region of the lattice as $\beta$ is increased. A linear $\beta \to \infty$ fit to the averaged data in the middle region (from $N_\tau/4$ to $3N_\tau/4$) gives us $-0.002(3)$. It is not possible to conclude whether SUSY is broken or not from this data alone. On the right panel, we show the Ward identity Eq. (2.28). It has large fluctuations around zero in the middle region of the lattice, and it does not diminish as $\beta$ is increased (temperature is decreased). A linear $\beta \to \infty$ fit to the averaged data in the middle

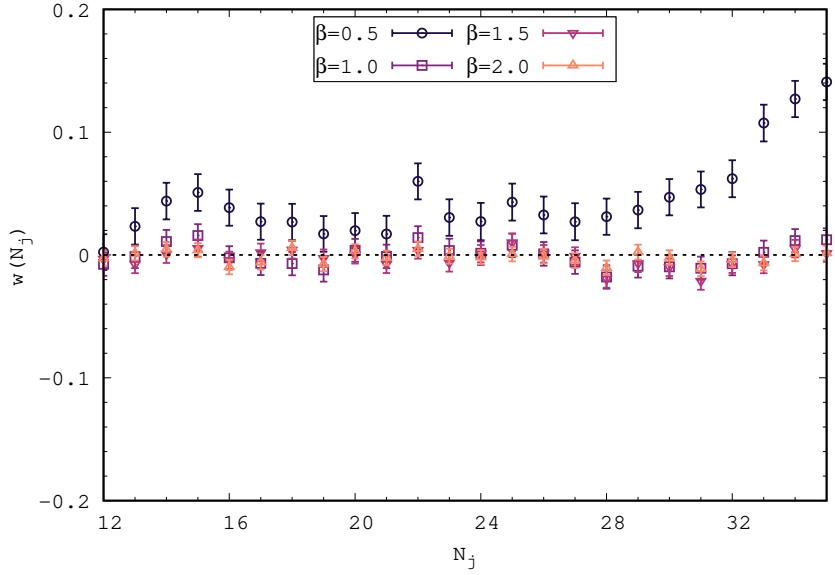

(a) Ward identity given by Eq. (2.27).

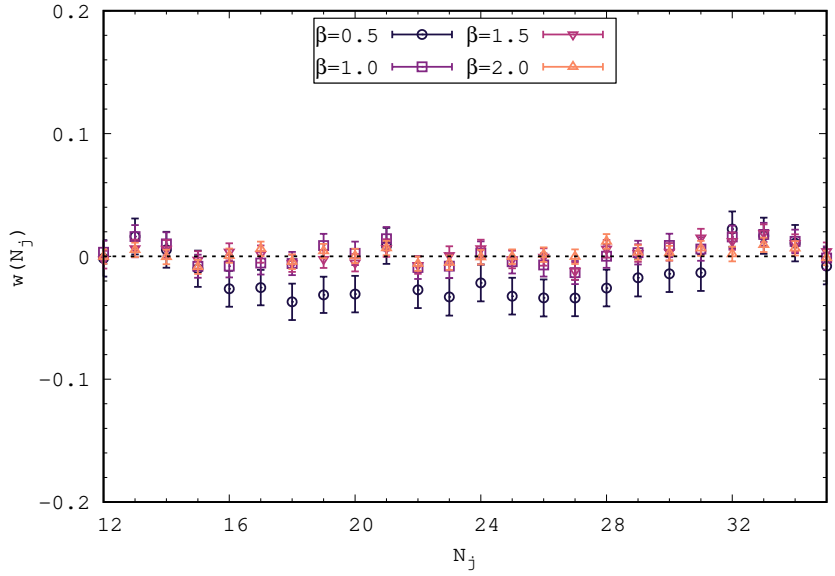

(b) Ward identity given by Eq. (2.28).

FIGURE 2.3: Supersymmetric anharmonic oscillator. Ward identities for different $\beta$ values on a lattice with $N_\tau = 48$ with $m_{phys} = 10$ and $g_{phys} = 100$. As $\beta$ is increased (temperature is decreased), the data points approach closer to zero in the middle region of the lattice, as expected for a theory with intact SUSY.

region (from $N_\tau/4$ to $3N_\tau/4$) gives us $-0.044(3)$. Though the Ward identity fits suggest an intact SUSY, the fluctuations in the Ward identities, in this case, are higher as compared to the degree-four superpotential (anharmonic oscillator) case.

Taking into consideration the overall trend from Figs. 2.4 and 2.5, we conclude that SUSY is broken in the model with degree-five superpotential.

## 2.4   Model with Scarf I Superpotential

This section shows the simulation results for the model with a shape-invariant potential, known as the Scarf-I superpotential [45, 46].

The potential has the form

$$W'(\phi) = A \tan(\alpha\phi) - B \sec(\alpha\phi), \quad -\frac{\pi}{2} \leq \alpha\phi \leq \frac{\pi}{2}, \tag{2.56}$$

where $A > B \geq 0$ and $\alpha > 0$.

We will focus on the case $B = 0$. Then

$$W'(\phi) = \lambda\alpha \tan(\alpha\phi), \quad -\frac{\pi}{2} \leq \alpha\phi \leq \frac{\pi}{2}. \tag{2.57}$$

The parameter $\alpha$ has the dimension of the square root of energy, and $\lambda$ is a dimensionless coupling.

This model was studied recently by Kadoh and Nakayama in Ref. [38], using a direct computational approach based on transfer-matrix formalism. It was shown there that SUSY was preserved in this model.

With this potential, we use the dimensionless parameters, $\alpha = \alpha_{\text{phys}} a^{1/2}$, $\lambda = \lambda_{\text{phys}}$ and $\phi = \phi_{\text{phys}} a^{-1/2}$ on the lattice. The simulations are performed for $\lambda_{\text{phys}} = 10$ and $\alpha_{\text{phys}} = \sqrt{60}$.

At a lattice site $k$ the derivative of the superpotential has the form $W'_k = \lambda\alpha \tan(\alpha\phi_k)$.

In Fig. 2.6 (left) we show $W'$ per site against $N_\tau$. As $\beta$ is increased (temperature is decreased), this observable approaches zero (data points with triangle symbols), suggesting that

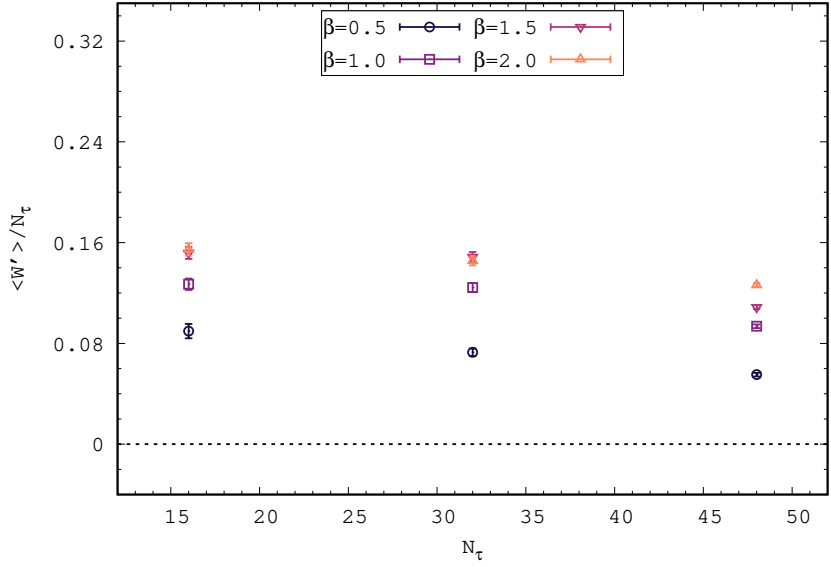

(a) $\langle W' \rangle$ per site against $N_\tau$.

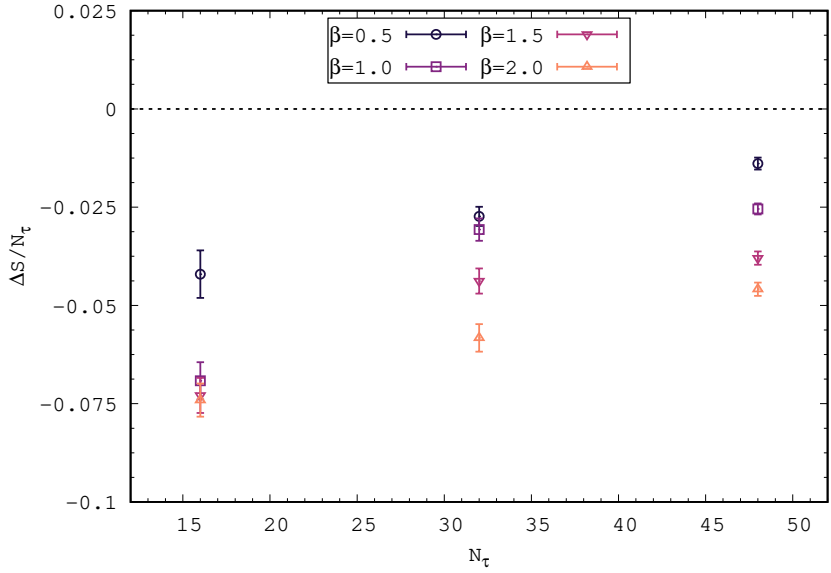

(b) $\Delta S$ per site against $N_\tau$.

FIGURE 2.4: Model with degree-five superpotential. (Top) The expectation value of $\langle W' \rangle$ per site against $N_\tau$. (Bottom) $\Delta S$ per site against $N_\tau$. We used $\beta = 0.5, 1.0, 1.5$, and $2.0$ with $m_{\text{phys}} = 10$ and $g_{\text{phys}} = 100$.

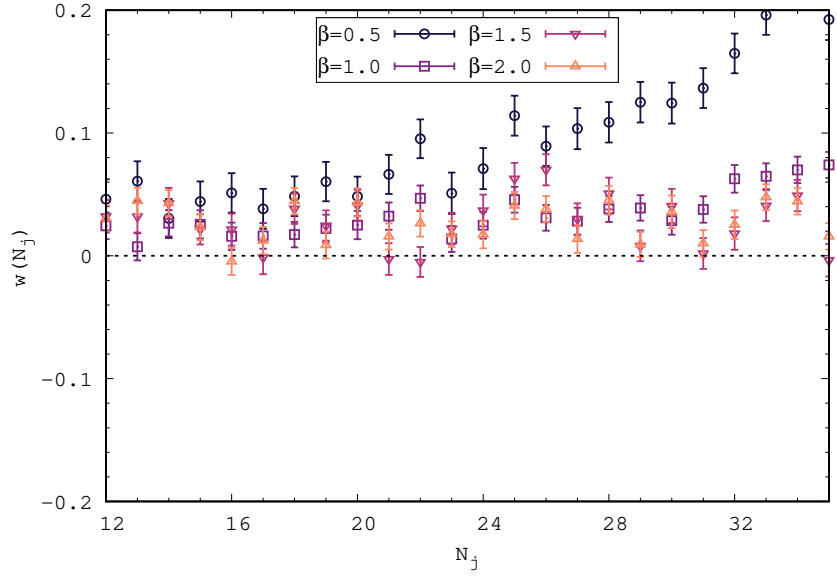

(a) Ward identity given by Eq. (2.27).

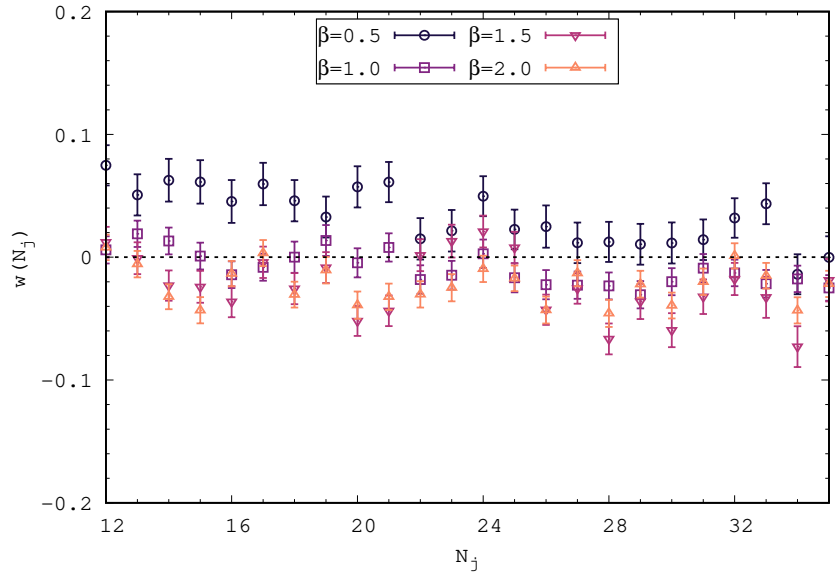

(b) Ward identity given by Eq. (2.28).

FIGURE 2.5: Model with degree-five superpotential. Ward identities for the model with different $\beta$ values on a lattice with $N_\tau = 48$, $m_{\text{phys}} = 10$, and $g_{\text{phys}} = 100$.

SUSY is preserved in the model. Fig. 2.6 (right) shows $\Delta S$ per site against $N_\tau$ for various $\beta$ values. Although all data points are close to zero, we see that they do not point toward a consistent trend suggesting an intact SUSY.

The Ward identities for various $\beta$ values are shown in Fig. 2.7 for a lattice with $N_\tau = 48$. Here also we discarded the contaminated data in the edge regions of the lattice. We see that in the middle region of the lattice, the Ward identities approach closer to zero as $\beta$ is increased. A linear $\beta \to \infty$ fit to the averaged data in the middle region gives us $0.003(1)$ for Fig. 2.7(a) and $-0.001(1)$ for Fig. 2.7(b). Thus, the data suggest that SUSY is intact in this model.

Taking into account the overall trend of the data from Figs. 2.6 and 2.7, we can conclude that SUSY is preserved in the model with Scarf I superpotential.

## 2.5 Models Exhibiting $PT$-Symmetry

Another interesting class of models we investigated on the lattice is the quantum mechanics with a particular type of $PT$-invariant superpotentials. Here, $P$ and $T$ denote the parity symmetry and time-reversal invariance, respectively. One of the motivations for considering $PT$-symmetric theories is the following. It is possible to obtain a real and bounded spectrum if we impose $PT$-symmetric boundary conditions on the functional-integral representation of the four-dimensional $-\lambda\phi^4$ theory [47]. In addition, the theory becomes perturbatively renormalizable and asymptotically free. These properties suggest that a $-\lambda\phi^4$ quantum field theory might help describe the Higgs sector of the Standard Model. It could also play a vital role in the supersymmetric versions of the Standard Model. Our one-dimensional model investigations can be considered the first step toward explorations in this direction.

The action of the supersymmetric quantum mechanics in Minkowski space is given by

$$S_M = \int dt \left( \frac{1}{2}(\partial_t \phi)(\partial_t \phi) + i\overline{\psi}\partial_t \psi - \overline{\psi}W''(\phi)\psi - \frac{1}{2}\left[W'(\phi)\right]^2 \right). \tag{2.58}$$

Under the Wick rotation

$$t \to -i\tau, \tag{2.59}$$

$$\frac{\partial}{\partial t} \to i\frac{\partial}{\partial \tau}, \tag{2.60}$$

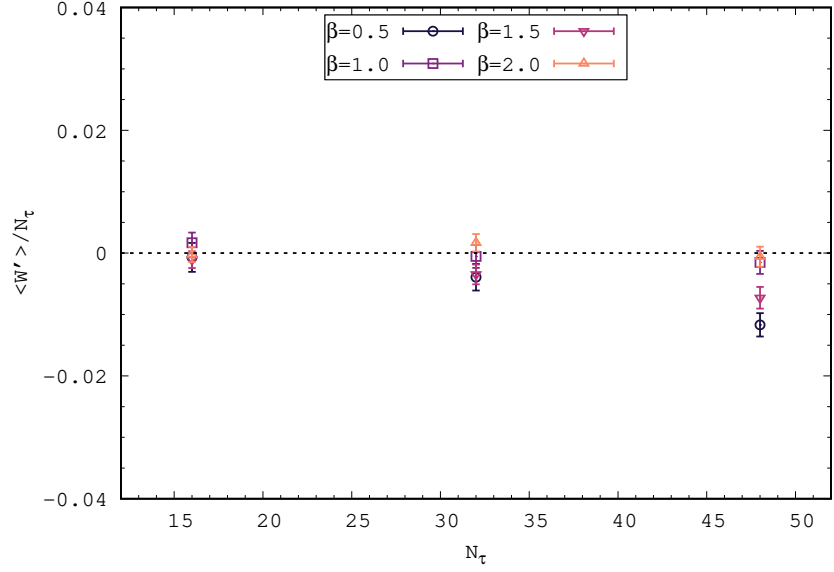

(a) $\langle W' \rangle$ per site against $N_\tau$.

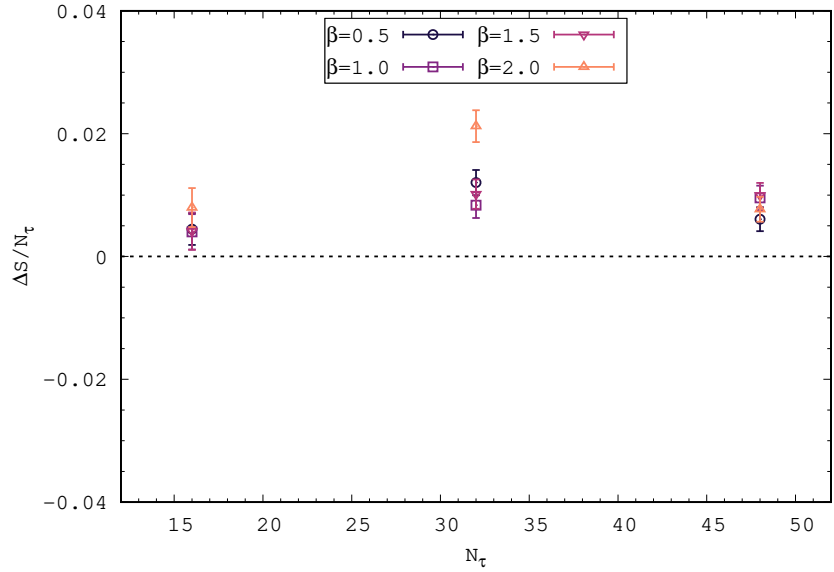

(b) $\Delta S$ per site against $N_\tau$.

FIGURE 2.6: Model with Scarf I superpotential. (Top) The expectation value of $\langle W' \rangle$ per site against $N_\tau$. It approaches zero as $\beta$ is increased (temperature is decreased), agreeing with what is expected for a model with intact SUSY. (Bottom) $\Delta S$ per site against $N_\tau$. We do not see a consistent trend suggesting intact SUSY from this observable. We used $N_\tau = 16, 32,$ and $48$, and for each $N_\tau$ we used $\beta = 0.5, 1.0, 1.5,$ and $2.0$ with $\lambda_{\text{phys}} = 10$ and $\alpha_{\text{phys}} = \sqrt{60}$.

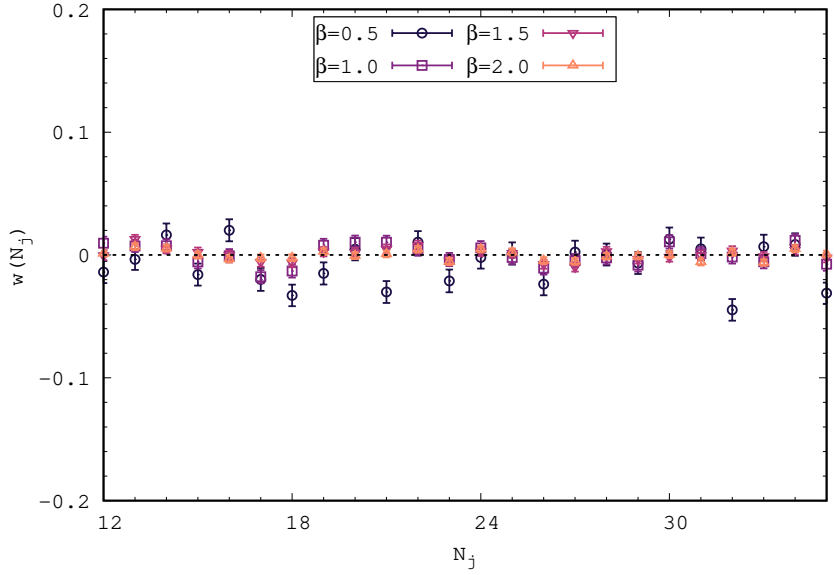

(a) Ward identity given by Eq. (2.27).

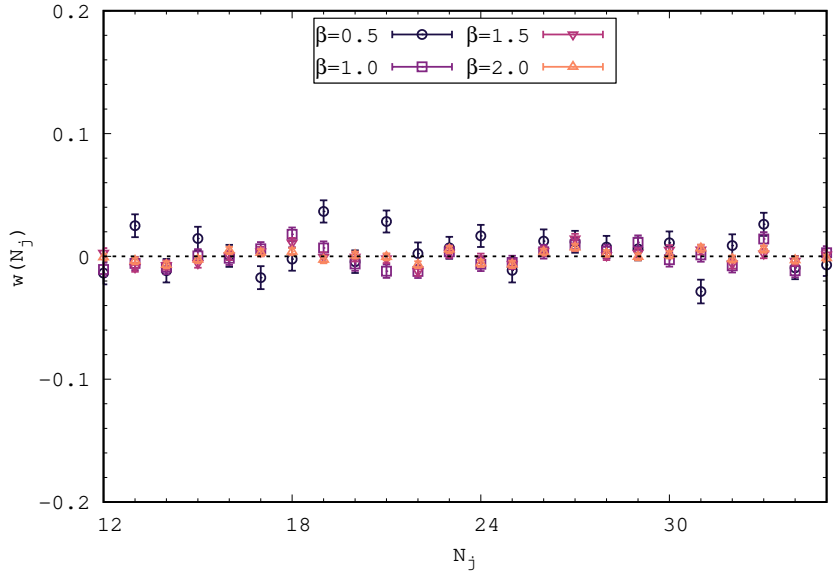

(b) Ward identity given by Eq. (2.28).

FIGURE 2.7: Model with Scarf I potential. Ward identities for various $\beta$ values on a lattice with $N_\tau = 48$. The dimensionless parameters used are $\lambda_{\text{phys}} = 10$ and $\alpha_{\text{phys}} = \sqrt{60}$. In the middle region of the lattice, the Ward identities approach closer to zero as $\beta$ is increased (temperature is decreased), indicating unbroken SUSY in this model.

the above action becomes

$$\begin{aligned}
S_M &= -i \int d\tau \left( (i)^2 \frac{1}{2}(\partial_t \phi)(\partial_t \phi) + i^2 \overline{\psi} \partial_t \psi - \overline{\psi} W''(\phi)\psi - \frac{1}{2}\left[W'(\phi)\right]^2 \right) \\
&= -i \int d\tau \left( -\frac{1}{2}(\partial_t \phi)(\partial_t \phi) - \overline{\psi} \partial_t \psi - \overline{\psi} W''(\phi)\psi - \frac{1}{2}\left[W'(\phi)\right]^2 \right) \\
&= i \int d\tau \left( \frac{1}{2}(\partial_t \phi)(\partial_t \phi) + \overline{\psi} \partial_t \psi + \overline{\psi} W''(\phi)\psi + \frac{1}{2}\left[W'(\phi)\right]^2 \right) \\
&= iS_E \equiv iS.
\end{aligned} \quad (2.61)$$

Under Parity transformation, $P$, we have

$$t \to t, \quad (2.62)$$
$$\phi(t), \psi(t), \overline{\psi}(t) \to \phi(t), \psi(t), \overline{\psi}(t). \quad (2.63)$$

Under time reversal symmetry, $T$, we have

$$t \to -t, \quad (2.64)$$
$$i \to -i, \quad (2.65)$$
$$\phi(t), \psi(t), \overline{\psi}(t) \to \phi(-t), \psi(-t), \overline{\psi}(-t). \quad (2.66)$$

Under $PT$ transformation

$$S_M = \int_{-\infty}^{\infty} dt \left( \frac{1}{2}(\partial_t \phi)(\partial_t \phi) + i\overline{\psi} \partial_t \psi - \overline{\psi} W''(\phi)\psi - \frac{1}{2}\left[W'(\phi)\right]^2 \right) \quad (2.67)$$

$$\to$$

$$\begin{aligned}
S_M &= \int_{-\infty}^{\infty} (-dt) \left( (-1)(-1)\frac{1}{2}(\partial_t \phi)(\partial_t \phi) + (-i)(-1)\overline{\psi} \partial_t \psi - \overline{\psi} \widetilde{W}''(\phi)\psi - \frac{1}{2}\left[\widetilde{W}'\right]^2 \right) \\
&= \int_{-\infty}^{\infty} dt \left( \frac{1}{2}(\partial_t \phi)(\partial_t \phi) + i\overline{\psi} \partial_t \psi - \overline{\psi} \widetilde{W}''(\phi)\psi - \frac{1}{2}\left[\widetilde{W}'\right]^2 \right),
\end{aligned} \quad (2.68)$$

where $\widetilde{W}$ is representing the transformed superpotential under $PT$ transformations. The action is $PT$ symmetric for certain classes of $W(\phi)$. In Ref. [48], it was shown that a two-dimensional supersymmetric field theory, with a $PT$-symmetric superpotential of the form

$$W(\phi) = \frac{-g}{2+\delta}(i\phi)^{2+\delta}, \tag{2.69}$$

with $\delta$, a positive parameter exhibited intact SUSY. We have

$$W'(\phi) = \frac{-g}{2+\delta}(i)^{2+\delta}(2+\delta)(\phi)^{1+\delta} = -ig(i\phi)^{1+\delta} \tag{2.70}$$

Under $PT$ transformation

$$\widetilde{W}'(\phi) = -(-i)g(-i\phi)^{1+\delta} = -(-i)(-1)(-1)^{\delta}g(i\phi)^{1+\delta} = -ig(-1)^{\delta}(i\phi)^{1+\delta} \tag{2.71}$$

Taking the square

$$\begin{aligned} \left[\widetilde{W}'(\phi)\right]^2 &= \left[-ig(-1)^{\delta}(i\phi)^{1+\delta}\right]^2 = [(-1)^2]^{\delta}\left[-ig(i\phi)^{1+\delta}\right]^2 \\ &= \left[-ig(i\phi)^{1+\delta}\right]^2 \\ &= \left[W'(\phi)\right]^2. \end{aligned} \tag{2.72}$$

Thus the term $-\frac{1}{2}\left[W'(\phi)\right]^2$ in the action is $PT$ invariant. Similarly we can show that $\widetilde{W}''(\phi) = W''(\phi)$ under $PT$ transformation, making the term $-\overline{\psi}W''(\phi)\psi$ in the action $PT$ symmetric.

A recent study using complex Langevin dynamics showed that SUSY is not dynamically broken in these models in zero and one dimensions [35, 36]. Since Monte Carlo is reliable only when the action is real, we simulated a subset of these $PT$-symmetric potentials with real actions. Complex Langevin dynamics or any other compatible method should be used for a full analysis of the model with a general $\delta$ parameter.

At a lattice site $k$, the $PT$-invariant superpotential has the following form

$$W_k = \frac{-g}{2+\delta}(i\phi_k)^{2+\delta}. \tag{2.73}$$

Its derivative is

$$W'_k = \sum_j K_{kj}\phi_j - ig\left(i\phi_k\right)^{1+\delta}. \tag{2.74}$$

Although we have massless theories as the continuum cousins, we need to introduce Wilson-type mass terms in the simulations, as shown in Eq. (2.74). We perform simulations for various mass values and then take the limit $m \to 0$. Since the action needs to be real for Monte Carlo simulations to be reliable, we have investigated only models with $\delta = 0, 2$, and 4. Numerical Monte Carlo analysis with such models in the non-perturbative quantum mechanical setup has not been done before, though in Ref. [48] using perturbative calculations, it was shown that in a two-dimensional supersymmetric field theory exhibiting $PT$ symmetry SUSY remains intact.

The superpotentials take the following forms for these $\delta$ values

$$\delta = 0: \quad W'_k = \phi_k + m\phi_k - \frac{1}{2}(\phi_{k-1} + \phi_{k+1}) + g\phi_k, \tag{2.75}$$

$$\delta = 2: \quad W'_k = \phi_k + m\phi_k - \frac{1}{2}(\phi_{k-1} + \phi_{k+1}) - g\phi_k^3, \tag{2.76}$$

$$\delta = 4: \quad W'_k = \phi_k + m\phi_k - \frac{1}{2}(\phi_{k-1} + \phi_{k+1}) + g\phi_k^5. \tag{2.77}$$

In Fig. 2.8 (left) we show $W'$ per site against $N_\tau$ for various $\delta$ values. In Fig. 2.8 (right) we show $\Delta S$ per site against $N_\tau$ for various $\delta$ values. Both plots are produced at $m_{\text{phys}} = 0$. For $\delta = 0$, we were able to simulate at $m_{\text{phys}} = 0$ value itself as the fermionic determinant is always positive definite in this case, and both $m$ and $g$ physically represent the same quantity. For $\delta = 2, 4$, larger mass values were needed to keep the fermionic determinant positive definite. These models were simulated with various $m_{\text{phys}}$ values, and then we took the $m_{\text{phys}} \to 0$ limit. The data in Fig. 2.8 (left) do not lie exactly on top of zero, making it difficult to conclude straight away that SUSY is preserved in this model, but the behavior improves as we move towards larger $N_\tau$ value. A linear $N_\tau \to \infty$ fit to the data with $\beta = 2.0$ for all $\delta$ values is given in Table 2.1. The data in Fig. 2.8 (right) fluctuate around zero within error bars for all $\delta$ values suggesting that SUSY is preserved in this model.

In Fig. 2.9, the Ward identities are plotted for all the mass parameter values used in the simulations. We used $m_{\text{phys}} = 0$ for $\delta = 0$; $m_{\text{phys}} = 20, 30, 40$ for $\delta = 2$; and $m_{\text{phys}} = 6, 8, 10$ for $\delta = 4$ in the simulations. We see that in Fig. 2.9, the data fluctuate around zero in the middle region of the lattice for all the $\delta$ values. In Table 2.1, we show the linear $m_{\text{phys}} \to 0$ fits to the

| $\delta$ | $\Delta S/N_\tau$ | $\langle W'\rangle/N_\tau$ | $\langle w_1\rangle$ | $\langle w_2\rangle$ |
|---|---|---|---|---|
| 0 | 0.0004(23) | −0.004(3) | 0.001(2) | −0.002(2) |
| 2 | 0.003(1) | −0.001(1) | 0.0005(15) | 0.0001(15) |
| 4 | −0.003(3) | 0.018(4) | 0.019(7) | 0.003(7) |

TABLE 2.1: Supersymmetric quantum mechanics with $PT$-symmetric superpotential. The fit values of various observables in the $N_\tau \to \infty$ limit taken linearly for different $\delta$ values for $\beta = 2.0$ and $g_{\text{phys}} = 10$.

Ward identities (for $\delta = 2, 4$). As mentioned above, for $\delta = 0$, we were able to simulate the model at the $m_{\text{phys}} = 0$ value itself.

Thus taking into consideration the overall behavior of the data in Figs. 2.8 and 2.9, we conclude that SUSY is preserved in these models with $PT$-invariant potentials.

## 2.6 Summary

In this chapter, we have investigated non-perturbative SUSY breaking in various quantum mechanics models using lattice regularized path integrals. We employed Hybrid Monte Carlo (HMC) algorithm to perform the field updates on the lattice. After reproducing the existing results in the literature for the supersymmetric anharmonic oscillator, we investigated SUSY breaking in a model with a degree-five superpotential and a shape invariant (Scarf I) superpotential. We then moved on to simulating the exciting case of models exhibiting $PT$-invariance. We simulated these models for various values of the parameter $\delta$ appearing in the $PT$-symmetric theory without violating the Markov chain Monte Carlo reliability criteria. Our simulations indicate that non-perturbative SUSY breaking is absent in quantum mechanics models exhibiting this type of $PT$ symmetry.

For the case of models with $PT$ symmetric superpotentials, an investigation that takes care of arbitrary $\delta$ values needs a simulation algorithm that can handle complex actions, such as the complex Langevin method. Ref. [35] shows that $PT$ symmetry is preserved in zero-dimensional supersymmetric theories. Recently, this study was extended to supersymmetric quantum mechanics models with arbitrary $\delta$ parameter [36] and there, with the help of complex Langevin simulations, it was shown that SUSY is preserved in these models. Our results in this paper complement these investigations.

It would be interesting to reproduce the conclusion in Ref. [48] using non-perturbative

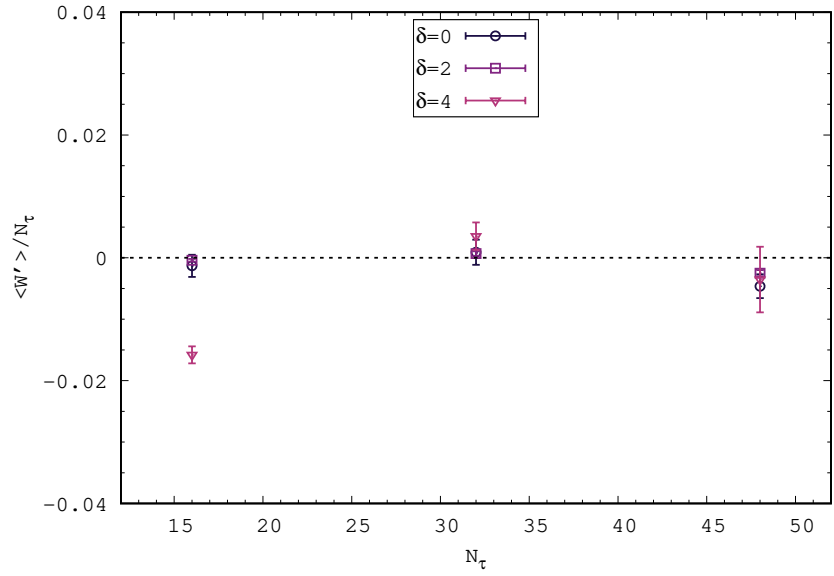

(a) $\langle W' \rangle$ per lattice site against $N_\tau$.

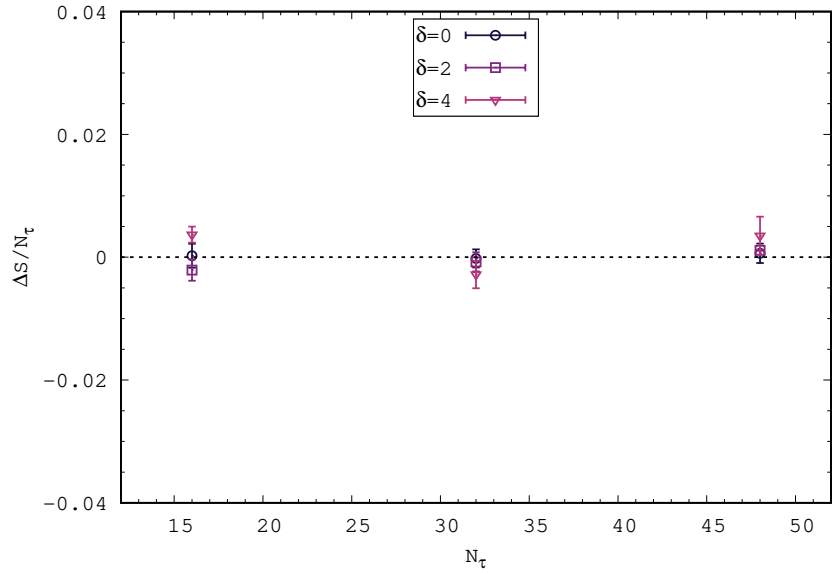

(b) $\Delta S$ per site against $N_\tau$.

FIGURE 2.8: Model with $PT$-symmetric superpotentials. (Top) The expectation value of $\langle W' \rangle$ per site against $N_\tau$. (Bottom) $\Delta S$ per site against $N_\tau$ with $\beta = 2.0$ and $g_{\text{phys}} = 10$.

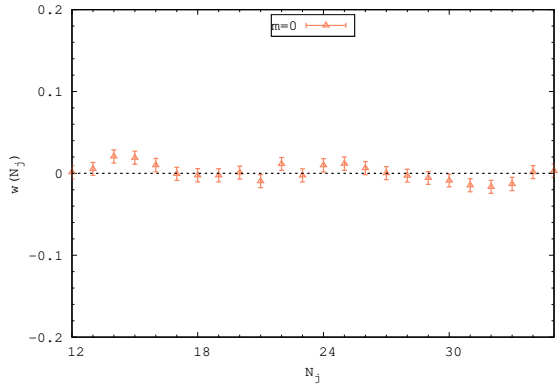
(a) Ward identity Eq. (2.27) for $\delta = 0$.

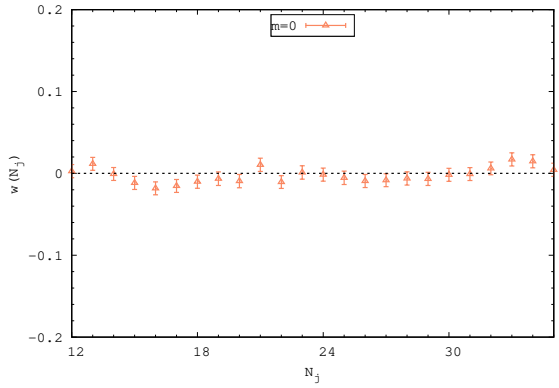
(b) Ward identity Eq. (2.28) for $\delta = 0$.

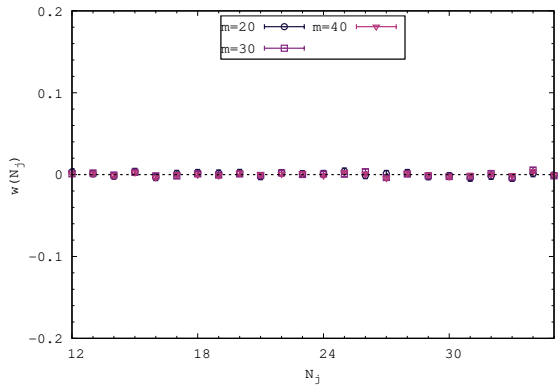
(c) Ward identity Eq. (2.27) for $\delta = 2$.

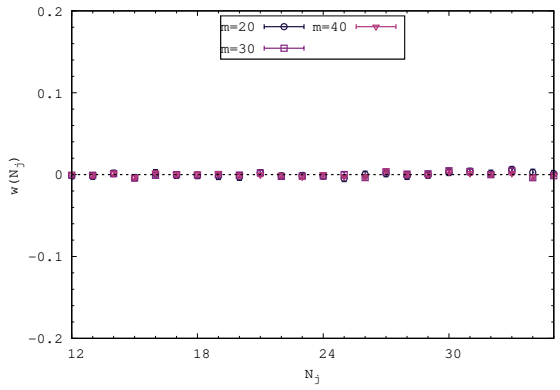
(d) Ward identity Eq. (2.28) for $\delta = 2$.

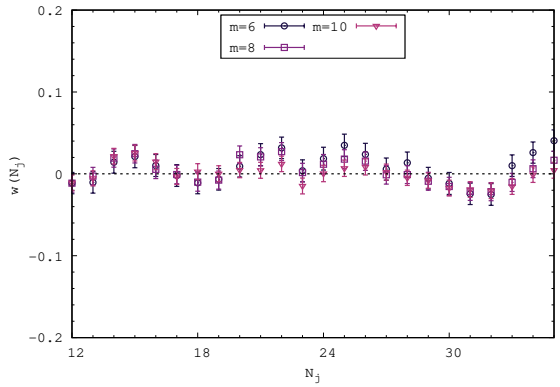
(e) Ward identity Eq. (2.27) for $\delta = 4$.

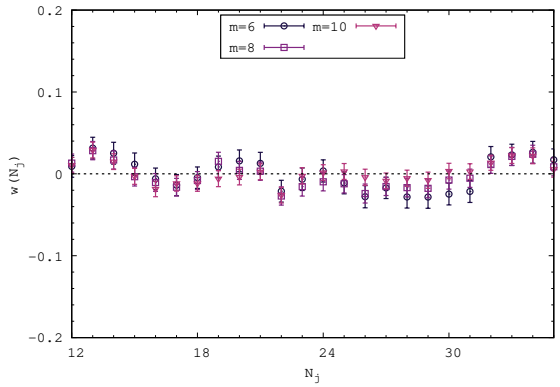
(f) Ward identity Eq. (2.28) for $\delta = 4$.

FIGURE 2.9: Models with $PT$-symmetric superpotentials. Ward identities on a $N_\tau = 48$ lattice. We used $\beta = 2.0$ and $g_{\text{phys}} = 10$ in the simulations.

methods such as Monte Carlo or complex Langevin dynamics. There, it was shown that SUSY remains unbroken, using perturbative calculations, in a two-dimensional supersymmetric field theory exhibiting $PT$ symmetry. It would also be interesting to perform simulations in four-dimensional supersymmetric models exhibiting $PT$ symmetry and thus comment on the nature of the spectrum of the theory and implications to Higgs physics.

In the next chapter, we will move towards slightly complicated systems in which point-like fields on a lattice site are replaced by $N \times N$ matrices. As it has been motivated, those models will take us towards understanding non-perturbative aspects of string theory.

# 3

# Matrix Models

Non-perturbative aspects of string theory can be captured by investigating the strongly coupled regimes of theories such as matrix models and supersymmetric Yang-Mills in various dimensions. In this chapter, we will discuss the general aspects of these theories.

## 3.1 Motivation

One of the major developments in various attempts to understand the features of quantum gravity was the observation that the lower-dimensional models of matrices could capture the dynamics of string/M-theory in an appropriate limit of the parameters. One of the first examples was the relation shown by Ref. [49] between the (non-supersymmetric) (0+1)-dimensional $c = 1$ matrix model and the two-dimensional bosonic string theory. This program of connecting quantum mechanical models to string/M-theory was extended by Banks, Fischler, Susskind, and Shenker (BFSS) through their proposal that the dimensional reduction of ten-dimensional $\mathcal{N} = 1$ super-Yang–Mills (SYM) with gauge group SU($N$) describes M-theory in the light-cone gauge, in the large-$N$ planar limit [50]. A few years later, Berenstein, Maldacena, and Nastase [51] extended this model by introducing a supersymmetry preserving one-parameter deformation. The resulting theory, known as the BMN matrix model, describes a certain limit of Type II string theory on a pp-wave background rather than the flat spacetime relevant to the BFSS model.

Though there has been excellent progress in understanding and verifying the gauge/gravity duality conjecture by studying $\mathcal{N} = 4$ SYM in four dimensions using ideas of integrability, the lower-dimensional non-conformal analogs of the four-dimensional theory have not attracted as much attention. This thesis is one attempt to understand the non-conformal analog of gauge/gravity correspondence. Only a handful of analytical attempts using certain approximations have been made so far [52, 53]. Since it is difficult to verify the duality conjecture in the finite-temperature setting relevant to these cases, we need a method that can provide information about their strongly coupled regimes. This opens up the possibility of exploring the dual field theories using the ideas and tools of lattice field theory. Though there are several approaches to putting SYM theories on a lattice, the lattice regularization of such theories breaks supersymmetry. This happens because supersymmetry transformations anti-commutes and connect with the generator of infinitesimal space-time translations, but on the lattice, no such infinitesimal translations exist, leading to broken SUSY. Hence on the lattice, all the supersymmetries are not preserved, but we are able to preserve a subset of supersymmetries. However, when we take the correct continuum limit, $\mathtt{a} \to 0$, by fine-tuning all the SUSY-breaking terms, we end up recovering all the supersymmetries.

In this regard, there has been good progress in understanding various aspects of $(0 + 1)$-dimensional matrix models as well as the thermodynamics of stacks of D$p$ branes, with $p = 1$ and 2, using the $(p + 1)$-dimensional dual supersymmetric theories in Euclidean lattice space-times [54–74].

## 3.2  SYM families

The idea of AdS/CFT correspondence from the field theory side can be investigated from the non-perturbative study of maximally supersymmetric Yang-Mills (MSYM) theories. Maximal theories with sixteen supercharges and even with lesser supersymmetries can be constructed from $\mathcal{N} = 1$ SYM in higher dimensions. The different families of SYM theories are as follows:

- $\mathcal{Q} = 16$

$$d = 10, \mathcal{N} = 1 \quad \to \quad d = 6, \mathcal{N} = 2 \quad \to \quad d = 4, \mathcal{N} = 4 \quad \to \quad d = 3, \mathcal{N} = 8$$
$$\to \quad d = 2, \mathcal{N} = (8, 8) \quad \to \quad d = 1, \text{BFSS} \quad \to \quad d = 0, \text{IKKT}$$

- $\mathcal{Q} = 8$

$$d = 6, \mathcal{N} = 1 \quad \to \quad d = 4, \mathcal{N} = 2 \quad \to \quad d = 3, \mathcal{N} = 4 \quad \to \quad d = 2, \mathcal{N} = (4, 4)$$

- $\mathcal{Q} = 4$

$$d = 4, \mathcal{N} = 1 \quad \to \quad d = 3, \mathcal{N} = 2 \quad \to \quad d = 2, \mathcal{N} = (2, 2)$$

We note that implementing all these theories on spacetime lattices would be a daunting task. In the next two chapters, we will discuss two important models from the above set of theories. One is a mass-deformed version of the BFSS matrix model, i.e., the BMN matrix model [51], but without fermions. In the BMN model, $SO(9)$ rotational symmetry of scalar fields gets explicitly broken into $SO(6) \times SO(3)$. The lattice setup for this model is significantly simpler as compared to the other model discussed in this thesis, which is the two-dimensional SYM theory with four supercharges. The latter theory is discretized on a lattice with the help of a procedure known as *twisting*. (The twisting method will be introduced in chapter 5.) The BMN matrix model has been the topic of a couple of studies on the lattice [69, 73, 75].

We can obtain the lower-dimensional SYM theories from the higher dimensional $N = 1$ SYM by using the Kaluza-Klein compactification. We can compactify ten-dimensional $\mathcal{N} = 1$ SYM to one dimension to obtain the BFSS model. (We will discuss the BFSS model in chapter 4.)

We note that when we compactify a specific dimension:

- There is no derivative term along that direction,

- Fields in the theory no longer are functions of that direction,

- Gauge field along that direction changes to a scalar field.

Hence for the BFSS model, we have to compactify nine spatial dimensions, which in turn gives nine scalars in the theory. For the IKKT matrix model, we have to compactify all ten dimensions; hence we get ten scalars in the theory.

## 3.3 Gauge Fields on a Lattice

The $d$-dimensional gauge field $A_a$, which is $N \times N$ traceless anti-hermitian matrix, with $a = 1, 2, \cdots, d$, is slightly different from bosons or fermions as far as its lattice implementation is concerned. If we put a gauge field on a lattice site, then it becomes difficult to preserve gauge invariance. Hence, the other option is to place the gauge field on a lattice link resulting in a setup where the gauge field is not the fundamental object, rather, its cousin, the group-valued Wilson link ($U_a$).

The Wilson link (or the link field) $U_a(n)$ attached to a lattice site $n$ has the form

$$U_a(n) = e^{A_a(n)}, \tag{3.1}$$

oriented from site $n$ to $n + \hat{\mu}_a$, with $\hat{\mu}_a$ denoting the unit vector along the $a$-th direction. It transforms in the following way under a gauge transformation on the lattice:

$$U_a(n) \to G(n) U_a(n) G^\dagger(n + \hat{\mu}_a). \tag{3.2}$$

Here, $G(n)$ is an element of the gauge group, say, SU($N$) of the lattice theory.

We consider the non-abelian fields of lattice theory on an anti-Hermitian basis. The gauge links in the continuum theory appear through the continuum covariant derivative as:

$$D_a \cdot = \partial_a \cdot + [A_a, \cdot]. \tag{3.3}$$

On a lattice, in order to preserve gauge invariance, the derivative operator changes to a difference operator. On a field, say, $X_a$, which is an $N \times N$ matrix sitting on a particular site of the lattice and transforms in the adjoint representation, this operator becomes

$$\mathcal{D}_+ X_a(n) \equiv U_a(n) X_a(n+1) U_a^\dagger(n) - X_a(n). \tag{3.4}$$

We can also consider higher orders of difference operators, which can take us to the continuum limit more quickly. In such a setup, we call the lattice action as *improved action*.

Now as we know how to put gauge fields on a lattice, along with bosons and fermions, we are ready to test our Monte Carlo simulations on matrix models. However, we need to be aware of some of the problems that we can encounter while simulating matrix models. We discuss two such problems, finite and large $N$ effects, and flat directions, in the next two sections.

## 3.4 Large $N$

The importance of a large $N$ limit, specifically in the case of SU($N$) gauge theories, can be understood from the fact that gauge/gravity duality correspondence can only be valid in the large $N$ limit. Even without gauge fields, the thermodynamics of matrix models (e.g., the IKKT model) have a strong dependence on $N$. We will see finite-$N$ effects in the various thermodynamic observables of the two models - the Gross-Witten-Wadia (GWW) model and the IKKT model.

The GWW model [76] is the gauge theory unitary matrix model in two spacetime dimensions with the unitary matrices mapped to the links on the lattice. As discussed in the original work, it is sufficient to consider only one plaquette action for the two-dimensional model to understand the thermodynamics of the system.

The action of the model is

$$S = -\frac{N}{\lambda} \operatorname{Tr}(U + U^\dagger), \tag{3.5}$$

where $U$ is an $N \times N$ unitary matrix, $\lambda$ is t' Hooft coupling given as $\lambda = g^2 N$. The Polyakov loop winding observable,

$$P_k = \left\langle \frac{1}{N} \operatorname{Tr}(U^k) \right\rangle \tag{3.6}$$

has the following analytical form

- First Polyakov loop winding $P \equiv P_{k=1}$:

$$P = \begin{cases} 1 - \frac{\lambda}{4} & , \lambda \leq 2.0 \\ \frac{1}{\lambda} & , \lambda \geq 2.0 \end{cases} \quad (3.7)$$

- Higher windings

$$P_k = \begin{cases} \left(1 - \frac{\lambda}{2}\right)^2 \frac{1}{k-1} \mathcal{P}_{k-2}^{(1,2)}(1-\lambda) & , \lambda \leq 2 \\ 0 & , \lambda \geq 2 \end{cases} \quad (3.8)$$

where $\mathcal{P}_{k-2}^{(1,2)}$ are the Jacobi polynomials.

Now if we simulate this model with the Monte Carlo method, we see that higher windings have a strong dependence on $N$ as can be seen in Fig. 3.1

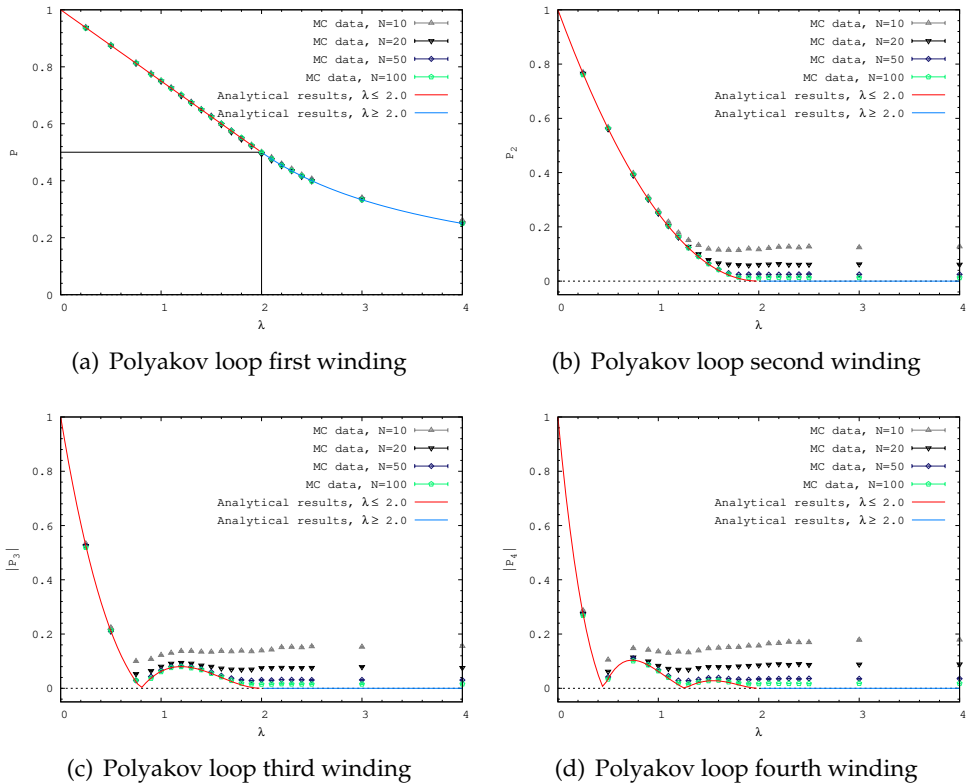

(a) Polyakov loop first winding  (b) Polyakov loop second winding

(c) Polyakov loop third winding  (d) Polyakov loop fourth winding

FIGURE 3.1: Polyakov loop windings in the GWW model for various values of $N$. It can be seen that higher Polyakov loop windings exhibit strong $N$ dependence for larger couplings.

Hence the finite $N$ effects can be easily seen in the higher windings of the Polyakov loop,

where at strong coupling, for lower $N$ values, the calculated values are away from the analytical calculation, and as $N$ is increased, the difference decreases. Therefore, to get valid results, one needs to work with different $N$ values and take the $1/N \to 0$ limit to approach the exact analytical results.

Another simple model in which we will see the finite $N$ effects is the IKKT model, which we can achieve by dimensionally reducing $\mathcal{N} = 1$ SYM in ten dimensions to zero dimensions. We will only work with the bosonic version of this model. The action for the same is

$$S_{\text{E}} = -\frac{N}{4\lambda} \sum_{i,j} \text{Tr}\big([X^i, X^j]^2\big), \tag{3.9}$$

where $X$'s are $N \times N$ hermitian traceless matrices with $i$ and $j = 1, 2, \ldots, 10$. The action has $SO(10)$ rotational symmetry as all the matrices can be rotated with each other internally without affecting the action. Since we are in zero dimensions, it is easy to see that the t' Hooft coupling can be absorbed in the scalars ($X$'s). This model has been studied extensively in the context of probing spontaneous symmetry breaking (SSB). With the help of Monte Carlo simulations, it has been argued that SO(10) symmetry stays intact in the bosonic model and the full model with phase-quenched simulations [77, 78]. Furthermore, there have been studies based on the complex Langevin method, which show that SO(10) symmetry gets broken into SO($d$) × SO(10 − $d$) [79, 80] due to SSB. However, our motivation for studying this model is actually to see how the finite-$N$ effects change the results of SSB in the bosonic IKKT case. We use the eigenvalues, $\xi_i$ with $i = 1, 2, \cdots, 10$, of a $10 \times 10$ tensor matrix, $I_{ij}$, constructed from the ten scalars, as the order parameter to probe SSB. We have

$$I_{ij} = \frac{1}{N} \text{Tr}\big(X^i X^j\big). \tag{3.10}$$

The variation of the eigenvalues against $N$ can be seen in Fig. 3.2. If all the eigenvalues for $N \to \infty$ stay clubbed together, then there is no SSB in theory. If for $N \to \infty$, the eigenvalues split such that $d$ eigenvalues are clubbed together with some finite value, and the rest $10 - d$ eigenvalues are almost zero, that will show that the $SO(10)$ symmetry has broken into $SO(d)$ and $SO(10 - d)$. This spontaneous symmetry breaking can tell us about the emergence of space-time. But we can only check with the full IKKT matrix model. Here, our motivation for studying this bosonic IKKT model is to show how finite $N$ can affect the analysis of a theory.

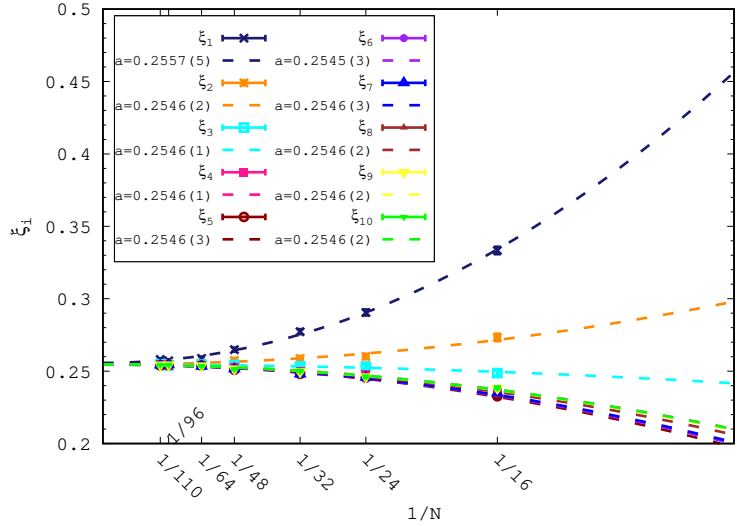

FIGURE 3.2: The eigenvalues of the tensor constructed from the ten scalars in the bosonic IKKT model given in Eq. (3.10). It can be seen that the scalars show finite $N$ effects in this model, and a consensus on the result can only be achieved in the large-$N$ limit.

We do see that symmetry remains intact in this model, but to see the same, we need large enough values for $N$. The nature of the $1/N$ effects can be checked with the help of the fit $a + bN^{-2}$. At $N \to \infty$, all the eigenvalues are clubbed together around $0.255$, indicating intact symmetry of the model.

## 3.5 Flat directions

In SUSY theories, the runaway behavior of scalars in the presence of flat directions is another issue we need to deal with in Monte Carlo simulations. Along the flat directions, the scalars commute, $[X_i, X_j] = 0$, but their eigenvalues keep on increasing since they have access to the continuum branch of the spectra [81]. This problem can be seen through the growth in the values of the scalar fields during a Monte Carlo calculation.

The behavior of the scalars can be treated in a manner equivalent to that of the D branes in the corresponding string theory. If the scalars clump around the origin, forming a bound state, then that corresponds to a bunch of D branes coming together. If the scalars show runaway behavior, then on the gravity side, it can be interpreted as the clump of D branes dispersing into a gas. To see how the flat directions affect the Monte Carlo simulations, we can look at one of the configurations in the supersymmetric BFSS model. Though the bosonic version of its mass deformation is discussed in the next chapter, this configuration is mentioned to show how the flat directions appear in the simulations of a particular theory. The extent of nine scalars in the supersymmetric BFSS model, without any mass deformation, is shown in Fig. 3.3.

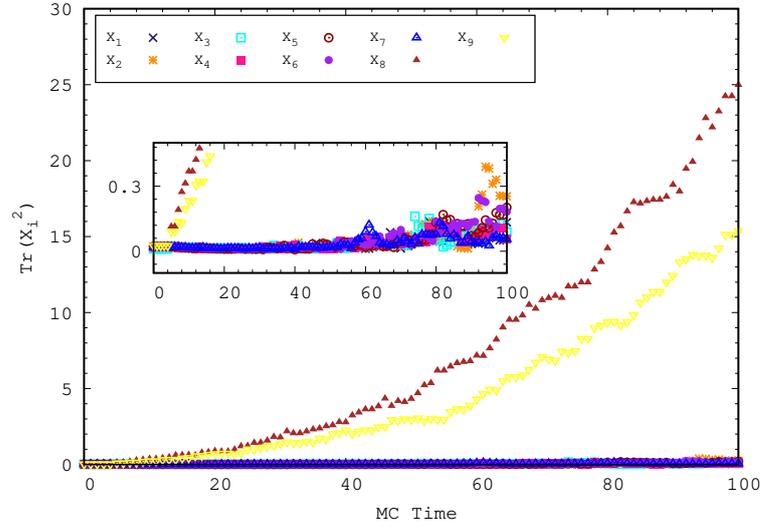

FIGURE 3.3: Extent of nine scalars in the supersymmetric BFSS model at dimensionless temperature $T = 0.2$ with $N = 12$ and $24$ lattice sites. It can be seen that two of the scalars run away because of the presence of flat directions, making the simulations unreliable.

In the simulations, to tackle this flat direction issue, we can take $N$ to be very large, which is difficult in supersymmetric theories as simulating fermions with large matrix sizes becomes computationally difficult. Another method for tackling the issue is to add a mass deformation to the scalar potential, such as $m^2 X^2$ [82], which controls the flat directions. We can work with different deformation values and then take the $m \to 0$ limit for different observables to get the thermodynamics of the original target theory.

In the next two chapters, the models discussed are the BMN model and $\mathcal{N} = (2,2)$ SYM in two dimensions. The BMN model is already a mass deformation of the BFSS model; hence flat directions in that model are already under control. In the other model, i.e., in the $\mathcal{N} = (2,2)$ SYM, we had to add an extra scalar potential term to curb the issue arising from flat directions.

# 4
# Non-perturbative phase structure of the bosonic BMN matrix model

*Content of this chapter is partially based on:*

- Pub. [3]: N. S. Dhindsa, R. G. Jha, A. Joseph, A. Samlodia, and D. Schaich, "Non-perturbative phase structure of the bosonic BMN matrix model", JHEP **05**, 169 (2022), arXiv:2201.08791 [hep-lat].

- Pub. [4]: N. S. Dhindsa, A. Joseph, A. Samlodia and D. Schaich, "Deconfinement Phase Transition in Bosonic BMN Model at General Coupling", arXiv:2308.02538 [hep-lat].

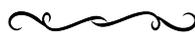

The gauge/gravity duality conjecture tells us that at large $N$ and finite temperature, there are often transitions between different quantum black hole solutions, which are dual to con-finement transitions in the field theory. In this regard, the D0 brane matrix model is an exception, with only a single deconfined phase at all temperatures in the planar limit. Though in Ref. [72] first results of the confined phase are discussed in the theory at finite $N$, whether the theory stays in confined phase as $N \to \infty$ still needs more investigation in supersymmetric BFSS model. Hence more studies are required around this topic before concluding that the D0 brane matrix model exhibits two different phases and hence a phase transition. However, this behavior of having only a deconfined phase at all couplings is drastically altered if we consider either a one-parameter deformation of the BFSS model, i.e., the BMN model, or if we decouple the fermions and study the bosonic sector of the BFSS model. In both cases, there is a well-defined confinement transition. The dual black hole solutions of the BMN model in the deconfined phase and the details of the phase transition were studied in Ref. [83]. It remains a challenge to understand the phase diagram for finite couplings and to verify the results obtained using gravity computations.

The BMN model is a one-parameter deformation of the BFSS model—the dimensional reduction of $(9+1)$-dimensional $\mathcal{N} = 1$ SYM with gauge group SU($N$) down to $0+1$ dimensions. In Euclidean time the action of the BFSS model is

$$S_{\text{BFSS}} = \frac{N}{4\lambda} \int_0^\beta d\tau \, \text{Tr}\Big\{ -(D_\tau X_i)^2 - \frac{1}{2} \sum_{i<j} [X_i, X_j]^2 \\ + \Psi_\alpha^T \gamma_{\alpha\sigma}^\tau D_\tau \Psi_\sigma + \Psi_\alpha^T \gamma_{\alpha\sigma}^i [X_i, \Psi_\sigma] \Big\}, \quad (4.1)$$

where $D_\tau \cdot = \partial_\tau \cdot + [A_\tau, \cdot]$ is the covariant derivative, $X_i$ are the nine scalars from the reduction of the ten-dimensional gauge field, and $\Psi_\alpha$ is a sixteen-component spinor. The indices $i, j = 1, \cdots, 9$ while $\alpha, \sigma = 1, \cdots, 16$. The degrees of freedom transform in the adjoint representation of the SU($N$) gauge group. The anti-Hermitan gauge group generators are normalized as $\text{Tr}(T^A T^B) = -\delta_{AB}$. The trace 'Tr' is taken over the gauge indices. In this $(0+1)$-dimensional model, the 't Hooft coupling $\lambda \equiv g_{\text{YM}}^2 N$ is dimensionful, $[\lambda] = 3$. The model is compactified on a circle with circumference $\beta = T^{-1}$, which corresponds to the inverse temperature because we impose thermal boundary conditions — periodic for the bosons and anti-periodic for the fermions.

The action of the BMN model is obtained by adding the following mass and scalar-trilinear

terms to Eq. (4.1)

$$S_\mu = -\frac{N}{4\lambda} \int_0^\beta d\tau \, \text{Tr} \left[ \left(\frac{\mu}{3} X_I\right)^2 + \left(\frac{\mu}{6} X_A\right)^2 + \frac{\mu}{4} \Psi_\alpha^T \gamma_{\alpha\sigma}^{123} \Psi_\sigma - \frac{\sqrt{2}\mu}{3} \epsilon_{IJK} X_I X_J X_K \right]. \quad (4.2)$$

Here $\mu$ is the deformation parameter, with dimension $[\mu] = 1$. We divide the indices $i, j$ into two sets: $I, J, K = 1, 2, 3$ and $A = 4, \cdots, 9$. The scalar mass terms break the SO(9) global symmetry of the BFSS model down to SO(3) × SO(6). As $\mu \to \infty$, the model reduces to a free supersymmetric Gaussian model, and it can be studied perturbatively for large $\mu$ [84, 85]. Since we are interested only in the bosonic sector, we can remove the fermions to obtain the action of the bosonic BMN (BBMN) model

$$S_{\text{BBMN}} = \frac{N}{4\lambda} \int_0^\beta d\tau \, \text{Tr} \left[ -(D_\tau X_i)^2 - \frac{1}{2} \sum_{i<j} [X_i, X_j]^2 \right. \\ \left. - \left(\frac{\mu}{3} X_I\right)^2 - \left(\frac{\mu}{6} X_A\right)^2 + \frac{\sqrt{2}\mu}{3} \epsilon_{IJK} X_I X_J X_K \right]. \quad (4.3)$$

Refs. [69, 72, 73] have numerically explored the phase structure of the full BMN model, with Ref. [69] reporting two different phase transitions — a confinement transition signaled by the Polyakov loop, and a 'Myers transition' signaled by the trilinear 'Myers term' — which merge into one for the dimensionless BMN deformation parameter $\widehat{\mu} \equiv \mu/\lambda^{1/3} \lesssim 3$ (where $\lambda$ is the dimensionful 't Hooft coupling). When these transitions are distinct, Ref. [69] observes the Myers transition to be between two deconfined phases, one where the system fluctuates around the trivial configuration and the other with fluctuations around expanded fuzzy spheres. More recently, as our work was in progress, Ref. [72] revisited the phase structure of the BMN model, introducing constraints on the Myers term to suppress fuzzy sphere contributions and focused on the confinement transition.

In this chapter, we report a detailed study of the phase structure that results from including the BMN deformation and decoupling of the fermions. While removing the fermions completely eliminates the $Q = 16$ supersymmetries of the BMN model and its holographic connection to quantum gravity, we take this step in order to accelerate numerical computations. This makes it easier to study larger lattice sizes and $N$ and thereby obtain more precise and reliable results for the phase structure. These robust results will provide a solid starting point for subsequent efforts to analyze the full supersymmetric theory.

The bosonic BMN model was investigated in Refs. [75, 86] for a fixed $\widehat{\mu} = 2$, finding a single first-order transition in the large-$N$ limit, at the dimensionless critical temperature $\widehat{T}_{\text{c}} \equiv T/\lambda^{1/3} = 0.915(5)$. It was not clear from this work whether different $\widehat{\mu}$ values might exhibit a Myers transition distinct from the confinement transition that was previously reported for the full BMN model by Ref. [69]. Addressing this question is part of the motivation for our investigations. While our work was underway, another work [72] appeared, reporting a single first-order phase transition for $0.375 \leq \widehat{\mu} \leq 3$ (the range where the two transitions had merged for the full BMN model [69]).[1] We pushed further into the large-$\widehat{\mu}$ regime, which allowed us to conclude that the bosonic BMN model features a single first-order transition for *all* values of the deformation parameter.

Our goal is to explore the functional form of the dependence of the bosonic BMN critical temperature $\widehat{T}_{\text{c}}$ on the deformation parameter $\widehat{\mu}$. To this end, we analyze twelve different values of $\widehat{\mu}$ spanning two orders of magnitude between the previously studied $\widehat{\mu} \to 0$ and $\widehat{\mu} \to \infty$ limits. As $\widehat{\mu} \to 0$, we recover the bosonic version of the BFSS model. Although early numerical and analytic bosonic BFSS investigations reported two near-by phase transitions [87, 88], more recent lattice calculations find only a single confinement transition with $\widehat{T}_{\text{c}}|_{\widehat{\mu}=0} = 0.8846(1)$ [72, 89]. In the $\widehat{\mu} \to \infty$ limit, the system reduces to a solvable gauged Gaussian model, and for large $\widehat{\mu}$, the critical temperature is found to scale as $\widehat{T}_{\text{c}} = (6\ln(3 + 2\sqrt{3}))^{-1}\widehat{\mu}$ [90, 91]. These two limits serve as consistency checks for our lattice computations.

The chapter is organized as follows. In Sec. 4.1, we discuss the lattice formulation and define the relevant observables we study. In Sec. 4.2, we present our results for a wide range of $0.5 \leq \widehat{\mu} \lesssim 45$ with $N = 16$, 32, and 48. The data leading to these results are available through Ref. [92]. We additionally discuss a *separatrix* method that we employ as a novel means to precisely estimate the critical temperature. We then study the $\widehat{\mu}$ dependence of these critical temperatures, fitting them to different functional forms for small and large $\widehat{\mu}$. In Sec. 4.6, we summarize our results for this model.

## 4.1 Bosonic BMN model on a lattice

We discretize the bosonic BMN model on a lattice with $N_\tau$ sites. The inverse temperature becomes $\beta = \text{a}N_\tau$, where 'a' is the lattice spacing with dimension $[\text{a}] = -1$. The integration becomes a summation over the lattice sites: $\int_0^\beta d\tau \longrightarrow \mathrm{a} \sum_{n=0}^{N_\tau - 1}$. The dimensionful gauge field $A_\tau$ is mapped to a dimensionless gauge link $U(n)$ connecting the sites $n$ and $n + 1$. We also work with dimensionless scalars $X_i(n) = \mathrm{a} X_i(\tau)$ once we are on the lattice. To discretize the covariant derivative $D_\tau X_i(\tau)$ we use the gauge link to define the finite-difference operator $\mathcal{D}_+ X_i(n) \equiv U(n) X_i(n + 1) U^\dagger(n) - X_i(n)$. Finally, we introduce the dimensionless lattice parameters $\mu_{\text{lat}} \equiv \mathrm{a}\mu$ and $\lambda_{\text{lat}} \equiv \mathrm{a}^3 \lambda$ to end up with a lattice action for the bosonic BMN model that has the same form as Eq. (4.3) while employing only dimensionless lattice quantities

$$S_{\text{lat}} = \frac{N}{4\lambda_{\text{lat}}} \sum_{n=0}^{N_\tau - 1} \mathrm{Tr}\left[ -(\mathcal{D}_+ X_i)^2 - \frac{1}{2} \sum_{i<j} [X_i, X_j]^2 \right.$$
$$\left. - \left(\frac{\mu_{\text{lat}}}{3} X_I\right)^2 - \left(\frac{\mu_{\text{lat}}}{6} X_A\right)^2 + \frac{\sqrt{2}\mu_{\text{lat}}}{3} \epsilon_{IJK} X_I X_J X_K \right]. \tag{4.4}$$

---

[1] Our $\widehat{\mu}$ is equivalent to $3\mu$ in the conventions of Ref. [72].

The following dimensionless combinations of parameters are particularly useful because they can be considered consistently in both the lattice and continuum theories:

$$\widehat{T} \equiv \frac{T}{\lambda^{1/3}} = \frac{1}{N_\tau \lambda_{\text{lat}}^{1/3}}, \qquad \widehat{\mu} \equiv \frac{\mu}{\lambda^{1/3}} = \frac{\mu_{\text{lat}}}{\lambda_{\text{lat}}^{1/3}}, \qquad \frac{\widehat{T}}{\widehat{\mu}} = \frac{T}{\mu} = \frac{1}{N_\tau \mu_{\text{lat}}}. \tag{4.5}$$

While taking the continuum limit, $N_\tau \to \infty$, we need to keep the quantities defined in the above equation fixed. Using this simple lattice action, we generate ensembles of matrix configurations using the hybrid Monte Carlo algorithm implemented by the publicly available parallel software presented in Ref. [93].[2] As our main goal is to analyze phase transitions, we concentrate our lattice calculations around the transition regions.

Let us look at first how to derive the internal energy, which is a useful observable in our calculations. To obtain the expression for the internal energy, we consider deforming the partition function by a small amount from $Z$ to $Z'$, which is expressed by the following set of transformations:

$$t' = \frac{\beta'}{\beta} t, \qquad A'(t') = \frac{\beta}{\beta'} A(t), \qquad X_i'(t') = \sqrt{\frac{\beta'}{\beta}} X_i(t). \tag{4.6}$$

Note that we have $[DX'] = [DX]$ and $[DA'] = [DA]$. For the bosonic BMN model, it is

---
[2]github.com/daschaich/susy

convenient to break up the action Eq. (4.3) into two pieces

$$S = S_0 + S_\mu, \tag{4.7}$$

$$S_0 = \frac{N}{4\lambda} \int_0^\beta d\tau \, \text{Tr}\left( -(D_\tau X_i)^2 - \frac{1}{2}\sum_{i<j}[X_i, X_j]^2 \right), \tag{4.8}$$

$$S_\mu = -\frac{N}{4\lambda} \int_0^\beta d\tau \, \text{Tr}\left( \frac{\mu^2}{9}X_I^2 + \frac{\mu^2}{36}X_A^2 - \frac{\sqrt{2}\mu}{3}\epsilon_{IJK}X_I X_J X_K \right). \tag{4.9}$$

Applying Eqs. (4.6), and defining $\Delta\beta \equiv \beta' - \beta$, we find

$$S' = S'_0 + S'_\mu, \tag{4.10}$$

$$S'_0 = S_0 + \frac{N}{4\lambda}\int_0^\beta d\tau \, \text{Tr}\left( -\frac{3}{2}\frac{\Delta\beta}{\beta}\sum_{i<j}[X_i, X_j]^2 \right) + \mathcal{O}(\Delta\beta^2), \tag{4.11}$$

$$S'_\mu = S_\mu - \frac{N}{4\lambda}\int_0^\beta d\tau \, \text{Tr}\left( 2\frac{\Delta\beta}{\beta}\left(\frac{\mu}{3}X_I\right)^2 + 2\frac{\Delta\beta}{\beta}\left(\frac{\mu}{6}X_A\right)^2 \right.$$
$$\left. - \frac{5\sqrt{2}\mu}{6}\frac{\Delta\beta}{\beta}\epsilon_{IJK}X_I X_J X_K \right) + \mathcal{O}(\Delta\beta^2). \tag{4.12}$$

The partition function therefore becomes

$$Z(\beta') = \int [DX']_{\beta'}[DA']_{\beta'}\, e^{-S'} = \int [DX]_\beta [DA]_\beta\, e^{-S} e^{E\Delta\beta + \mathcal{O}(\Delta\beta^2)}$$
$$= Z(\beta)\left[1 + E\Delta\beta + \mathcal{O}(\Delta\beta^2)\right]. \tag{4.13}$$

Hence, we can write

$$\frac{\widehat{E}}{N^2} = \frac{E}{\lambda^{1/3}N^2} = \frac{1}{Z(\beta)\lambda^{1/3}N^2}\lim_{\Delta\beta\to 0}\frac{Z(\beta') - Z(\beta)}{\Delta\beta}. \tag{4.14}$$

Using Eqs. (4.7)–(4.12) in Eq. (4.14) we get

$$\frac{\widehat{E}}{N^2} = \frac{1}{\lambda^{1/3}\beta}\left\langle \frac{1}{4N\lambda}\int_0^\beta d\tau \, \text{Tr}\left( -\frac{3}{2}\sum_{i<j}[X_i, X_j]^2 - 2\left(\frac{\mu}{3}X_I\right)^2 - 2\left(\frac{\mu}{6}X_A\right)^2 \right.\right.$$
$$\left.\left. + \frac{5\sqrt{2}\mu}{6}\epsilon_{IJK}X_I X_J X_K \right) \right\rangle. \tag{4.15}$$

Upon discretizing the bosonic BMN model, as discussed in Sec. 4.1, the dimensionless lat-tice internal energy takes the form reported in Eq. (4.17)

$$\frac{\widehat{E}}{N^2} = \frac{1}{4N\lambda_{\text{lat}}^{4/3} N_\tau} \left\langle \sum_{n=0}^{N_\tau-1} \text{Tr}\left( -\frac{3}{2}\sum_{i<j}[X_i, X_j]^2 - \frac{2\mu_{\text{lat}}^2}{9}X_I^2 - \frac{\mu_{\text{lat}}^2}{18}X_A^2 \right. \right.$$
$$\left.\left. + \frac{5\sqrt{2}\mu_{\text{lat}}}{6}\epsilon_{IJK}X^I X^J X^K \right) \right\rangle. \quad (4.16)$$

We focus our analyses on the following four observables, again employing dimensionless quantities that connect smoothly between the continuum and lattice theories:

- The internal energy. As derived above, on the lattice, this is

$$\frac{\widehat{E}}{N^2} \equiv \frac{E}{\lambda^{1/3}N^2} = \frac{1}{4N\lambda_{\text{lat}}^{4/3} N_\tau} \left\langle \sum_{n=0}^{N_\tau-1} \text{Tr}\left( -\frac{3}{2}\sum_{i<j}[X_i, X_j]^2 - \frac{2\mu_{\text{lat}}^2}{9}X_I^2 - \frac{\mu_{\text{lat}}^2}{18}X_A^2 \right. \right.$$
$$\left.\left. + \frac{5\sqrt{2}\mu_{\text{lat}}}{6}\epsilon_{IJK}X^I X^J X^K \right) \right\rangle. \quad (4.17)$$

- The scalar-trilinear term, also known as the Myers term. In the continuum, we define this as the dimensionless quantity

$$\widehat{M} \equiv \frac{M}{\lambda} = \frac{\sqrt{2}}{12N} \frac{1}{\lambda\beta} \left\langle \int d\tau\, \epsilon_{IJK}\, \text{Tr}\left(X_I X_J X_K\right) \right\rangle. \quad (4.18)$$

On the lattice, it takes the form

$$\widehat{M} = \frac{\sqrt{2}}{12N\lambda_{\text{lat}} N_\tau} \left\langle \sum_{n=0}^{N_\tau-1} \epsilon_{IJK}\, \text{Tr}\left(X_I X_J X_K\right) \right\rangle. \quad (4.19)$$

- The Polyakov loop magnitude. The Polyakov loop is the holonomy around the time direction and is the order parameter for the confinement transition in the large-$N$ limit. We have

$$|P| = \left\langle \left| \frac{1}{N}\text{Tr}\, \mathcal{P}\exp\left[ -\int_0^\beta d\tau\, A_\tau \right] \right| \right\rangle, \quad (4.20)$$

where $\mathcal{P}\exp$ is the path-ordered exponential. Translating this on to the lattice,

$$|P| = \left\langle \left| \frac{1}{N}\text{Tr}\left( \prod_{n=0}^{N_\tau-1} U(n) \right) \right| \right\rangle. \quad (4.21)$$

- The 'extent of space' — terminology motivated by the holographic dual of the full BMN

model — which is given by the sum of the squared scalars. In the continuum, we consider the dimensionless quantity

$$\widehat{R}^2 \equiv \frac{R^2}{\lambda^{2/3}} = \frac{1}{2N\lambda^{2/3}\beta} \left\langle \int d\tau \, \text{Tr}\left(X_i^2\right) \right\rangle. \tag{4.22}$$

On the lattice, this becomes

$$\widehat{R}^2 = \frac{1}{2N\lambda_{\text{lat}}^{2/3} N_\tau} \left\langle \sum_{n=0}^{N_\tau-1} \text{Tr}\left(X_i^2\right) \right\rangle. \tag{4.23}$$

In addition, to help identify and characterize transitions, we also consider two further observables related to those above:

- The susceptibility of the Polyakov loop magnitude,

$$\chi \equiv N^2 \left( \langle |P|^2 \rangle - \langle |P| \rangle^2 \right). \tag{4.24}$$

- The specific heat, which on the lattice takes the form

$$C_V \equiv \frac{\lambda_{\text{lat}}^{2/3} N_\tau^2}{N^2} \left\langle \left(\widehat{E} - \left\langle \widehat{E} \right\rangle\right)^2 - \widehat{E}' \right\rangle, \tag{4.25}$$

with $\widehat{E}$ from Eq. (4.17) and

$$\widehat{E}' = \frac{N}{4\lambda_{\text{lat}}^{5/3} N_\tau^2} \left\langle \sum_{n=0}^{N_\tau-1} \text{Tr}\left( -3\sum_{i<j}[X_i, X_j]^2 - \frac{2\mu_{\text{lat}}^2}{9} X_I^2 - \frac{\mu_{\text{lat}}^2}{18} X_A^2 \right.\right.$$
$$\left.\left. + \frac{5\sqrt{2}\mu_{\text{lat}}}{4} \epsilon_{IJK} X^I X^J X^K \right) \right\rangle. \tag{4.26}$$

The behavior of these six observables in different regimes can be understood as follows:

- Internal energy (given by Eq. (4.17)), Extent of space (given by Eq. (4.23)) and Myers term (given by Eq. (4.19)) behave similarly. Before transition, all these observables are constant. As the temperature is increased, these observables showed a sharp change in their value, indicating a phase transition. After the transition, there is no sharp change in the observables. Rather, they change as a smooth function of temperature. For BFSS (i.e, $\widehat{\mu} = 0$), this behavior at high temperatures is studied using non-lattice calculations [94].

For the BMN model, though, the exact dependence on temperature for these observables needs probing.

- Polyakov loop (given by Eq. (4.21)) before transition represents the configuration in the confined phase. Hence $|P| = 0$ is expected in this phase. But as in the numerical simulations, this magnitude of the Polyakov loop is obtained from its real and imaginary parts i.e., $|P| = \sqrt{P_{\text{real}}^2 + P_{\text{imag}}^2}$, which are not zero rather they wander around zero. Hence in the numerical runs confined phase can be observed when $P = 1/N$. Around the transition, the Polyakov loop shows a sharp transition, and it slowly moves towards the deconfined phase where $|P| = 1$. For high temperatures when the theory is not confined, the behavior of the Polyakov loop can again be predicted from non-lattice simulations [94].

- Internal energy, Extent of space, Myers term, and Polyakov loop shows sharp change around transition. Hence we can check for the exact transition point by tuning their susceptibilities. At the transition point, their susceptibilities will peak, and that will be the critical transition point. The susceptibility of the Polyakov loop is one such observable (given by Eq. (4.24)). The expression of internal energy already contains the contribution from the scalars quadratic term and the trilinear term. Hence we will not tune susceptibilities of scalars extent and Myers term independently. The susceptibility of internal energy can be interpreted from the specific heat term (given by Eq. (4.25)).

Considering both the Polyakov loop susceptibility and the specific heat will allow us to search separately for the confinement transition and the Myers transition reported by Ref. [69] (for the full BMN model). The latter is signaled by the energy (in addition to the Myers term) and hence by the specific heat.

In contrast to lattice studies of finite-temperature transitions in higher-dimensional field theories, the absence of any spatial volume in lattice discretizations of matrix models means that the thermodynamic limit corresponds to increasing the number of colors, $N \to \infty$. Prior studies of supersymmetric matrix models [65, 66] have found that $N \geq 16$ is needed to control finite-$N$ artifacts in this context [70]. We, therefore, consider $N = 16$, 32, and 48, finding that we need to increase $N$ as $\widehat{\mu}$ decreases in order to keep finite-$N$ artifacts negligible compared to our statistical precision.

In addition to the $N \to \infty$ thermodynamic limit, lattice analyses also need to consider the $N_\tau \to \infty$ continuum limit. We did some preliminary calculations with $N_\tau = 16, 24$ and $32$ for $N = 16$ and $\widehat{\mu} = 0.0, 6.0$. These calculations indicated that $N_\tau = 24$ appears sufficient to keep finite-$N_\tau$ artifacts negligible compared to our statistical precision. Therefore we proceed by fixing $N_\tau = 24$ in our calculations. This is also consistent with earlier lattice studies [69, 75, 86, 89].

| $\widehat{\mu}$ | $N$ | $\widehat{T}_c$ | $\Delta$ |
|---|---|---|---|
| 0.5 | 48 | 0.900(2) | 0.000(3) |
| 1.0 | 48 | 0.903(1) | 0.000(2) |
| 2.0 | 48 | 0.912(1) | 0.000(2) |
| 4.0 | 32 | 0.949(2) | 0.001(4) |
| 6.0 | 32 | 1.016(4) | 0.004(6) |
| 9.0 | 16 | 1.158(8) | 0.000(20) |
| 10.0 | 16 | 1.213(8) | 0.010(26) |
| 11.0 | 16 | 1.275(3) | 0.000(10) |
| 13.0 | 16 | 1.398(8) | 0.012(18) |
| 15.0 | 16 | 1.531(9) | 0.012(23) |
| 21.54 | 16 | 2.04(3) | 0.01(6) |
| 44.66 | 16 | 4.00(3) | – |

TABLE 4.1: The critical temperature $\widehat{T}_c$ for the 12 $\widehat{\mu}$ values considered in this work. As $\widehat{\mu}$ decreases, larger $N$ is needed. We determine $\widehat{T}_c$ either from the peak of the Polyakov loop susceptibility or from the separatrix method. The last column shows the values of the $\Delta$ parameter defined by Eq. (4.28). $\lambda_{\text{lat}}$ for the runs is given by $\lambda_{\text{lat}} = (\hat{T} N_\tau)^{-3}$, where $N_\tau$ is fixed as 24 throughout the runs.

## 4.2  Determination of the critical temperature

As a first look at our lattice results, we collect some representative plots for the six observables summarized above. In Fig. 4.1, we consider $\widehat{\mu} = 1$ with $N_\tau = 24$, scanning the small range $0.896 \leq \hat{T} \leq 0.906$ around the transition and observing clear growth in the Polyakov loop susceptibility and specific heat peaks as the number of colors increases from $N = 32$ to $48$. We see similar behavior in Fig. 4.2 for $\widehat{\mu} = 6$ and $N_\tau = 24$ with $1 \leq \hat{T} \leq 1.03$, here comparing $N = 16$ and $32$. Ref. [92] provides a comprehensive release of our data, including full accounting of statistics, auto-correlation times, and other observables computed in addition to the six highlighted in Figs. 4.1 and 4.2. The wider temperature range plot for the configurations with $\widehat{\mu} = 1$ with $N_\tau = 24$ are shown in Figure 4.3 with $N = 32$.

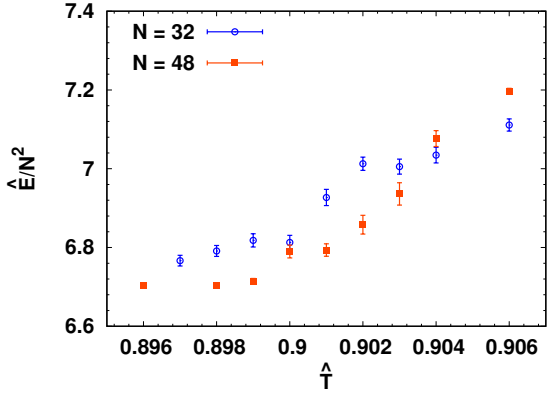

(a) Internal energy, Eq. (4.17)

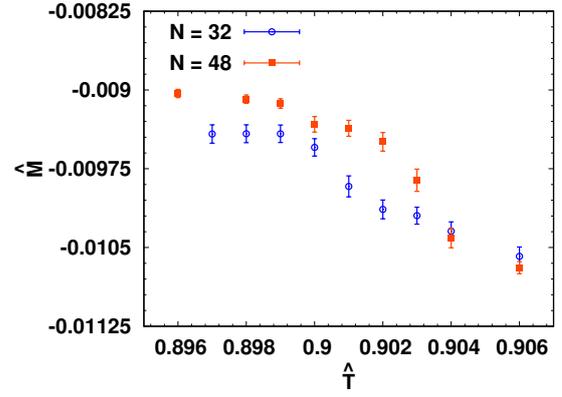

(b) Myers term, Eq. (4.19)

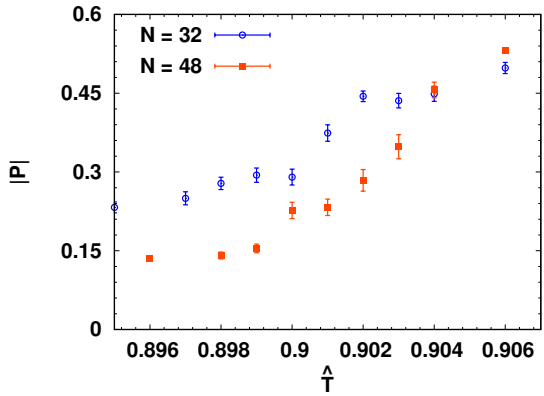

(c) Polyakov loop, Eq. (4.21)

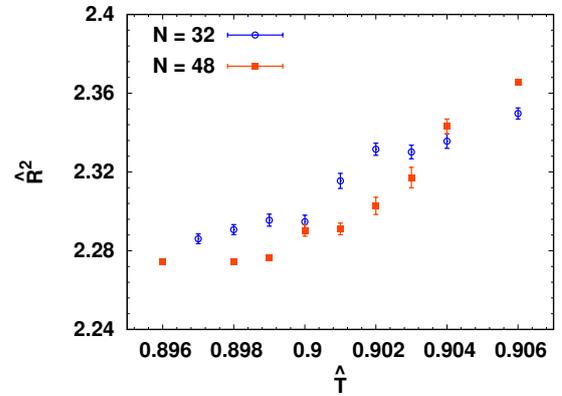

(d) Extent of space, Eq. (4.23)

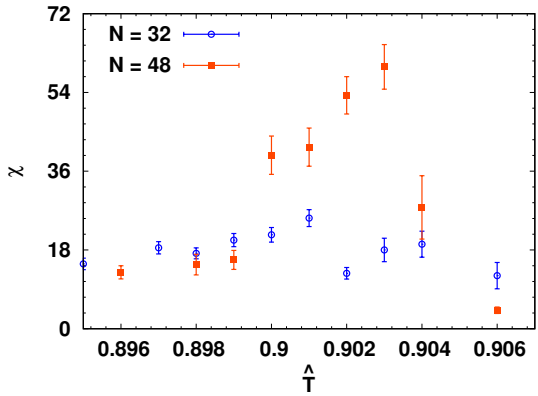

(e) Polyakov loop susceptibility, Eq. (4.24)

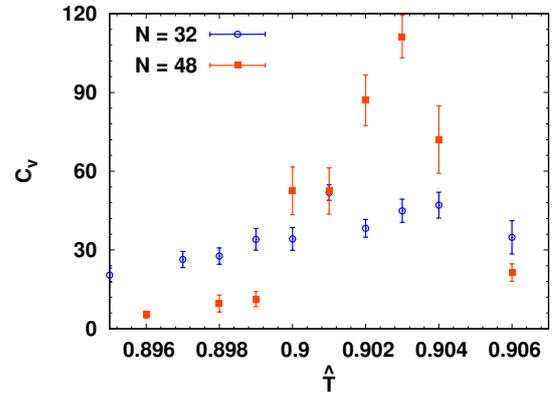

(f) Specific heat, Eq. (4.25)

FIGURE 4.1: The six observables discussed in the text, for $\widehat{\mu} = 1$ and $N_\tau = 24$, in a small range of temperatures $0.896 \leq \widehat{T} \leq 0.906$ around the transition. As $N$ increases from 32 to 48, there is clear growth in the Polyakov loop susceptibility and the specific heat peaks in the final row.

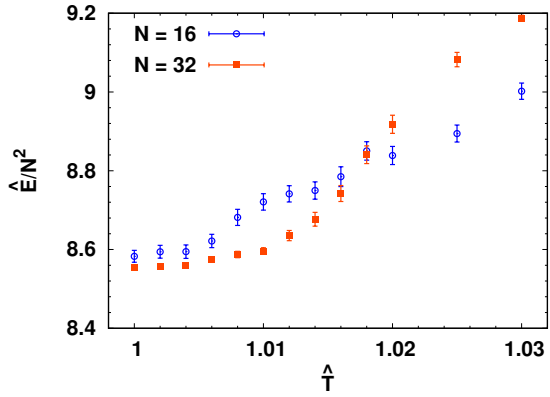

(a) Internal energy, Eq. (4.17)

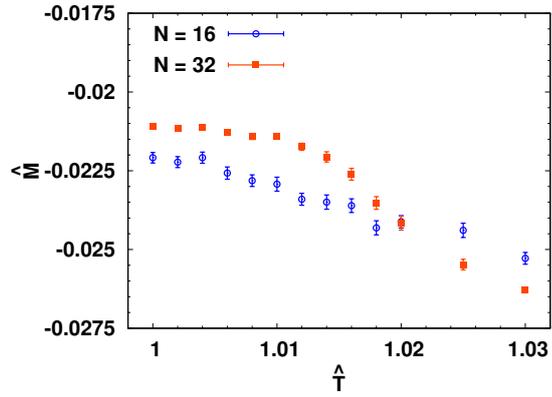

(b) Myers term, Eq. (4.19)

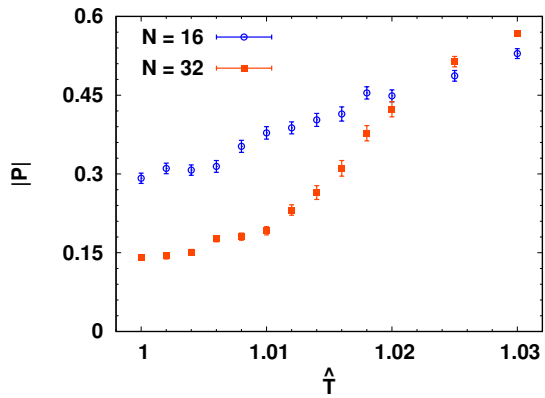

(c) Polyakov loop, Eq. (4.21)

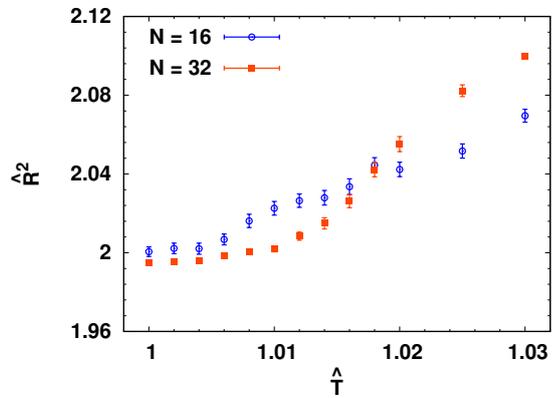

(d) Extent of space, Eq. (4.23)

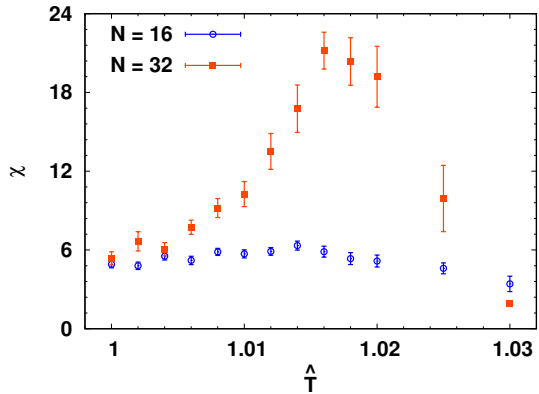

(e) Polyakov loop susceptibility, Eq. (4.24)

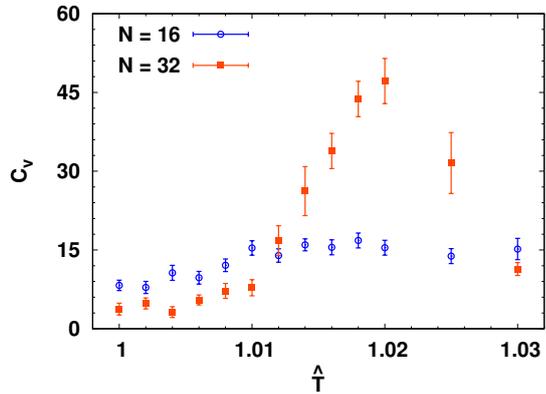

(f) Specific heat, Eq. (4.25)

FIGURE 4.2: The six observables discussed in the text, for $\widehat{\mu} = 6$ and $N_\tau = 24$, in a small range of temperatures $1 \leq \widehat{T} \leq 1.03$ around the transition. As in Fig. 4.1, increasing $N$ from 16 to 32 produces clear growth in the Polyakov loop susceptibility and the specific heat peaks in the final row.

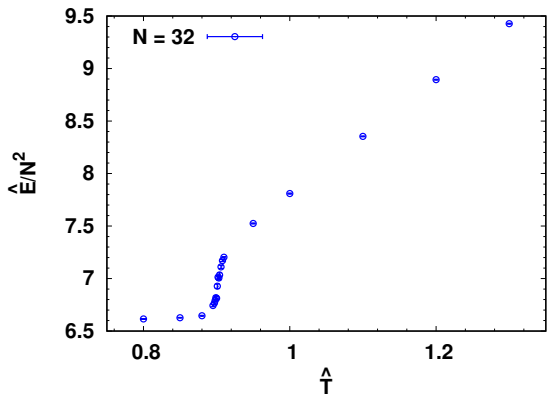

(a) Internal energy, Eq. (4.17)

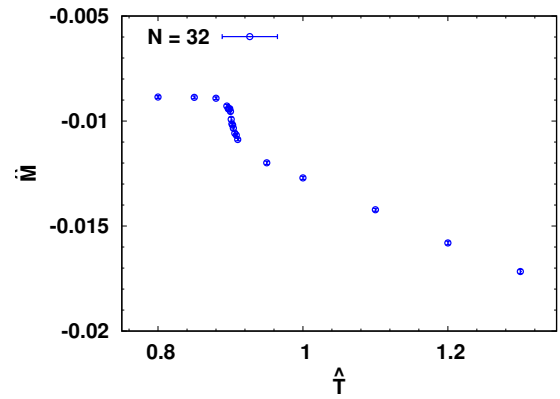

(b) Myers term, Eq. (4.19)

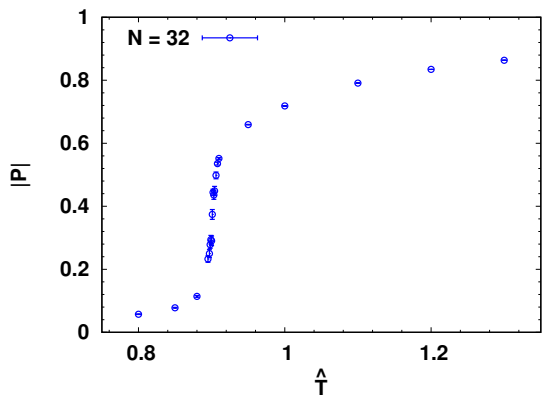

(c) Polyakov loop, Eq. (4.21)

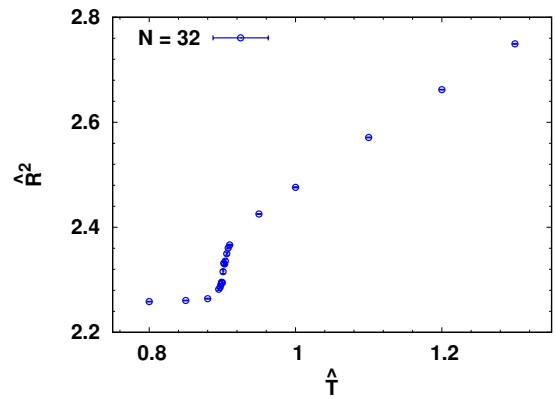

(d) Extent of space, Eq. (4.23)

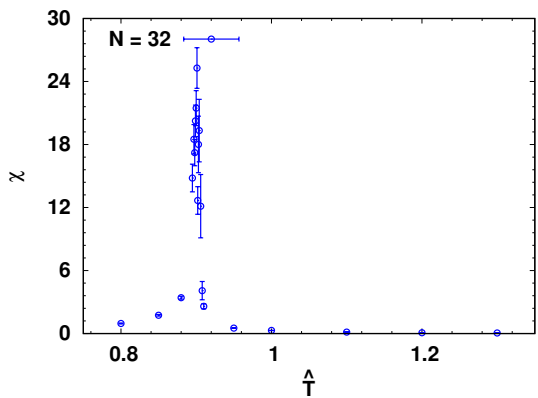

(e) Polyakov loop susceptibility, Eq. (4.24)

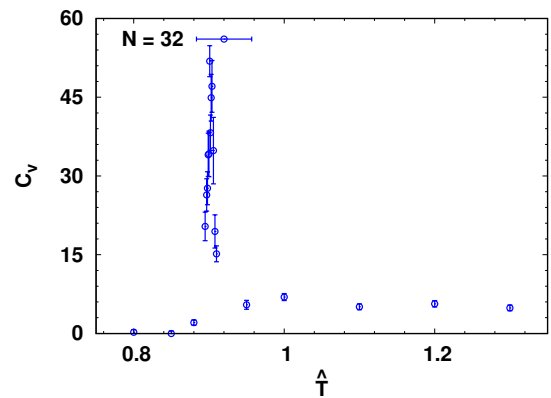

(f) Specific heat, Eq. (4.25)

FIGURE 4.3: The six observables discussed in the text, for $\widehat{\mu} = 1$ and $N_\tau = 24$, with a wide range of temperatures for $N = 32$.

Using $N_\tau = 24$, we performed similar scans in the temperature for all 12 values of $0.5 \leq \widehat{\mu} \leq 44.66$ listed in Table 4.1. In addition to identifying the critical temperature $\widehat{T}_c$ from the peak in the Polyakov loop susceptibility, we also carry out analyses using the *separatrix* method. This is a novel way to determine the critical temperature away from the thermodynamic limit, which can work well even when susceptibility peaks are difficult to resolve. While Ref. [95] introduced the Polyakov loop separatrix method specifically for (non-supersymmetric) SU(3) Yang–Mills theory, it generalizes to $N > 3$. The idea is to consider the unit disk in the plane of the real and imaginary parts of the Polyakov loop and separate this into two regions such that Polyakov loop measurements for deconfined ensembles fall predominantly in one region while those from confined ensembles fall predominantly in the other. A simple ratio $S(\widehat{T})$ is defined, which is given by configurations inside the disk by the total number of configurations.

- In the confined phase, all the configurations are centered around the zero value of the scatter plot, hence the separatrix ratio, $S(\hat{T}) \approx 1$.

- In the deconfined phase, all the configurations become deconfined, and they sit in one of the vacua, leaving no configurations inside the disk, hence then the separatrix ratio, $S(\hat{T}) \approx 0$.

- During the transition, configurations start moving from the confined to the deconfined phase, and hence when half of the configurations are deconfined, we consider it as the transition point. At that point $S(\hat{T}) = 0.5$.

Therefore a simple ratio $S(\widehat{T})$ then changes from 0 deep in the deconfined phase to 1 deep in the confined phase, with the transition identified as the (interpolated) point where this ratio crosses $0.5$.[3] Fig. 1 in Ref. [95] illustrates the equilateral triangle used as the SU(3) separatrix. As we are working with different $N$ values, our separating disk will be a polygon with $N$ sides. For example, if we are working with $N = 4$, the separating disk will be in the form of a square. For very large values of $N$, a polygon can be roughly replaced by a circle. For our $N \geq 16$, we approximate the corresponding $N$-gon by a circle of radius $r_S < 1$, so that we just have to consider the Polyakov loop magnitude $|P|$. This circle with radius $r_S$ will act as a separator disk between confined and deconfined configurations.

We treat the radius $r_S$ as an adjustable parameter,[4] which introduces a systematic uncer-tainty from our choice of $r_S$. For all $\widehat{\mu}$ we consider, we find $r_S = 0.4$ provides stable and reliable results. We also observe that the systematic dependence on our choice of $r_S$ becomes less significant as $\widehat{\mu}$ increases. In Figs. 4.4 and 4.5, we show representative Polyakov loop scatter plots, separatrices, and the resulting $S(\widehat{T})$ for $\widehat{\mu} = 1$ with $N = 32$ and $\widehat{\mu} = 44.66$ with $N = 16$, respectively. In Fig. 4.6, we have plotted the separatrix ratio with the different radii of the circle that is used as a separator. The plot is for $\widehat{\mu} = 2.0$ and $N = 32$. It shows how with different sizes of circles, we can still get good precision of critical transition point up to certain orders.

---

[3]Ref. [75, 86] considers a similar ratio but identifies the transition as the point where it becomes non-zero.
[4]In Ref. [95], the separatrix is defined using the position of the minimum between two peaks in the distribution

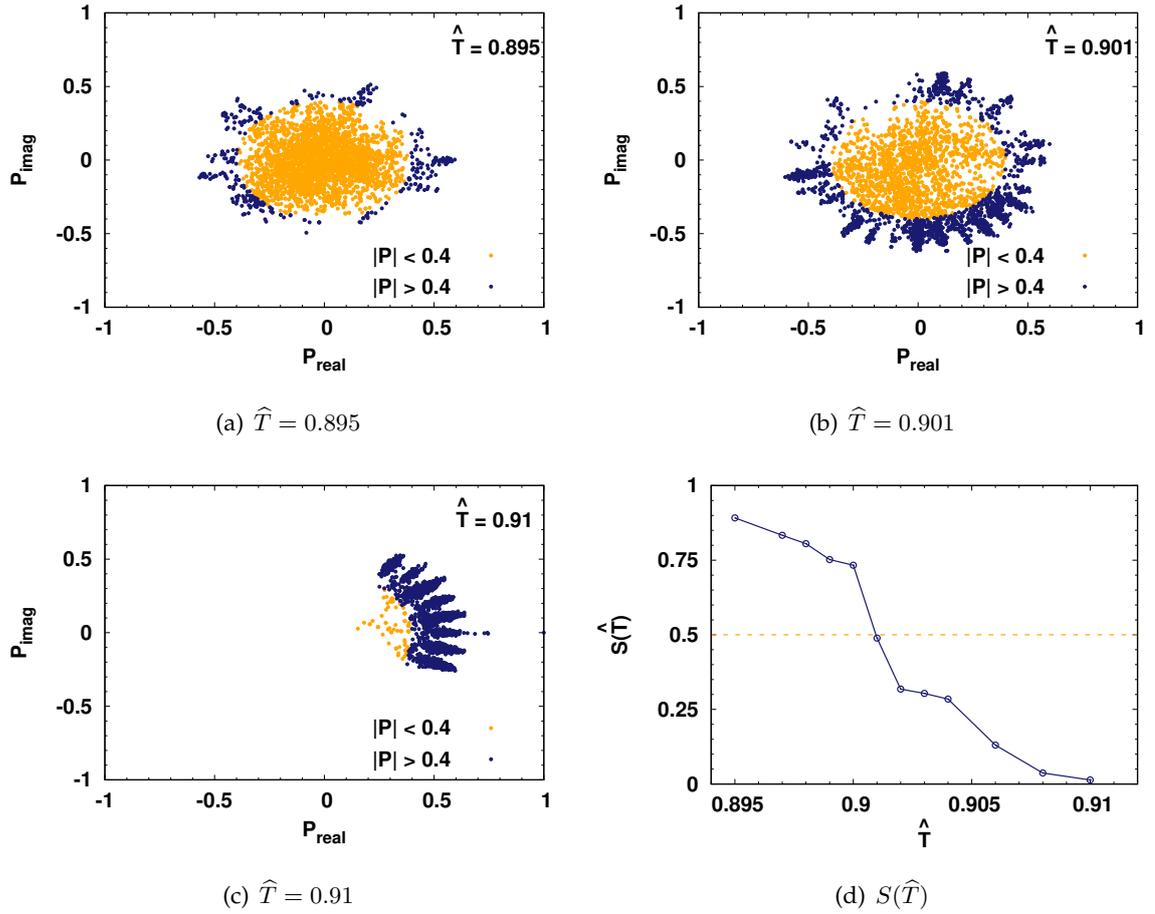

FIGURE 4.4: Polyakov loop scatter plots for $N = 32$ with $\widehat{\mu} = 1$ for three values of $\widehat{T}$. The separatrix is a circle of radius $r_S = 0.4$ shown by the two colors. Panel (d) shows the resulting $S(\widehat{T})$ that identifies the critical temperature $\widehat{T}_c \approx 0.901$. (Note the $\widehat{T}_c = 0.903(1)$ in Table 4.1 comes from $N = 48$.)

The $N$ eigenvalues of the Polyakov loop provide yet another means both to estimate the critical temperature and to characterize the phases between which the system transitions. Deep in the confined phase, the angular distribution of these eigenvalues is uniform around the unit circle, while deep in the deconfined phase, their distribution is localized around some angle,

---

of Polyakov loop measurements.

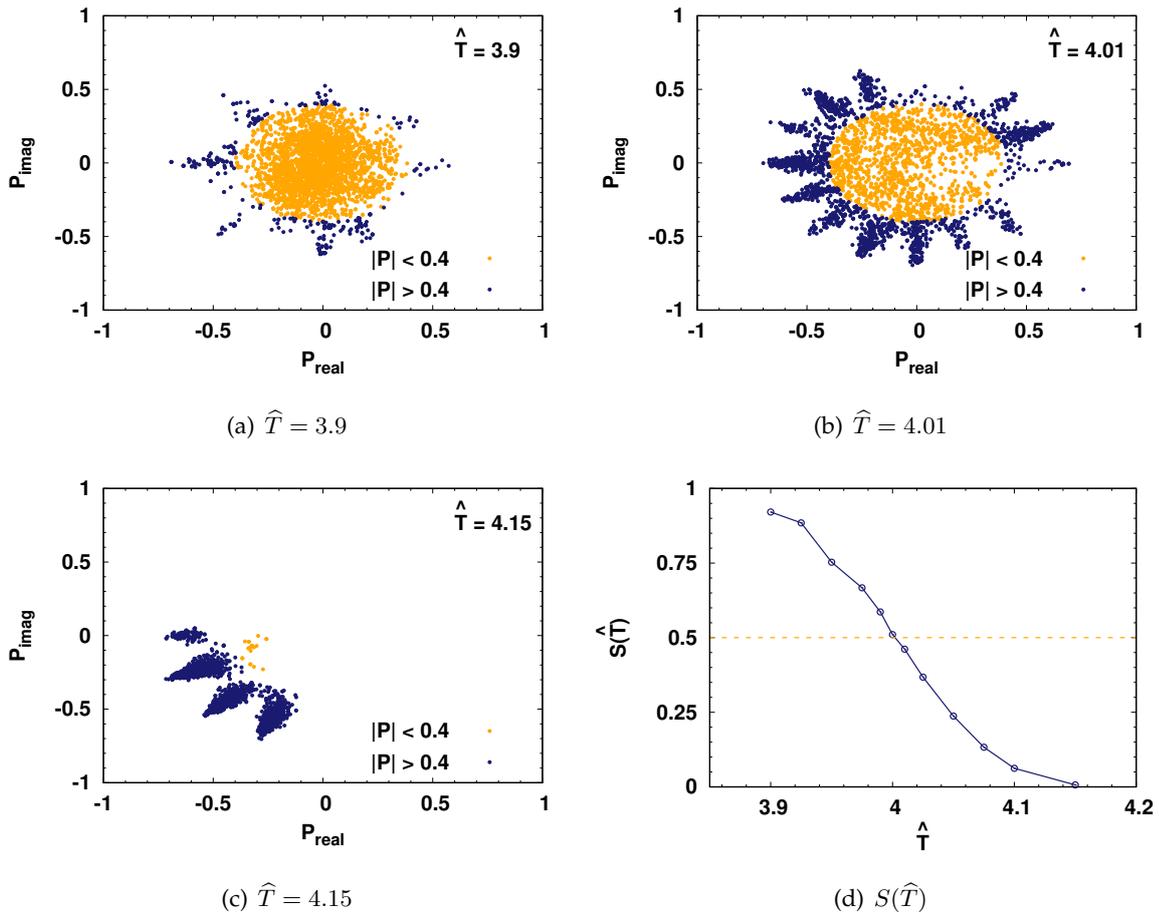

FIGURE 4.5: Polyakov loop scatter plots for $N = 16$ with $\widehat{\mu} = 44.66$ for three values of $\widehat{T}$ and the same $r_S = 0.4$ circular separatrix as Fig. 4.4. The resulting $S(\widehat{T})$ in panel (d) identifies the critical temperature $\widehat{T}_c \approx 4$.

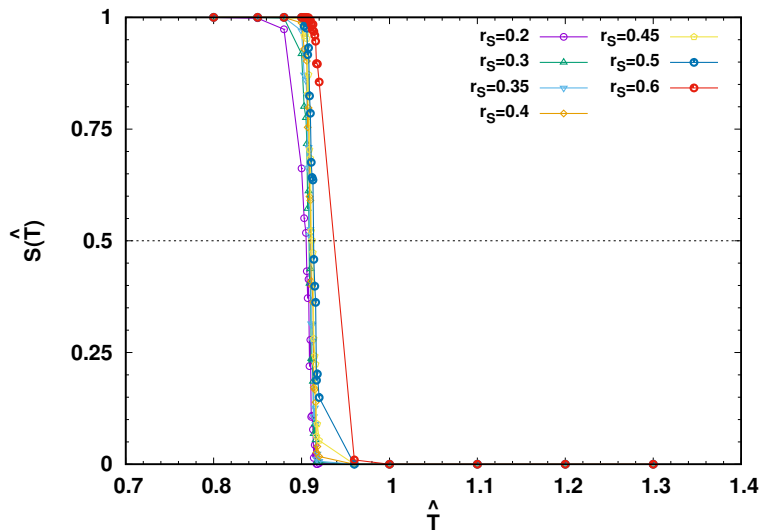

FIGURE 4.6: Sepatarix ratio vs temperature for $\widehat{\mu} = 2.0$ and $N = 32$ for various radii. It can be seen that the radius of the circle used as a separator has less effect on finding critical temperature up to certain order of precision.

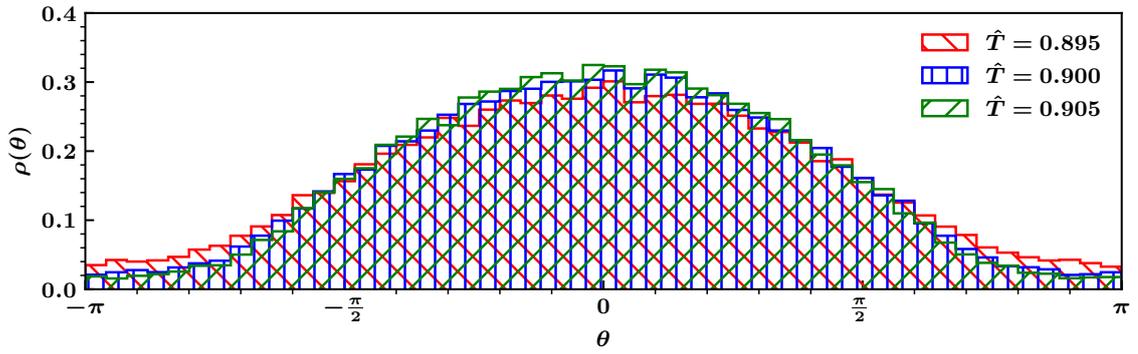

(a) $N = 16$

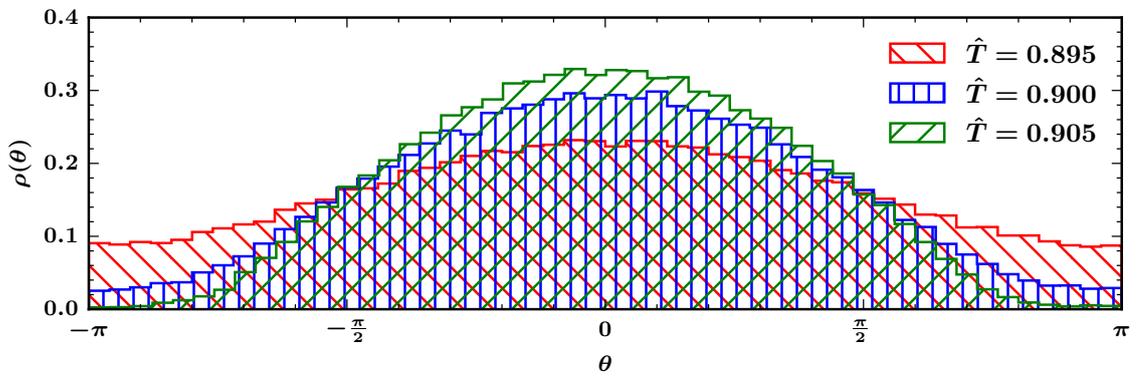

(b) $N = 32$

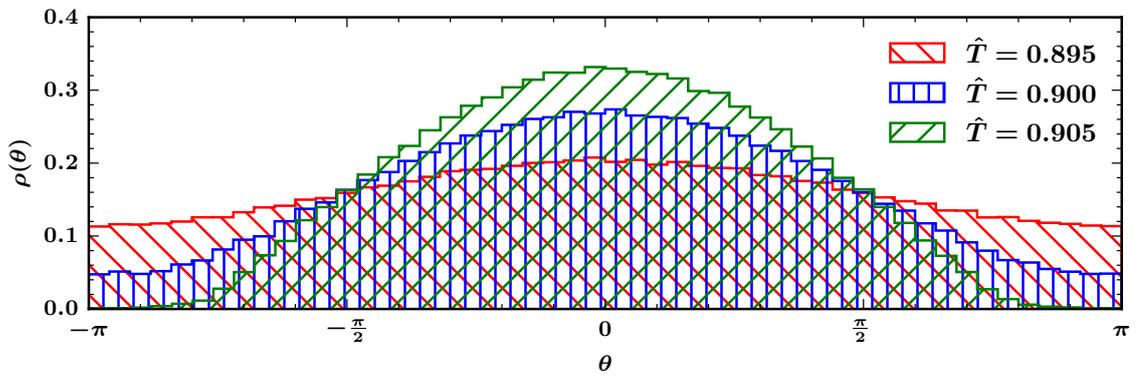

(c) $N = 48$

FIGURE 4.7: Angular distributions of Polyakov loop eigenvalues for $\widehat{\mu} = 0.5$ with $N = 16$, 32 and 48, considering three temperatures around the critical $\widehat{T}_{\mathrm{c}} = 0.900(2)$. A gap appears for $\widehat{T} = 0.905 > \widehat{T}_{\mathrm{c}}$, while the distribution for $\widehat{T} = 0.895 < \widehat{T}_{\mathrm{c}}$ becomes more uniform as $N$ increases.

which we can set to $\theta = 0$ by convention. This behavior can be modeled as

$$\rho(\theta) = \frac{1}{2\pi} + \frac{1}{q\pi} \cos\theta \qquad -\pi \leq \theta < \pi, \qquad (4.27)$$

where the positive parameter $q \to \infty$ in the uniform limit, while a gap opens for $q < 2$. In Fig. 4.7, we show a representative example of this gap opening at the critical $\widehat{T}_c$ identified from the Polyakov loop susceptibility and separatrix, confirming that the corresponding transition is between the uniform confined phase and the gapped deconfined phase. In this figure, we consider our most challenging data set with the smallest $\widehat{\mu} = 0.5$ and three $N = 16, 32$, and $48$. Comparing these three values of $N$ allows us to confirm that the transition becomes sharper as $N$ increases: the distribution for $\widehat{T} = 0.895 < \widehat{T}_c$ becomes more uniform for larger $N$ while the size of the gap for $\widehat{T} = 0.905 > \widehat{T}_c$ also increases.

Contrary to the challenging $\widehat{\mu} = 0.5$ simulations with different $N$ values, if we try to look at the scatter plot of the Polyakov loop distribution and its corresponding eigenvalues distribution, the different phases can be easily distinguished. Let us look at the scatter plot and Polyakov loop eigenvalues with $\widehat{\mu} = 2.0$ for three different temperatures. One temperature is near the computed transition temperature $\widehat{T} = 0.913$, one significantly before the transition $\widehat{T} = 0.8$, and the other significantly after the transition $\widehat{T} = 1.3$. It can be seen from Figs. 4.8 and 4.9 that as the temperature changes from low to high and passes via the transition point, the scatter plot easily captures the confined and deconfined phase information, and the same can be seen from eigenvalues distribution plot which changes from uniform phase to non-uniform phase and finally to gapped phase.

Finally, we also check that the Myers transition and the confinement transition occur at the same critical temperature by defining

$$\Delta \equiv \left| \widehat{T}_c(C_V) - \widehat{T}_c(\chi) \right| \qquad (4.28)$$

to quantify the difference between the locations of the specific heat and Polyakov loop susceptibility peaks. Our results for $\Delta$ in Table 4.1 vanish within uncertainties for all $\widehat{\mu} < 40$, consistent with the existence of only a single phase transition. For $\widehat{\mu} = 44.66$, we do not observe well-defined peaks and rely on the separatrix method to determine $\widehat{T}_c$. If there were separate phase transitions signaled by these observables, in order to be consistent with these results, their critical temperatures would need to be too close to resolve with $N \leq 48$.

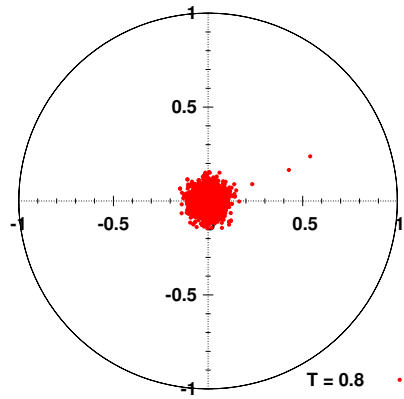

(a) $\widehat{T} = 0.8$

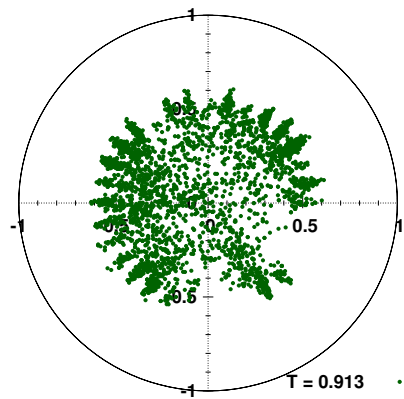

(b) $\widehat{T} = 0.913$

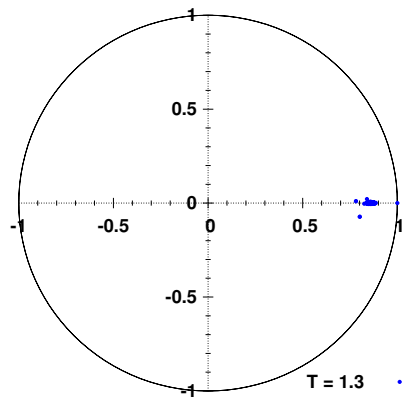

(c) $\widehat{T} = 1.3$

FIGURE 4.8: Polyakov loop scatter plot for $\widehat{\mu} = 2$ and $N_\tau = 24$, with three different temperatures for $N = 32$. a) Temperature well before transition, b) Temperature around transition, c) Temperature way after the transition.

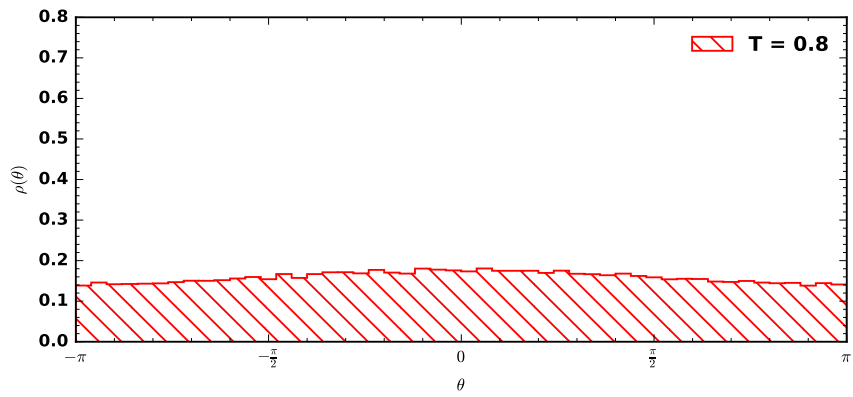

(a) $\widehat{T} = 0.8$

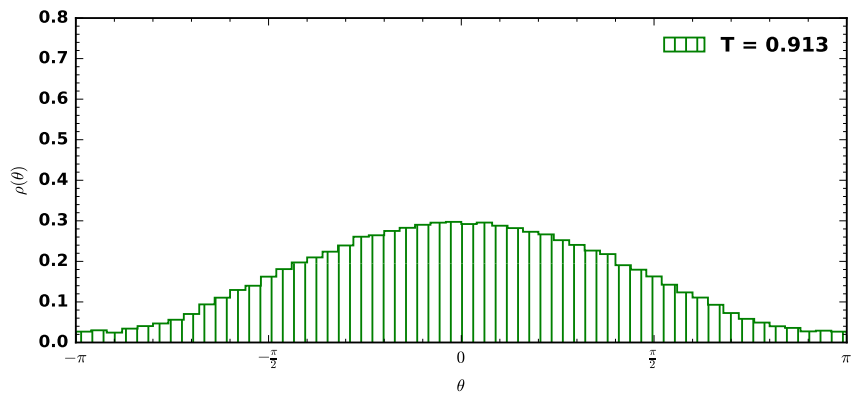

(b) $\widehat{T} = 0.913$

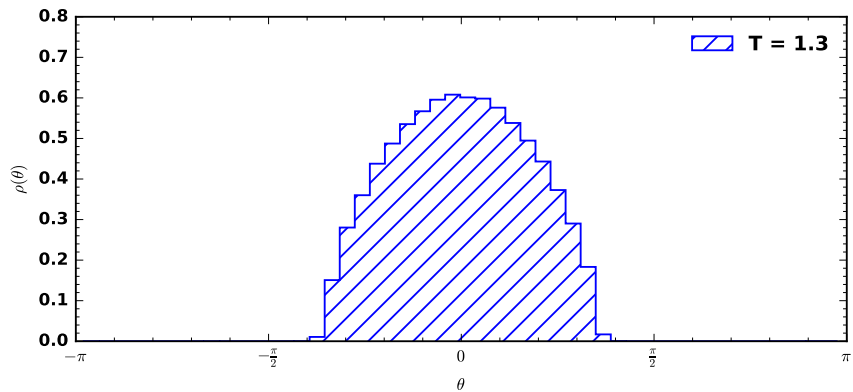

(c) $\widehat{T} = 1.3$

FIGURE 4.9: Polyakov loop eigenvalues plot for $\widehat{\mu} = 2$ and $N_\tau = 24$, with three different temperatures for $N = 32$. a) Temperature well before transition, b) Temperature around transition, c) Temperature way after the transition.

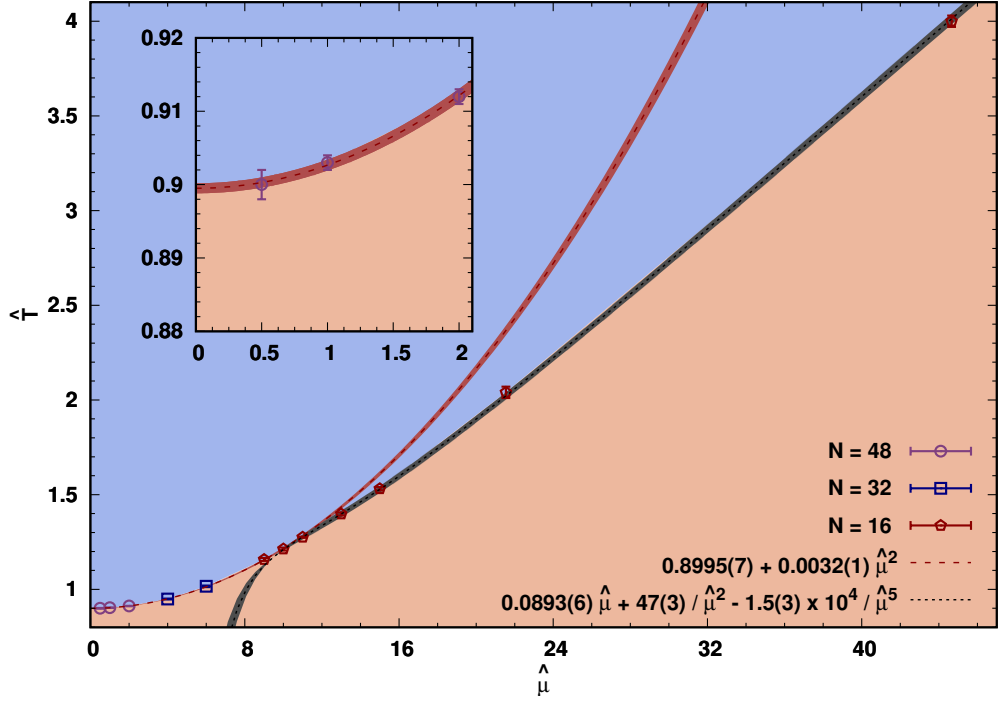

FIGURE 4.10: The $\widehat{T}$–$\widehat{\mu}$ phase diagram of the bosonic BMN model from our $N_\tau = 24$ results for $\widehat{T}_c$ computed with $N = 16$, 32 and 48. The red curve shows our fit of the $\widehat{\mu} \geq 10$ results to Eq. (4.33), while the blue curve shows our fit of the $\widehat{\mu} \leq 10$ results to Eq. (4.34). Our results self-consistently identify the $\widehat{\mu}_\star \sim 10$ separating the small- and large-$\widehat{\mu}$ regimes. The inset zooms in on the $\widehat{\mu} \to 0$ limit. The shaded regions indicate two different phases, the confined phase (orange shade) and the deconfined phase (blue shade).

## 4.3 Critical temperature dependence on deformation parameter

Figure 4.10 plots our critical temperature results obtained above to visualize the phase diagram of the bosonic BMN model in the $\widehat{T}$–$\widehat{\mu}$ plane. We now analyze the dependence of $\widehat{T}_c$ on the deformation parameter. Considerations of the simplified $\widehat{\mu} \to \infty$ and $\widehat{\mu} \to 0$ limits suggest that this dependence must differ for large vs. small $\widehat{\mu}$.

Our results confirm this and also identify the $\widehat{\mu}_\star \sim 10$ separating these two regimes.

First, for large $\widehat{\mu}$ the expected critical temperature is easy to calculate using the argument [90, 91] that for a model with $D > 1$ matrices (bosonic or fermionic), of masses $\omega_j > 0$ with $j = 1, \cdots, D$, the inverse critical temperature $\beta_c$ is given by the solution of

$$\sum_{j=1}^{D} e^{-\beta \omega_j} = 1. \tag{4.29}$$

For the case of the bosonic BMN model with $D = 3 + 6$, the large-$\mu$ critical temperature is the solution of

$$3e^{-\beta\mu/3} + 6e^{-\beta\mu/6} - 1 = 0, \tag{4.30}$$

which gives

$$\frac{1}{\mu\beta_c} = \frac{T_c}{\mu} = \frac{\widehat{T}_c}{\widehat{\mu}} = \frac{1}{6\ln(3 + 2\sqrt{3})} = 0.089305\ldots \tag{4.31}$$

A similar analysis can be done for the full BMN model, which reduces to a supersymmetric Gaussian model in the $\widehat{\mu} \to \infty$ limit. In this case, Refs. [91, 96, 97] have perturbatively computed the critical temperature of the confinement transition up to next-to-next-to-leading order in $1/\widehat{\mu}^3 \ll 1$

$$\widehat{T}_c = \frac{\widehat{\mu}}{12\ln 3}\left[1 + \frac{320}{3}\frac{1}{\widehat{\mu}^3} - \left(\frac{458321}{12} + \frac{1765769\ln 3}{144}\right)\frac{1}{\widehat{\mu}^6} + \mathcal{O}\left(\frac{1}{\widehat{\mu}^9}\right)\right]. \tag{4.32}$$

In the $\widehat{\mu} \to \infty$ limit, this produces a smaller $T_c/\mu \approx 0.076$ compared to the bosonic BMN result in Eq. (4.31). Motivated by the functional form of this perturbative result, we adopt the following ansatz to fit our $\widehat{T}_c$ results for sufficiently large $\widehat{\mu}$

$$\widehat{T}_c = \widehat{\mu}\left[C + H\frac{1}{\widehat{\mu}^3} + F\frac{1}{\widehat{\mu}^6}\right], \tag{4.33}$$

where we expect $C = 1/[6\ln(3 + 2\sqrt{3})] \approx 0.0893$ from Eq. (4.31). The fit to our results for $\widehat{\mu} \geq 10$ shown in Fig. 4.10 indeed produces $C = 0.0893(6)$, providing a good check of our numerical setup and code.

Also, from Fig. 4.10, we can observe that this ansatz fails for $\widehat{\mu}_\star \lesssim 10$, consistent with the expected breakdown of the perturbative expansion when the coupling $1/\widehat{\mu}^3$ is too large. In the small-$\widehat{\mu}$ regime, we must use a different fit form to describe the dependence of the critical temperature on the deformation parameter. While it might be possible to carry out a strong-coupling expansion in this regime, for the purposes of this work, we will simply employ an empirical expansion in powers of $\widehat{\mu}^2$. The recent Ref. [72] takes the same approach, employing the quadratic ansatz

$$\widehat{T}_c = A + B\widehat{\mu}^2, \tag{4.34}$$

where $A$ is the constant critical temperature of the bosonic BFSS model.

| Coefficient | Value | Reference |
|---|---|---|
| $A$ | 0.8846(1) | Ref. [72] |
| $A$ | 0.8995(7) | This work |
| $B$ | 0.00330(2) | Ref. [72] |
| $B$ | 0.0032(1) | This work |
| $C$ | 0.0893 | Refs. [90, 91] |
| $C$ | 0.0893(6) | This work |
| $H$ | 47(3) | This work |
| $F$ | $-15(3) \times 10^3$ | This work |

TABLE 4.2: Comparison of the values of the fit parameters appearing in Eqs. (4.33) and (4.34).

We will also use Eq. (4.34) to fit our $\widehat{T}_c$ results for small $\widehat{\mu}$. Using our conventions, Ref. [72] reports $A = 0.8846(1)$ and $B = 0.00330(2)$ from a fit to their critical temperature results for $0.375 \leq \widehat{\mu} \leq 3$ (i.e., $0.125 \leq \mu \leq 1$ in their conventions). The fit to our results for $0.5 \leq \widehat{\mu} \leq 10$ shown in Fig. 4.10 produces $A = 0.8995(7)$ and $B = 0.0032(1)$, with purely statistical uncertainties. While our result for $B$ agrees with Ref. [72], there is a clear tension in $A$. The inset in Fig. 4.10 makes it clear that our numerical results demand $A > 0.89$ regardless of the range of $\widehat{\mu}$ we include in our fit. Finite-$N$ and discretization artifacts could play a role in this disagreement. So far, we have considered only $N \leq 48$ rather than the $N \leq 64$ that Ref. [72] was able to reach. Although we use the same $N_\tau = 24$ as Ref. [72], our first-order lattice finite-difference operator in Eq. (4.4) differs from the second-order discretization they employ. The lattice action of Ref. [72] also includes a Faddeev–Popov term from gauge fixing to the static diagonal gauge, though we do not expect this gauge fixing to affect the critical temperature.

Table 4.2 summarizes our findings for the coefficients in Eqs. (4.33) and (4.34). We note that $H$ and $F$ are new predictions from this work. Comparing these with the perturbative computation for the full BMN model in Eq. (4.32), we see that our non-perturbative results for the bosonic case are both a few times larger: $H_{\text{BBMN}} \simeq 47$ compared to $H_{\text{BMN}} \simeq 8$, while $F_{\text{BBMN}} \simeq -15 \times 10^3$ compared to $F_{\text{BMN}} \simeq -4 \times 10^3$.

## 4.4 Order of the phase transition

We expect that the phase transition we observe is of the first order. We confirm this in two ways. First, in the left panel of Fig. 4.11, we plot the distribution of the Polyakov loop magnitude for three temperatures around the critical $\widehat{T}_c \approx 0.949$, for a representative $\widehat{\mu} = 4$ with

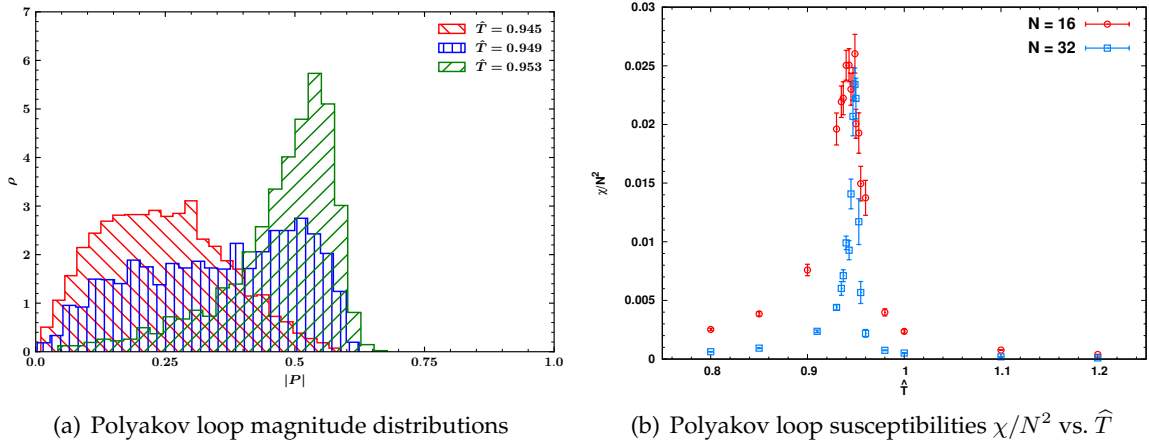

(a) Polyakov loop magnitude distributions  (b) Polyakov loop susceptibilities $\chi/N^2$ vs. $\widehat{T}$

FIGURE 4.11: Signs that the phase transition is first order, for $\widehat{\mu} = 4$. (a) The distribution of the Polyakov loop magnitude has support across two different regions for $\widehat{T}_c \approx 0.949$, suggesting a developing two-peak structure. (b) Normalizing the Polyakov loop susceptibility by $N^2$ produces the same peak height for $N = 16$ and $32$, indicating that the maximum $\chi$ is scaling with the number of degrees of freedom $\propto N^2$.

$N = 32$. While the $|P|$ distributions for $\widehat{T} = 0.945$ and $\widehat{T} = 0.953$ each have a single peak in two different regions, respectively corresponding to the confined and deconfined phases, the distribution for $\widehat{T} = 0.949$ has support across both of these regions. This suggests that a developing two-peak structure would be visible for larger $N > 32$, which is characteristic of phase coexistence at a first-order transition. The same can also be seen from the Monte Carlo time history of the Polyakov loop for temperature $\widehat{T} = 0.949$ as shown in Figure 4.13. To illustrate a more developed two-peak structure, Polyakov loop magnitude distribution is added for $\widehat{\mu} = 2.0$ with $N = 48$ in Figure 4.12.

Second, in the right panel of Fig. 4.11, we check the scaling of the Polyakov loop susceptibility with $N$. Because the thermodynamic limit for the bosonic BMN matrix model corresponds to $N^2 \to \infty$, this scaling can distinguish between first- and higher-order phase transitions [98, 99]. Considering the same $\widehat{\mu} = 4$, we plot $\chi/N^2$ against $T$ for both $N = 16$ and $32$. Within uncertainties, both values of $N$ produce the same peak height, again suggesting a first-order transition where the maximum $\chi$ would scale with the number of degrees of freedom.

## 4.5 Dependence of the internal energy on $\widehat{T}$ and $\widehat{\mu}$

Finally, we comment on the internal energy of the bosonic BMN model. Let us begin in the $\widehat{\mu} \to \infty$ limit, where this system reduces to a gauged Gaussian model with $D = 9$ scalar

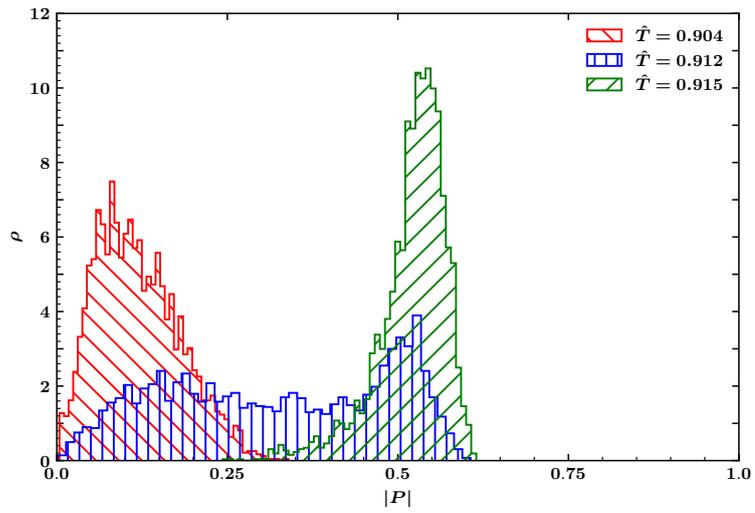

FIGURE 4.12: Polyakov loop magnitude distribution at three different temperatures for $\widehat{\mu} = 2.0$ with $N = 48$. A two-peak structure appears to develop more clearly as compared with lower $N$ values.

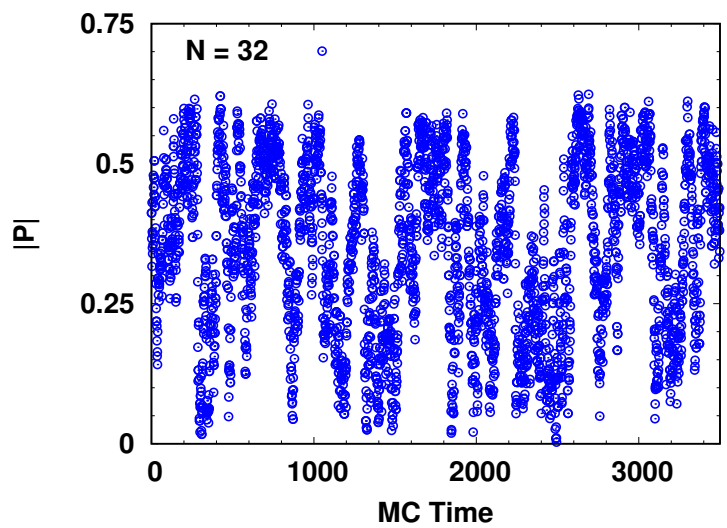

FIGURE 4.13: Polyakov loop time history over certain Monte Carlo simulation period for the configuration with $\widehat{T} = 0.949, N = 32, N_\tau = 24$. As can be seen from this history, the Polyakov loop continuously moves from one phase to other, indicating a first-order phase transition in this theory.

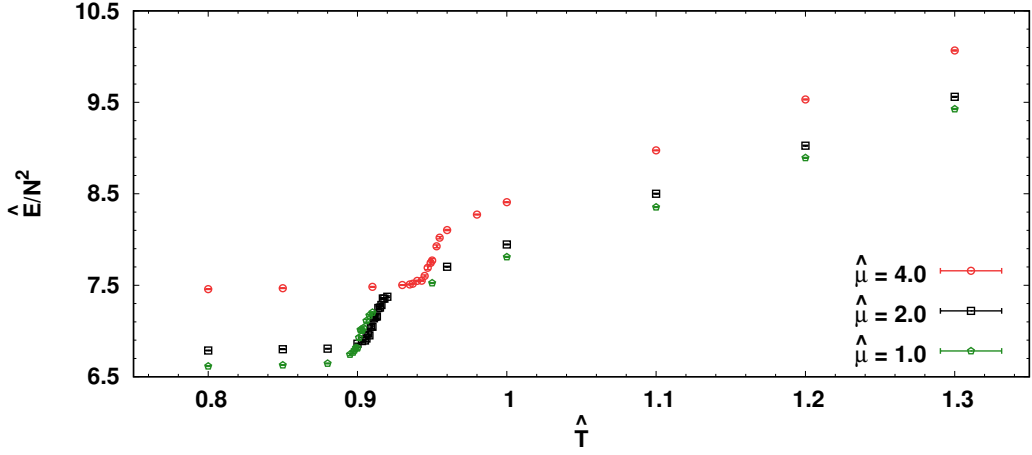

FIGURE 4.14: Internal energy vs. temperature for various $\widehat{\mu}$ values and $N = 32$. We see that for $\widehat{T} < \widehat{T}_c$, the energy has no temperature dependence, and after the transition, it depends linearly on temperature.

matrices. For general $D$, Ref. [52] computed that the internal energy of this Gaussian model is

$$\frac{1}{N^2}\widehat{E} = \frac{3}{4}(D-1)\widehat{T} + \mathcal{O}\Big(\frac{1}{N^2}\Big) \tag{4.35}$$

for high temperatures and large $N$. Plugging in the $D = 9$ relevant for the bosonic BMN model, we want to explore how this result will be modified for finite $\widehat{\mu}$. We expect that the relevant parameter will be $\frac{\widehat{T}}{\widehat{\mu}} = \frac{T}{\mu}$, so that

$$\frac{1}{N^2}\widehat{E} = 6\widehat{T}\Big[1 + f\left(\frac{T}{\mu}\right)\Big] + \mathcal{O}\Big(\frac{1}{N^2}\Big) \tag{4.36}$$

for some as-yet-unknown function $f$.

In Fig. 4.14, we show our lattice results for the temperature dependence of the energy for $\widehat{\mu} = 1$, 2, and 4 with $N = 32$. Although our lattice calculations focus on the transition regions, we do consider enough $\widehat{T} > \widehat{T}_c$ points in the deconfined phase to clearly see the leading-order linear dependence predicted by Eq. (4.35). This is in contrast to the $\widehat{T}$-independent energy in the $\widehat{T} < \widehat{T}_c$ confined phase. Although the energy depends on $\widehat{\mu}$ in both phases, we observe that the high-temperature slope is insensitive to the deformation parameter within the range $1 \leq \widehat{\mu} \leq 4$. A precise determination of $f(T/\mu)$ will be interesting to pursue through future generations of bosonic BMN lattice calculations.

## 4.6  Summary

In this chapter, we have presented a lattice study of the non-perturbative phase structure of the bosonic BMN matrix model. Our main results for the transition temperatures $\widehat{T}_c$ for twelve $0.5 \leq \widehat{\mu} \leq 44.66$, collected in Fig. 4.10, show how our numerical investigations smoothly connect the bosonic BFSS model in the $\widehat{\mu} \to 0$ limit to the known behavior of the $\widehat{\mu} \to \infty$ gauged Gaussian model. In addition to monitoring several standard observables and susceptibilities, we have also applied a novel separatrix method to determine these critical temperatures. We observed only a single transition, finding evidence that it is of the first order and analyzing the Polyakov loop eigenvalues to confirm that it separates the uniform confined phase from the gapped deconfined phase.

Using our results for $\widehat{T}_c$, we have investigated the functional forms of the dependence of the critical temperature on the deformation parameter in both the small- and large-$\widehat{\mu}$ regimes. Our results for the parameters of these functional forms, collected in Table 4.2, agree with some existing values in the literature and also provide some new predictions. However, there is a disagreement with the results from Refs. [72, 89] for $\widehat{T}_c$ in the $\widehat{\mu} \to 0$ bosonic BFSS limit, which deserves further investigation.

For these future investigations, we are particularly interested in exploring smaller $\widehat{\mu}$, which will require larger $N > 48$ to overcome challenges associated with flat directions and metastable vacua that make the numerical calculations more difficult. We also have lattice investigations of the full BMN model underway [73], with similar plans to pursue smaller $\widehat{\mu}$ with larger $N$.

We have already mentioned our ambition to determine the function $f(T/\mu)$ that modifies the dependence of the internal energy on the temperature and deformation parameter in Sec. 4.5. In that section, we also raised the possibility of generalizing the bosonic BMN model to a different number of scalar matrices, $D \neq 9$. It would be interesting to explore how the phase diagram and the order of phase transitions depend on $D$. Ref. [100] recently addressed this problem for the analogous generalization of the $\widehat{\mu} = 0$ bosonic BFSS model, investigating that system for a range of $D$ and concluding that the transition changes from first order to second order for $D \sim 36$. In the future, we hope to report how the BMN deformation affects this phenomenon. In the next chapter, we move towards another non-conformal theory, which is slightly more complicated than this one, the $\mathcal{N} = (2, 2)$ SYM in two dimensions with fermions.

# 5

# Non-perturbative study of Yang-Mills theory with four supercharges in two dimensions

*Content of this chapter is partially based on:*

- Pub. [2]: N. S. Dhindsa, R. G. Jha, A. Joseph, and D. Schaich, "Large-$N$ limit of two-dimensional Yang–Mills theory with four supercharges ", PoS **LATTICE2021** (2022) 433, arXiv:2109.01001 [hep-lat].

- Pub. [5]: N. S. Dhindsa, R. G. Jha, A. Joseph and D. Schaich, "Phase diagram of two-dimensional SU($N$) super-Yang–Mills theory with four supercharges", arXiv:2312.04980 [hep-lat].

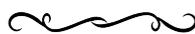

## 5.1 Motivation

In this chapter, we will present the study of SYM in two dimensions with four supersymmetries. The motivation to study this particular theory is to understand how close this theory is to its cousin which has a holographic dual.

It can be extremely hard to study strongly coupled supersymmetric Yang–Mills (SYM) theories using perturbative methods. A lattice formulation of these theories provides an inherently non-perturbative way to investigate them at strong coupling and finite $N$. A naive lattice regularization of SYM theories lacks supersymmetry at finite lattice spacing[1]. However, it is possible to construct supersymmetric lattice actions at finite lattice spacing for certain classes of SYM theories if they have a sufficient amount of supersymmetry in the target continuum theory. It is fortuitous that these lattice constructions can be done for all theories with holographic dual for all $d \leq 4$.

The holographic duality imposes a strict condition on the number of supersymmetries in the gauge theory, and it is often an interesting question how the reduction of supersymmetry can affect the holographic features of the theory. In this sense, our main goal is to investigate the thermal phase structure of the $1+1$-dimensional SU($N$) Yang–Mills theory with four supercharges and compare it with that of the $1+1$-dimensional SU($N$) Yang–Mills theory with sixteen supercharges. Although it is known that this theory does not have a holographic dual, we would like to understand how close it is to its sixteen supercharge counterpart, which has a well-defined dual. In order to pursue this question, we focus on the presence of a possible deconfinement transition which is a characteristic feature of the sixteen-supercharge theory [67, 68, 108, 109] at finite temperatures. This large $N$ transition is captured by the spatial Wilson loop as the order parameter.

In this chapter, we will discuss the existence of bound states at large $N$, as captured by the clumping of the scalars of the theory around the origin, despite the presence of classical flat directions [2]. Such bound states have already been observed in the $0+1$-dimensional sixteen-supercharge theory [54]. We continue the discussion about the bound states in this work. We also revisit the absence of dynamical supersymmetry breaking of this theory using energy density and Ward identities [24, 110] with a relatively larger $N$.

---

[1] See Ref. [101–107] for detailed constructions of these theories.

The theory under consideration does not exhibit the sign problem when the continuum limit is taken properly [24, 111–114]. The coupling regimes we have probed and the sufficiently larger lattice volumes used ensure that our numerical calculations are free from this notorious sign problem [24].

The chapter is organized as follows: In Sec. 5.2, we present the supersymmetric lattice construction of the two-dimensional theory with four supercharges. Results of the numerical calculations are discussed in Sec. 5.4. There we discuss the energy density, Ward identities, behavior of scalars, presence of the spatial deconfinement transition, and the dependence of the critical coupling $r_\tau^c$ on the aspect ratio $\alpha$. In Sec. 5.5, we provide the conclusions and discuss the directions for future work.

## 5.2 Two-dimensional theory with four supercharges

The theory we are interested in, the $1 + 1$-dimensional Yang–Mills theory with four supercharges, can be obtained by dimensionally reducing the four-dimensional $\mathcal{N} = 1$ SYM. The theory can be put on a lattice with the help of either twisting or orbifolding procedures. We will take the help of the twisting procedure to put it on a lattice. The parent four-dimensional theory, in Euclidean spacetime, has the following global symmetry group

$$G = \mathrm{SO}(4)_E \times \mathrm{U}(1), \tag{5.1}$$

with $\mathrm{SO}(4)_E$ and $\mathrm{U}(1)$ denoting the Euclidean rotation symmetry and chiral symmetry, respectively. After dimensional reduction, the global symmetry group becomes

$$G = \mathrm{SO}(2)_E \times \mathrm{SO}(2)_{R_1} \times \mathrm{U}(1)_{R_2}, \tag{5.2}$$

with $\mathrm{SO}(2)_E$ denoting the Euclidean rotation symmetry, $\mathrm{SO}(2)_{R_1}$ the rotation symmetry along the reduced dimensions, and $\mathrm{U}(1)_{R_2}$ the chiral symmetry. But $\mathrm{U}(1)_{R_2}$ can be treated locally equivalent to $\mathrm{SO}(2)_{R_2}$, i.e.,

$$\mathrm{SO}(2)_E \times \mathrm{SO}(2)_{R_1} \times \mathrm{U}(1)_{R_2} \sim \mathrm{SO}(2)_E \times \mathrm{SO}(2)_{R_1} \times \mathrm{SO}(2)_{R_2}. \tag{5.3}$$

In order to have a lattice construction that preserves one supersymmetry charge, we need to twist the theory by combining the $R$-symmetry group with the Euclidean rotation group. The twisting process is nothing but a change of variables in flat Euclidean spacetime. We have the following two possible twist options:

- $A$ twist:
$$\text{SO(2)}' = \text{diag}\Big(\text{SO(2)}_E \times \text{U(1)}_{R_2}\Big). \tag{5.4}$$

- $B$ twist:
$$\text{SO(2)}' = \text{diag}\Big(\text{SO(2)}_E \times \text{SO(2)}_{R_1}\Big). \tag{5.5}$$

We work with the $B$ twist in our calculations. After twisting, the fermions and supercharges of the theory decompose into integer-spin representations of the twisted rotation group SO(2)'. The model under consideration has four bosonic and four fermionic degrees of freedom. It also has four real supercharges.

After the twist, the gauge field ($A_a$, $a = 1, 2$) and the scalars ($X_a$) of the theory transform identically under the twisted rotation group, and both are $N \times N$ anti-hermitian matrices. Thus, we can combine them to form a complexified gauge field $\mathcal{A}_a = A_a + iX_a$. These complexified gauge fields, $\mathcal{A}_a$ in the continuum are mapped to link fileds living between $n$ and $n + \hat{\mu}_a$ on lattice as: $\mathcal{U}_a(n) = e^{\mathcal{A}_a(n)}$.

The action of the twisted theory is composed of a complexified gauge field $\mathcal{A}_a$ and twisted fermions (Grassmann valued fields), which we denote as $\eta$, $\psi_a$, and $\chi_{ab} = -\chi_{ba}$. The supercharges also undergo a decomposition similar to that of the fermions: $\mathcal{Q}$, $\mathcal{Q}_a$, and $\mathcal{Q}_{ab}$.

The supercharge $\mathcal{Q}$ acts on the twisted fields in the following way

$$\begin{aligned}
\mathcal{Q}\mathcal{A}_a &= \psi_a, \\
\mathcal{Q}\overline{\mathcal{A}}_a &= 0, \\
\mathcal{Q}\psi_a &= 0, \\
\mathcal{Q}\chi_{ab} &= -\overline{\mathcal{F}}_{ab}, \\
\mathcal{Q}\eta &= d, \\
\mathcal{Q}d &= 0.
\end{aligned} \tag{5.6}$$

In the above equation, an extra bosonic degree of freedom can be seen, the auxiliary field

$d$, which is introduced to close the algebra. The equation of motion for this auxiliary field is

$$d = \left[\overline{\mathcal{D}}_a, \mathcal{D}_a\right]. \tag{5.7}$$

The complexified covariant derivatives have the following form

$$\begin{aligned}\mathcal{D}_a &= \partial_a + \mathcal{A}_a, \\ \overline{\mathcal{D}}_a &= \partial_a + \overline{\mathcal{A}}_a.\end{aligned} \tag{5.8}$$

The complexified field strength is given as

$$\begin{aligned}\mathcal{F}_{ab} &= [\mathcal{D}_a, \mathcal{D}_b], \\ \overline{\mathcal{F}}_{ab} &= \left[\overline{\mathcal{D}}_a, \overline{\mathcal{D}}_b\right].\end{aligned} \tag{5.9}$$

Hence the twisted action involving all the above-discussed fields can be written in a $\mathcal{Q}$-exact form as

$$S = \frac{N}{4\lambda} \mathcal{Q} \int d^2x\, \mathrm{Tr}\left(\chi_{ab}\mathcal{F}_{ab} + \eta\left[\overline{\mathcal{D}}_a, \mathcal{D}_a\right] - \frac{1}{2}\eta d\right). \tag{5.10}$$

Here, $\lambda$ is the 't Hooft coupling. After performing the $\mathcal{Q}$ variation on the action and integrating over the auxiliary field $d$, we get the twisted action

$$S = \frac{N}{4\lambda} \int d^2x\, \mathrm{Tr}\left(-\overline{\mathcal{F}}_{ab}\mathcal{F}_{ab} + \frac{1}{2}\left[\overline{\mathcal{D}}_a, \mathcal{D}_a\right]^2 - \chi_{ab}\mathcal{D}_{[a}\psi_{b]} - \eta\overline{\mathcal{D}}_a\psi_a\right). \tag{5.11}$$

This action can be discretized on a two-dimensional square lattice spanned by two orthogonal unit vectors with the help of the geometrical discretization scheme [115] as shown in Fig. 5.1.

Let us take $N_\tau$ and $N_x$ as the number of lattice sites along temporal and spatial directions, respectively. Denoting $\beta$ and $L$ as dimensionful temporal and spatial extents, respectively, and $\mathtt{a}$ as the lattice spacing, we define

$$\beta \equiv \mathtt{a} N_\tau, \quad L \equiv \mathtt{a} N_x. \tag{5.12}$$

Periodic boundary conditions are imposed for all fields along spatial and temporal direc-tions, except for fermions, where anti-periodic boundary conditions are introduced along the temporal direction. All fields and variables on the lattice are made dimensionless using $\lambda$. We can define dimensionless temporal and spatial extents as

$$r_\tau \equiv \sqrt{\lambda}\beta = 1/t, \quad r_x \equiv \sqrt{\lambda}L, \tag{5.13}$$

with $t$ denoting the dimensionless temperature. We also introduce the aspect ratio $\alpha$, defined as

$$\alpha \equiv \frac{r_x}{r_\tau} = \frac{N_x}{N_\tau}. \tag{5.14}$$

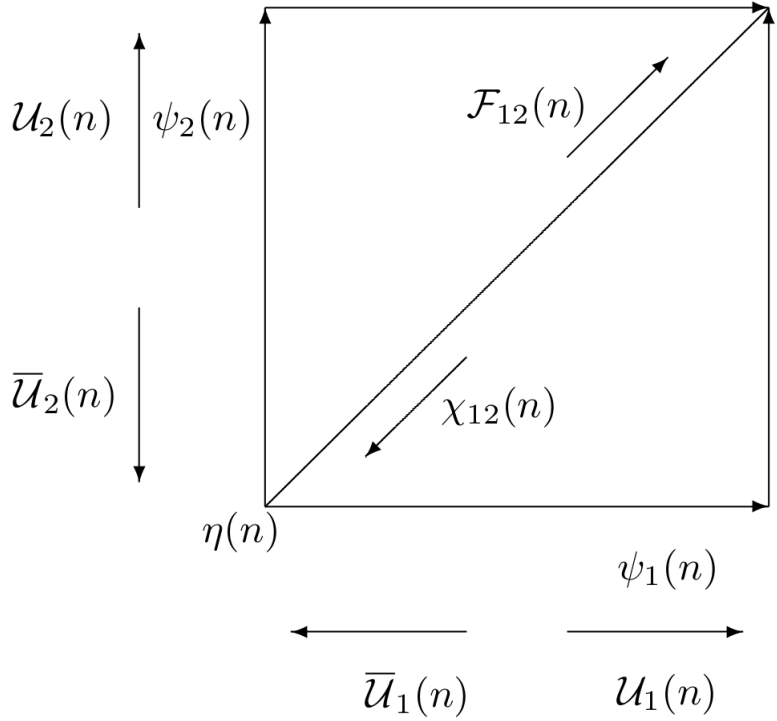

FIGURE 5.1: Orientations of fields on a unit cell for two-dimensional SYM theory with four supercharges.

The complexified gauge fields are mapped to complexified Wilson links as

$$\mathcal{A}_a(x) \to \mathcal{U}_a(n), \tag{5.15}$$

living on the links (n, n + $\hat{\mu}_a$) of a square lattice, where $\hat{\mu}_a$ denotes the unit vector in the $a$-th direction. They transform under $U(N)$ lattice gauge transformations as:

$$\mathcal{U}_a(n) \to G(n)\mathcal{U}_a(n)G^\dagger(n+\hat{\mu}_a). \tag{5.16}$$

Supersymmetry invariance then implies that $\psi_a(n)$ also lives on the same links and transforms identically. The scalar fermion $\eta(n)$ is associated with a site and transforms the following way under gauge transformations

$$\eta(n) \to G(n)\eta(n)G^\dagger(n). \tag{5.17}$$

To have gauge invariance, we associate $\chi_{ab}(n)$ with a plaquette. In practice, we introduce diagonal links running through the center of the plaquette and choose $\chi_{ab}(n)$ to lie with opposite orientation along those diagonal links. The placements and orientations of the fields ensure a gauge invariant lattice action. $\chi_{ab}(n)$ transforms in the following way

$$\chi_{ab}(n) \to G(n+\hat{\mu}_a+\hat{\mu}_b)\chi_{ab}(n)G^\dagger(n). \tag{5.18}$$

Covariant difference operators replace the covariant derivatives [115] as

$$\overline{\mathcal{D}}_a^{(-)} f_a(n) = f_a(n)\overline{\mathcal{U}}_a(n) - \overline{\mathcal{U}}_a(n-\hat{\mu}_a)f_a(n-\hat{\mu}_a),$$
$$\mathcal{D}_a^{(+)} f_b(n) = \mathcal{U}_a(n)f_b(n+\hat{\mu}_a) - f_b(n)\mathcal{U}_a(n+\hat{\mu}_b). \tag{5.19}$$

The lattice field strength is given by $\mathcal{F}_{ab}(n) = \mathcal{D}_a^{(+)}\mathcal{U}_b(n)$, and is anti-symmetric. It yields a gauge invariant loop on the lattice when contracted with $\chi_{ab}(n)$. Similarly, the term involving the covariant backward difference operator $\overline{\mathcal{D}}_a^{(-)}\mathcal{U}_a(n)$ can be contracted with the site field $\eta(n)$ to yield a gauge invariant expression. The lattice action is then written as

$$S = \frac{N}{4\lambda_{\text{lat}}} \sum_n \text{Tr}\left[ -\overline{\mathcal{F}}_{ab}(n)\mathcal{F}_{ab}(n) + \frac{1}{2}\left(\overline{\mathcal{D}}_a^{(-)}\mathcal{U}_a(n)\right)^2 \right.$$
$$\left. - \chi_{ab}(n)\mathcal{D}_{[a}^{(+)}\psi_{b]}(n) - \eta(n)\overline{\mathcal{D}}_a^{(-)}\psi_a(n) \right], \tag{5.20}$$

where $\lambda_{\text{lat}}$ is the dimensionless t' Hooft coupling. In addition to this action, we also add a scalar-potential term with a tunable parameter $\mu$ to control the flat directions in the theory.

Thus the complete action is given by

$$S_{\text{total}} = S + \frac{N\mu^2}{4\lambda_{\text{lat}}} \sum_{n,a} \text{Tr} \left( \overline{\mathcal{U}}_a(n) \mathcal{U}_a(n) - \mathbb{I}_N \right)^2. \tag{5.21}$$

In order to take the proper supersymmetric limit, we tune $\mu$ in the form of another variable $\zeta$, which is related to $r_\tau$ as

$$\mu = \zeta \frac{r_\tau}{N_\tau} = \zeta \sqrt{\lambda} \mathtt{a} = \zeta \sqrt{\lambda_{\text{lat}}}. \tag{5.22}$$

Here $\zeta$ is introduced because when we take the continuum limit i.e. $N_\tau \to \infty$, by keeping the temporal circle size and lattice spacing fixed, the deformation term automatically goes to zero as $\mu \propto N_\tau^{-1}$. Hence we can better control the flat directions with finite $\zeta$ and still can get to the continuum limit without any numerical runaway.

As the scalar potential term is purely bosonic and it explicitly breaks supersymmetry depending upon the value of $\mu$, we need to keep track of the effect of this breaking by a suitable observable. Discussed in the next section is one of the observable energy density, which is just a particular normalization of bosonic action. As discussed in chapter 2, another observable that tunes SUSY breaking effects is Ward identity. In order to measure Ward identity $\langle \mathcal{QO} \rangle$, we need to choose an appropriate operator $\mathcal{O}$ which is purely fermionic, such that $\mathcal{Q}$ variation gives the bosonic term. We discuss that particular term and the Ward identity in the next section.

## 5.3 Simulation details

The numerical calculations are performed for various configurations using publicly available software, github.com/daschaich/susy, which is described in Ref. [93]. It is a parallel code that can be used to study Yang–Mills theories numerically in various dimensions. The theory we are focusing on in this chapter is the two-dimensional four-supercharge theory. Because this theory does not suffer from any sign problem in the coupling regime, we wish to investigate [24, 114]. Hence we do not perform phase-quenched calculations. After integrating out the fermions, we have to deal with pseudofermions and Pfaffian, hence with the negative fractional powers of the determinant of the fermionic matrix. These negative fractional powers are calculated in the setup with the help of the Rational Hybrid Monte Carlo (RHMC) algorithm.

| Gauge group | $\alpha$ | Lattice size, $N_x \times N_\tau$ | Range of $r_\tau$ | $\zeta$ |
|---|---|---|---|---|
| SU(2) | 1 | $24 \times 24$ | 5.0 | 0.3, 0.4, 0.5 |
| SU(4) | 1 | $24 \times 24$ | 5.0 | 0.3, 0.4, 0.5 |
| SU(8) | 1 | $24 \times 24$ | 5.0 | 0.3, 0.4, 0.5 |
| SU(12) | $\frac{1}{2}$ | $12 \times 24$ | 0.6 - 2.0 | 0.3, 0.4, 0.5 |
| | 1 | $12 \times 12$ | 0.6 - 1.6 | 0.2, 0.3, 0.4 |
| | 1 | $16 \times 16$ | 3.0 - 5.0 | 0.3 |
| | 1 | $24 \times 24$ | 0.333 - 7.0 | 0.3, 0.4, 0.5 |
| | 1 | $32 \times 32$ | 1.0 - 7.0 | 0.3, 0.4, 0.5 |
| | $\frac{3}{2}$ | $24 \times 16$ | 0.3 - 0.9 | 0.3, 0.4, 0.5 |
| | 2 | $24 \times 12$ | 0.2 - 1.4 | 0.3, 0.4, 0.5 |
| | 4 | $24 \times 6$ | 0.05 - 0.2 | 0.6, 0.7, 0.8 |
| SU(16) | 1 | $12 \times 12$ | 0.6 - 1.6 | 0.3 |
| SU(20) | 1 | $12 \times 12$ | 0.6 - 1.6 | 0.3 |

TABLE 5.1: List of different ensembles used in the numerical setup.

More about the algorithmic details can be looked at from Ref. [93].

We have worked in different coupling ($r_\tau$) regimes, and the numerical runs become expensive for stronger couplings with larger lattices. We worked with a couple of lattices initially and concluded that the $24 \times 24$ lattice is close enough to the continuum limit. We leave continuum extrapolations of this work for the near future, which can be achieved by taking $N_\tau \to \infty$ by keeping lattice spacing (a) and temporal circle size ($r_\tau$) fixed. Unless otherwise specified, our results are obtained on a $24 \times 24$ lattice, with gauge group SU(12) and three values for the $\zeta$ parameter: $\zeta = 0.3, 0.4, 0.5$. Apart from this, the different gauge groups and lattices used in our work are collated in Table 5.1.

## 5.4 Simulation results

### 5.4.1 Energy density

The energy density on the lattice takes the following form

$$E = \frac{3}{\lambda_{\text{lat}}} \left(1 - \frac{2}{3N^2} S_B \right), \tag{5.23}$$

where $S_B$ is the bosonic action averaged over the lattice volume.

In an earlier study [24], it was concluded that in the two-dimensional four supercharge theory, with gauge groups U(2) and U(3), the energy density vanishes within uncertainties in the zero-temperature limit. We performed numerical calculations on a $24 \times 24$ lattice with $N = 12$. The behavior of the energy density $E$, as $r_\tau$ is varied, is shown in Fig. 5.2. The simulations were performed for various $\zeta$ values ($\zeta = 0.3, 0.4, 0.5$), and the data points represent the $\zeta \to 0$ extrapolated values. We see that for larger $r_\tau$, the energy density approaches zero within errors, suggesting that dynamical SUSY breaking is absent in this model. In the low-temperature region, $3.0 \leq r_\tau \leq 7.0$, performing linear and quadratic fits to the energy density gives similar values, which is close to zero within errors. We note that

$$E \propto t^2 \quad \text{for } t \geq 1, \tag{5.24}$$

$$E \propto t^0 \quad \text{for } t \ll 1. \tag{5.25}$$

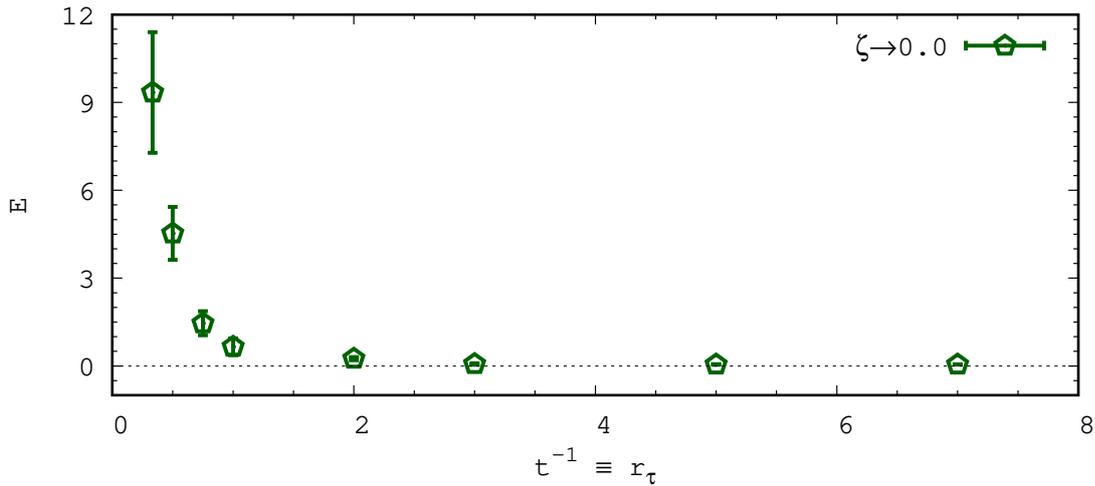

FIGURE 5.2: Energy density $E$ for various $r_\tau$ values in the $\zeta \to 0$ limit. The simulations were performed on a $24 \times 24$ lattice with $N = 12$.

The behavior of the energy density of the four-supercharge theory at strong coupling ($t \ll 1$) is drastically different from that of its maximally supersymmetric cousin. In the maximally supersymmetric theory, we have [24, 53]

$$E \propto \begin{cases} t^2 & \text{for } t \geq 1, \\ t^3 & \text{for } t \ll 1. \end{cases} \tag{5.26}$$

Though a precise calculation in the four-supercharge theory that shows the temperature dependence of the energy density at strong couplings does not exist, it might be possible to estimate the temperature dependence with the approach taken in Ref. [53], and we leave this for future work.

### 5.4.2 Ward identities

We can consider the normalized action, which is related to the bosonic action $S_B$, as the simplest Ward identity. In Fig. 5.3, we show the plot of the normalized action $2S_B/3N^2$ against $r_\tau$ in the $\zeta \to 0$ limit. This observable fluctuates around one, as expected for a theory with unbroken SUSY.

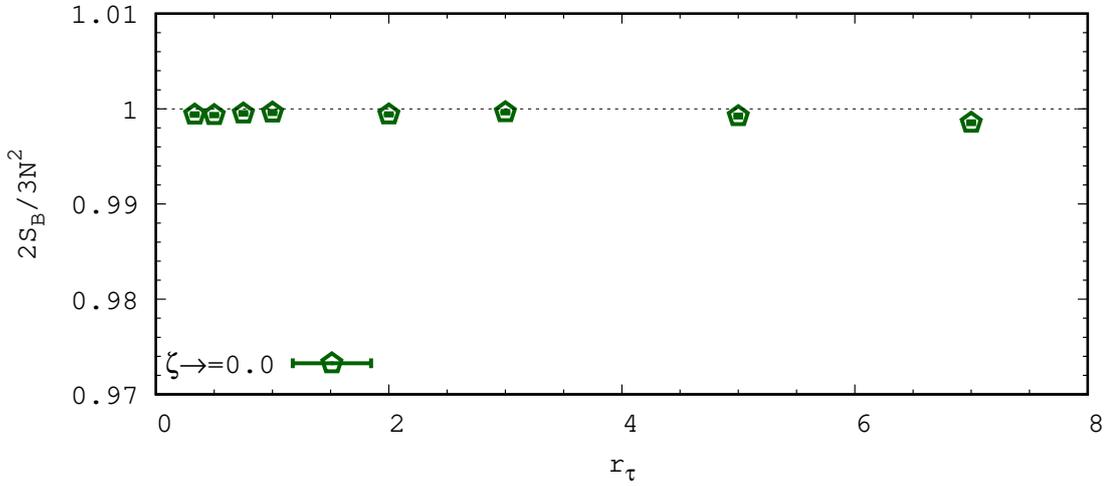

FIGURE 5.3: Normalized action for various $r_\tau$ values in the $\zeta \to 0$ limit.

As the lattice is large enough, and we work with different finite $\zeta$ values, there is no need to worry about sign problem and flat directions. Another Ward identity is given by

$$\begin{aligned} W &= \mathcal{Q} \sum_a \left( \eta \mathcal{U}_a \bar{\mathcal{U}}_a \right) \\ &= \text{Tr} \left( \sum_b \bar{\mathcal{D}}_b \mathcal{U}_b \sum_a \mathcal{U}_a \bar{\mathcal{U}}_a \right) - \sum_a \text{Tr} \left( \eta \psi_a \bar{\mathcal{U}}_a \right). \end{aligned} \quad (5.27)$$

In Fig. 5.4, we show the simulation results for this Ward identity. We see that for smaller $r_\tau$ values, the Ward identity is well satisfied for all $\zeta$ values. For larger $r_\tau$ and smaller $\zeta$ values, the Ward identity deviates from zero significantly. However, we should also keep in mind that data at smaller temperatures have a strong dependence on lattice volume. This can be seen in Fig. 5.5 where we show the volume dependence for $r_\tau = 3, 5$. As we take $N_\tau \to \infty$ and $\zeta \to 0$, we see that the Ward identity is satisfied for the entire temperature range.

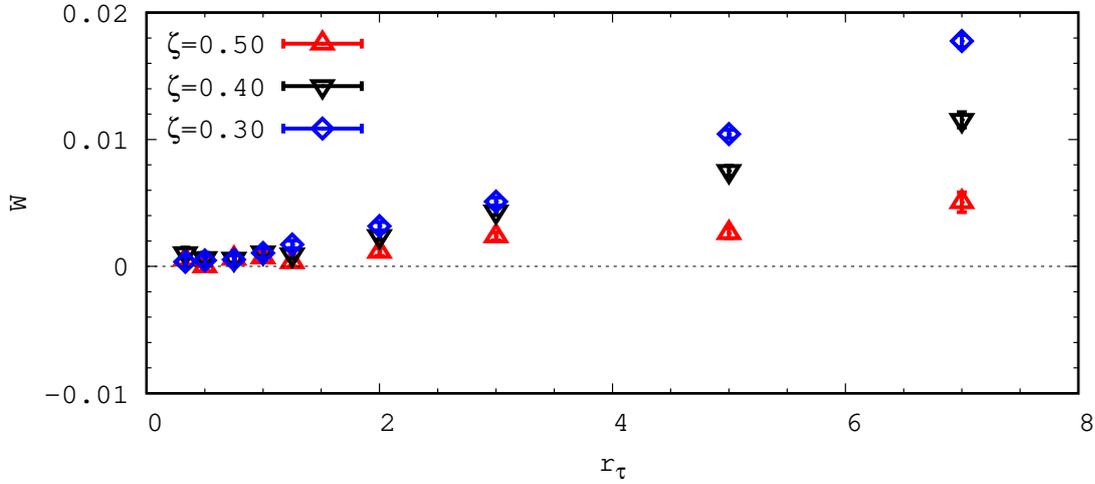

FIGURE 5.4: The Ward identity (Eq. (5.27)) for various $r_\tau$ values with different $\zeta$ values.

### 5.4.3 Behavior of scalars

In the sixteen-supercharge theory, the scalars behave in such a way that they assemble around the origin, even in the presence of flat directions, for large $N$. This assembly of scalars can be interpreted as a *bound state* in the gauge theory.

The scalar fields in the gauge theory carry the information about the collective coordinates of D-branes in the dual gravitational theory. From a pure field theoretical point of view, this scalar bound state's very existence is nontrivial, which is also why we resort to numerical computation. In a previous study of the four supercharge theory [110], with periodic boundary conditions for fermions along the temporal direction, it was observed that the scalars clumped around the origin even in the presence of flat directions. However, they found that the scalar bound states were not observed for smaller $N$ with thermal boundary conditions.

In order to monitor the scalar bound states, we define an observable known as the extent of scalars

$$\langle \text{Scalar}^2 \rangle \equiv \left\langle \frac{1}{2NN_\tau N_x} \sum_{i=1}^{N_\tau N_x} \text{Tr}(X_{1,i}^2 + X_{2,i}^2) \right\rangle. \quad (5.28)$$

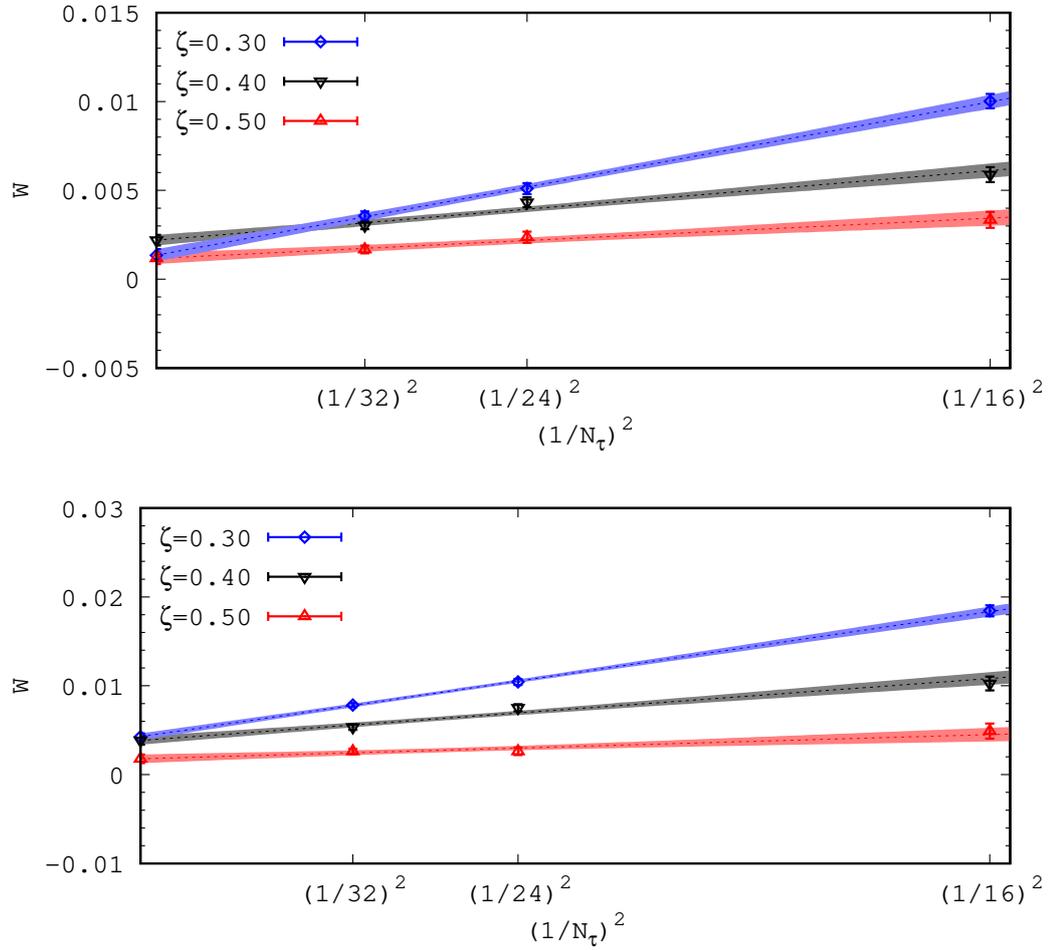

FIGURE 5.5: The Ward identities (Eq. (5.27)) for $r_\tau = 3.0$ (upper panel) and $r_\tau = 5.0$ (lower panel) with different $\zeta$ values. As we take the $1/N_\tau \to 0$ limit, we see that the Ward identities for all $\zeta$ values go to zero, suggesting that SUSY is preserved in the model. The plots also suggest that the Ward identities have strong volume dependence at lower temperatures.

The information about scalars can be extracted from complexified link fields by decomposing them as [111]

$$\mathcal{U}_a(n) = P_a(n) U_a(n), \tag{5.29}$$

where $U_a(n)$ is a unitary matrix and $P_a(n)$ is a positive semidefinite hermitian matrix, the logarithm of it gives us the scalar as

$$X_a(n) = \log(P_a(n)). \tag{5.30}$$

In Fig. 5.6, we show our simulation results for the temperature dependence of the extent of scalars on a $24 \times 24$ lattice with $N = 12$. The simulations were performed for $\zeta = 0.3, 0.4, 0.5$, and then $\zeta \to 0$ extrapolation is taken. We see that the best fit in the range $3.0 \leq r_\tau \leq 7.0$ has the form

$$b + c r_\tau^3,$$

with $b = 0.0090(3), c = 0.0004(0)$, and $\chi^2/\text{d.o.f.} = 0.0529$.

In the range $0.333 \leq r_\tau \leq 1.0$, the best fit has the form

$$b + c r_\tau,$$

with $b = 0.0002(1), c = 0.0030(1)$, and $\chi^2/\text{d.o.f.} = 5.6023$.

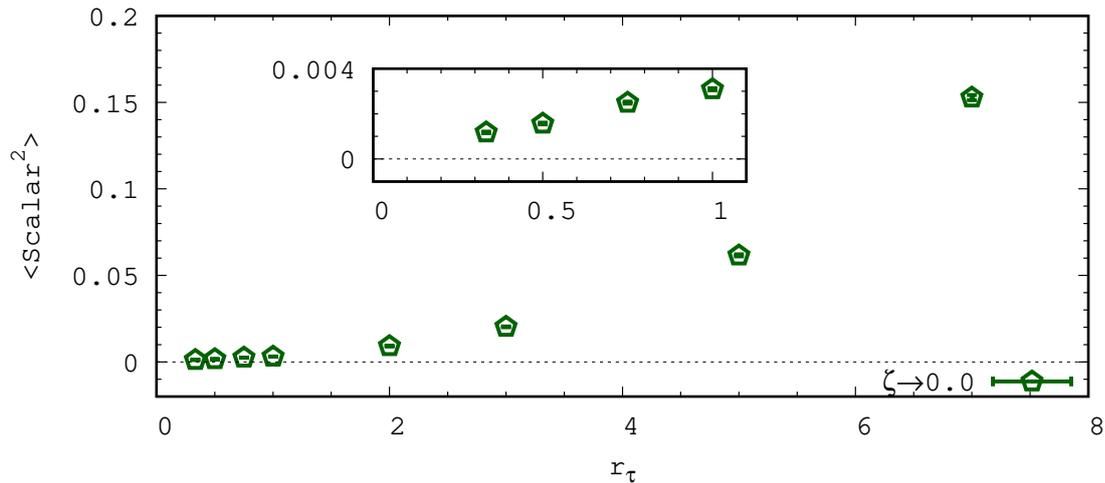

FIGURE 5.6: The extent of scalars for various $r_\tau$ values in the $\zeta \to 0$ limit. The simulations were performed on a $24 \times 24$ lattice with $N = 12$.

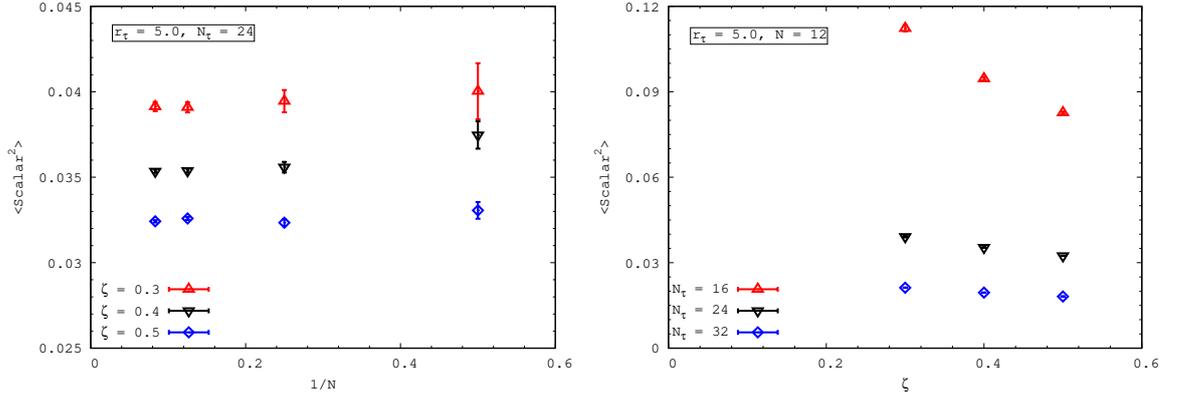

FIGURE 5.7: The dependence of the scalar extent (Eq. (5.28)) on $N \in \{2, 4, 8, 12\}$ for different $\zeta$ at fixed $r_\tau = 5$ on $24^2$ lattices (left), and its dependence on $\zeta$ for different lattice sizes at fixed $r_\tau = 5$ and $N = 12$ (right).

Our simulations show the following temperature dependence for the extent of scalars

$$\left\langle \text{Scalar}^2 \right\rangle \propto \begin{cases} t^{-1} & \text{for } t \geq 1, \\ t^{-3} & \text{for } t \ll 1. \end{cases} \quad (5.31)$$

In the sixteen supercharge theory, the dependence is

$$\left\langle \text{Scalar}^2 \right\rangle \propto t \quad \text{for } t \ll 1. \quad (5.32)$$

We report that there is no visible $N$ dependence for the extent of scalars for a relatively larger lattice with $N_\tau = 24$. These results suggested that the scalars clump around the origin even for small U($N$) gauge groups, provided we are working with sufficiently large lattices. There, it is also observed that for smaller lattices, say, $N_\tau = 16$, the extent of scalars may not converge to a definite value on the $\zeta \to 0$ limit.

Our analyses of $\left\langle \text{Scalar}^2 \right\rangle$ at $r_\tau = 5$ for different $N$ and $N_\tau$ values are shown in Fig. 5.7.

Considering the large-$N$ limit in the left-panel plot of Fig. 5.7, we observe no visible dependence on $N$ for a relatively larger lattice with $N_\tau = 24$. These results suggest that the scalars seem to clump around the origin even for small U($N$) gauge groups, provided we are working with sufficiently large lattices. For smaller lattices, say $N_\tau = 16$, the right-panel plot in Fig. 5.7 suggests that the scalars may not converge to a definite value in the $\zeta \to 0$ limit.

### 5.4.4 Spatial deconfinement transition

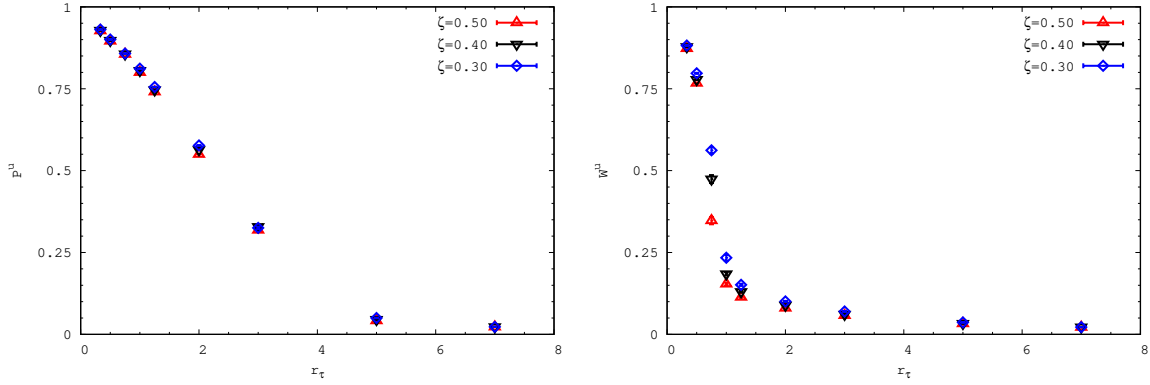

FIGURE 5.8: The $r_\tau$ dependence of the Wilson loops $P^u$ (left plot) and $W^u$ (right plot) on a $24 \times 24$ lattice with $N = 12$.

The holographic dual to the maximal SYM theory has the first-order Gregory–Laflamme phase transition [116], and the same can be expected in its gauge theory dual. This phase transition in the gauge theory can be interpreted as the deconfinement transition, and it is shown to exist in the sixteen-supercharge theory [67, 108, 117]. One interesting question is whether such a transition exists in the four-supercharge theory.

Our simulations suggest a possible spatial deconfinement transition in the four-supercharge theory only at weak couplings. In order to track the deconfinement transitions, we look at the unitarized Wilson loops

$$W^u_{\beta, L} = \frac{1}{N} \left\langle \left| \text{Tr} \left[ \mathcal{P} e^{\iota \oint_{\beta, L} A} \right] \right| \right\rangle, \tag{5.33}$$

along the temporal and spatial circles. We will denote the temporal and spatial loops as $P^u \equiv W^u_\beta$ and $W^u \equiv W^u_L$, respectively.

In order to monitor the spatial deconfinement transition, we will measure the susceptibility of the spatial Wilson loop $W^u$. It is given as

$$\chi_{W^u} \equiv N^2 \left( \langle |W^u|^2 \rangle - \langle |W^u| \rangle^2 \right). \tag{5.34}$$

In Fig. 5.8 we show the $r_\tau$ dependence of $P^u$ and $W^u$ for $\zeta = 3, 4, 5$ on a $24 \times 24$ lattice with $N = 12$. We see that both $P^u$ and $W^u$ undergo deconfinement transitions around $r_\tau \sim 1.0$. The scatter plot for the Wilson loops, both in spatial and temporal directions for three different couplings are shown in Figure 5.9.

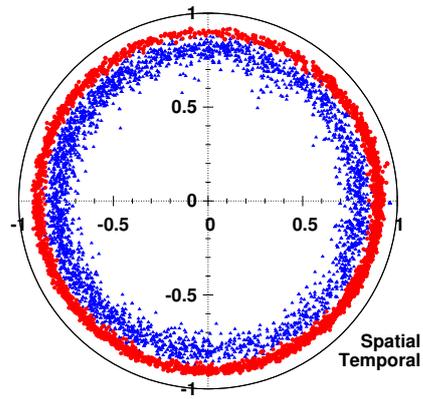

(a) $r_\tau = 0.5$

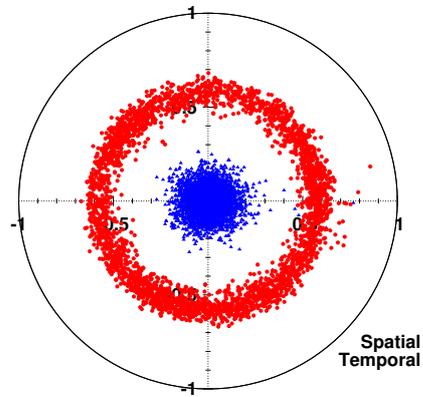

(b) $r_\tau = 2.0$

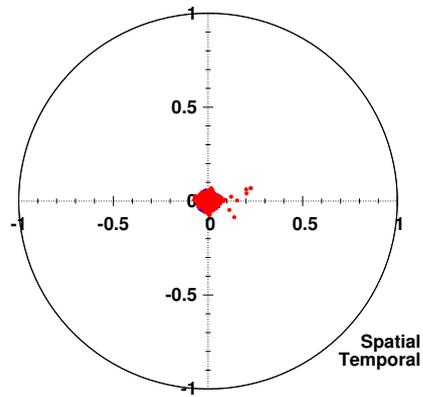

(c) $r_\tau = 7.0$

FIGURE 5.9: Wilson loop scatter plot on a $24 \times 24$ lattice with $N = 12$, with three different couplings for $\zeta = 0.30$. a) Coupling in the deconfined phase, b) Coupling around transition, c) Coupling in the confined phase.

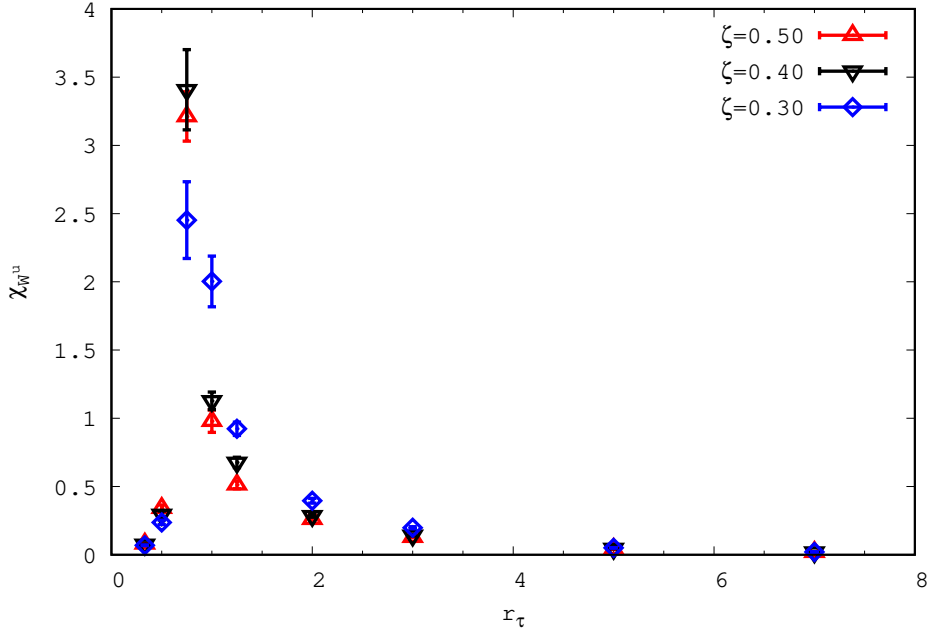

FIGURE 5.10: The $r_\tau$ dependence of the susceptibility of the spatial Wilson loop on a $24 \times 24$ lattice with $N = 12$.

In Fig. 5.10, we show the susceptibility of the spatial Wilson loop for various coupling values on a $24 \times 24$ lattice $N = 12$. A clear peak for $\chi_{W^u}$ is visible for $r_\tau \sim 1$, again indicating a potential spatial deconfinement in this theory. It is clear from Fig. 5.10 that the susceptibility peak persists for the $\zeta$ values used.

To determine the $r_\tau$ value and the order of the transition, we carried out further simulations on a $12 \times 12$ lattice ($\alpha = 1$ for this lattice as well) with $N = 12, 16$, and $20$. In Fig. 5.11 we show the $r_\tau$ dependence of $W^u$ and $\chi_{W^u}$ for $N = 12$ and $\zeta = 0.2, 0.3, 0.4$. There is a clear $\zeta$ dependent shift in the peak value of susceptibility. Scanning the different peak values with different $\zeta$ values used, we conclude that $r_\tau^c = 0.9(1)$ is independent of deformation strength.

In Fig. 5.12 we show the $r_\tau$ dependence of $W^u$ and $\chi_{W^u}$ for $N = 12, 16, 20$ at $\zeta = 0.3$. There is no $N$ dependence on critical $r_\tau$ value at this fixed $\zeta$. Also, this plot suggests that the transition is of second order as the susceptibility peak does not show an $N^2$ scaling.

From Fig. 5.12, we see that the transition gets sharper with larger $N$, but the scaling of susceptibility of the Wilson loop is significantly slower than $N^2$. This hints at a continuous phase transition in this theory [99] unlike the case in the maximally supersymmetric theory, where a first-order transition is seen at both weak and strong couplings.

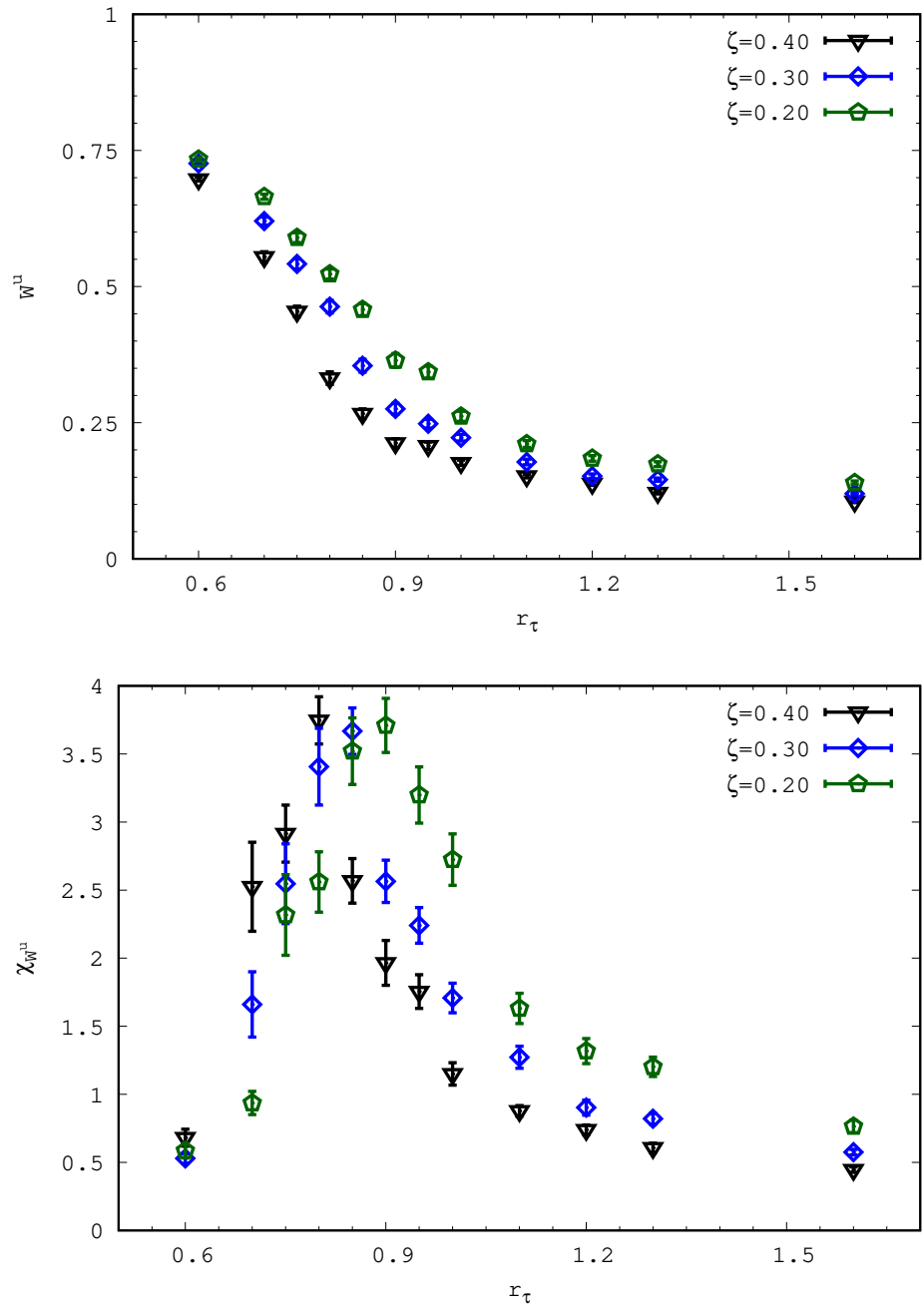

FIGURE 5.11: The $r_\tau$ dependence of the Wilson loop along the spatial direction (top plot) and its susceptibility (bottom plot) on a $12 \times 12$ lattice with $N = 12$.

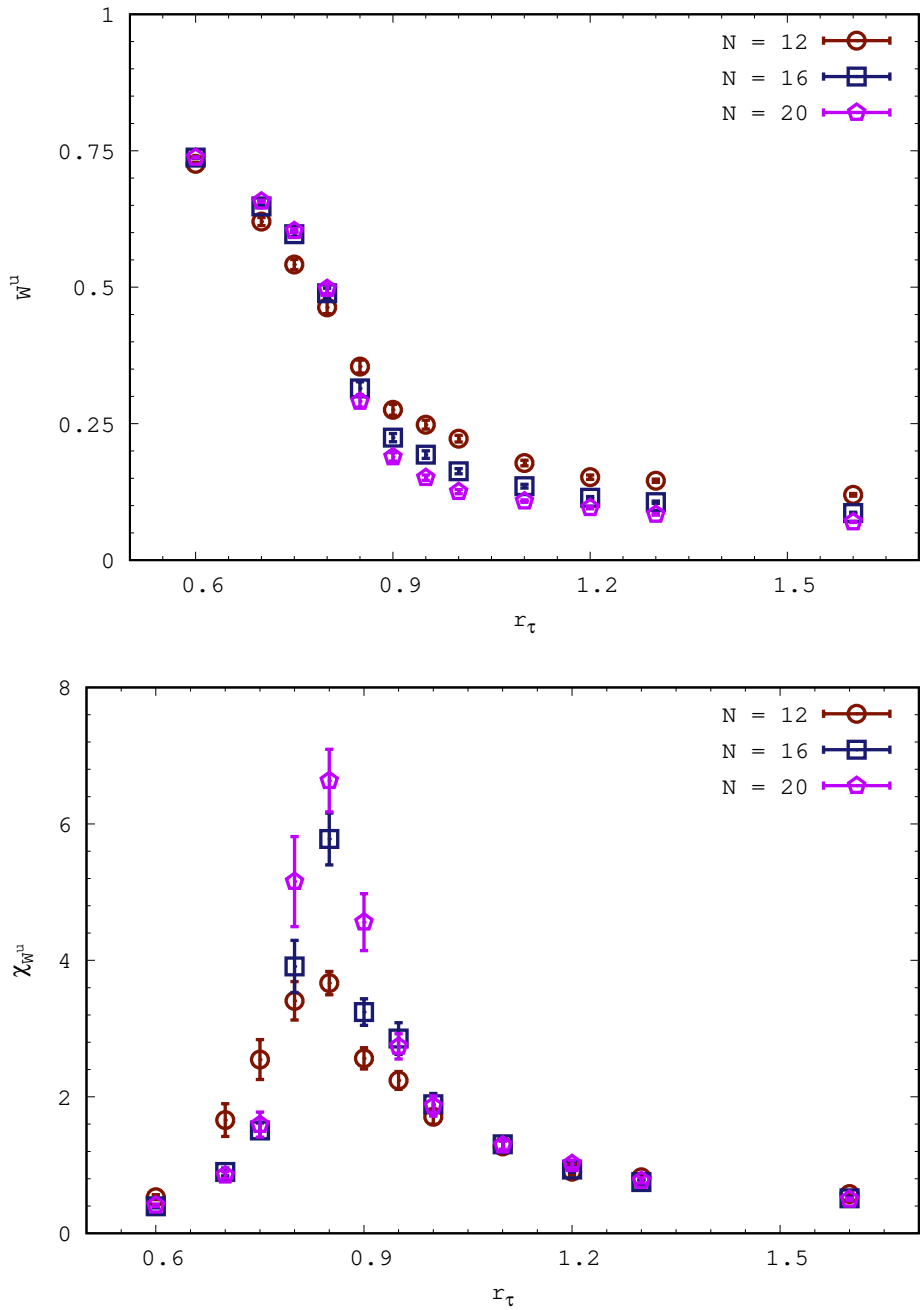

FIGURE 5.12: The $r_\tau$ dependence of the Wilson loop along the spatial direction (top plot) and its susceptibility (bottom plot) on a $12 \times 12$ lattice with $\zeta = 0.3$ for various $N$ values.

### 5.4.5 Dependence of $r_\tau^c$ on $\alpha$

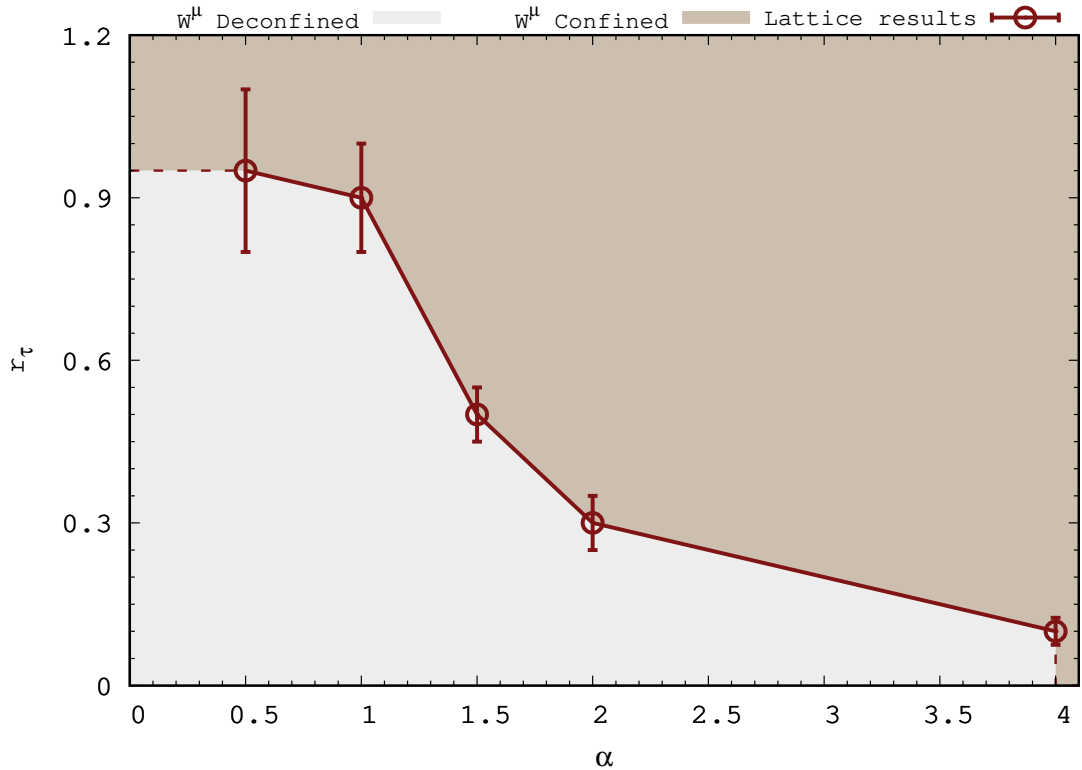

FIGURE 5.13: The $\alpha$ dependence of $r_\tau$ for $N = 12$ in $\zeta \to 0$ limit.

In order to further explore the phase structure of the theory, we performed simulations with $\alpha = 0.5, 1.5, 2.0, 4.0$. The data leading to these results are available through Ref. [118]. Several $\zeta$ values were used for a fixed $\alpha$, and $\zeta \to 0$ scanning of different peaks was taken to compute $r_\tau^c$ values. For larger $\alpha$ values, we do not see strong dependence on $\zeta$ for critical transition $r_\tau^c$ as compared to low $\alpha$ values.

In Fig. 5.13, we show a summary of our simulation results as the dependence of $r_\tau^c$ on $\alpha$. Similar to the sixteen-supercharge theory, here also $r_\tau^c$ shows a nontrivial dependence on $\alpha$.

The phase structure is on the same terms as the maximal theory with sixteen supercharges, but for smaller aspect ratios, it does not go towards stronger couplings. The five aspect ratios shown are $0.5, 1.0, 1.5, 2.0$, and $4.0$. For smaller aspect ratios than $0.5$, it was getting difficult to resolve the order parameter peak.

In the $r_\tau - r_x$ plane, Fig. 5.13 can be represented as shown in Fig. 5.14. It should be noted that critical values of $r_x$ are achieved by using the relationship between $r_\tau, r_x$ and $\alpha$, which is given in Eq. (5.14).

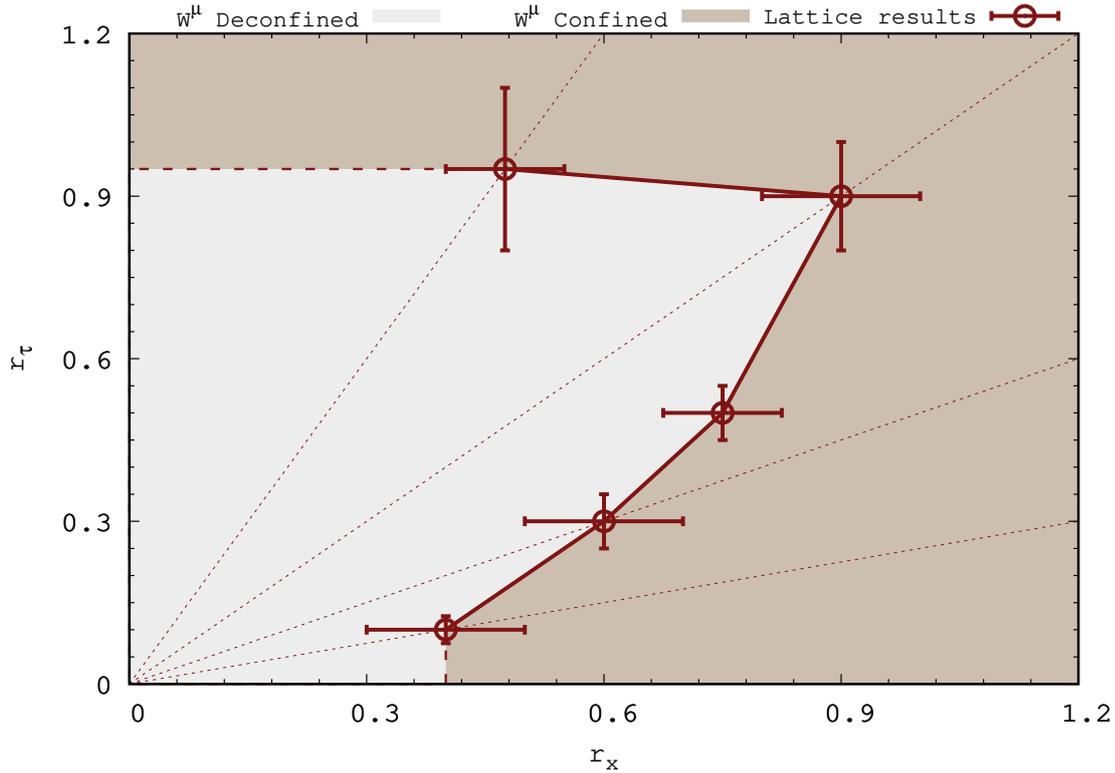

FIGURE 5.14: The phase structure corresponding to phase transition in $r_\tau - r_x$ plane for $N = 12$ in $\zeta \to 0$ limit. The dotted lines represent different aspect ratio lines.

## 5.5 Conclusions

The analysis reported in this work indicates the existence of the bound states of scalars at finite temperatures for two-dimensional $\mathcal{N} = (2,2)$ SYM. Even though we need large lattices to construct the bound state (and to control the sign problem [24, 111, 114]), we observe that there is no significant dependence on the gauge group U($N$). Even modest values of $N \gtrsim 4$ appear to provide access to the large-$N$ limit of the theory, given that we have data on large enough lattices.

By looking at the temperature dependence of the energy density and the distribution of the scalars, we can say that the two-dimensional $\mathcal{N} = (2,2)$ SYM theory seems to behave very differently at low temperatures compared to its maximal $\mathcal{N} = (8,8)$ counterpart.

We also wanted to investigate whether there is a deconfinement transition in the $\mathcal{N} = (2,2)$ theory as we move from small to large lattice sizes. In the sixteen-supercharge theory, there is a phase transition which is dual to a transition between different black hole solutions [67, 68]. And we found that there is a phase transition in this four supercharge theory.

Though the phase transition starts looking similar to the maximally supersymmetric case in the weak coupling regime and depends on the lattice geometry, it does not, however, continue to strong couplings according to our numerical results. While this result is consistent with what was expected, it does point out that there might not be any classical supergravity dual in this model. This work also seems to suggest that extended supersymmetry and field content are important factors for holographic interpretation.

One important future direction might be to consider the eight-supercharge model and understand whether the strong coupling results obtained here and in Ref. [67] admit the feature of an interpolating function.

# 6

# Conclusions and Future Outlook

In this thesis, we have investigated several low-dimensional field theory models with the help of non-perturbative lattice formalism and Monte Carlo simulations. We successfully simulated supersymmetric quantum mechanics with various superpotentials, bosonic BMN matrix model, and two-dimensional $\mathcal{N} = (2, 2)$ Yang-Mills with four supercharges.

We tested our lattice setup by studying dynamical supersymmetry breaking in supersymmetric quantum mechanics models with various superpotentials. In our next work, we produced the phase diagram for the bosonic BMN matrix model and confirmed that the phase transition is of the first order for all the couplings by interpolating between bosonic BFSS and the gauged Gaussian model. We also confirmed a spatial deconfinement transition in the two-dimensional four-supercharge Yang-Mills theory. This model was regularized on a lattice with the help of the twisting procedure.

In the near future, it would be interesting to connect the four supercharge and sixteen supercharge Yang-Mills theories in two dimensions by simulating the eight supercharge Yang-Mills theory. As the sixteen-supercharge theory has a well-defined holographic description, more analysis of the four- and eight-supercharge theories can give insight into the gauge/gravity duality from theories other than the maximally supersymmetric ones. We also plan to study the 'ungauged' version of the BMN matrix model, with and without fermions, again building upon prior investigations of the $\widehat{\mu} = 0$ BFSS model [119, 120]. A recent study in this direction has also investigated this version numerically [121]. In addition to exploring the effects of the deformation parameter in this context, it will be interesting to see the extent to which holographic arguments carry over to the non-supersymmetric bosonic BMN matrix model. In lattice Monte Carlo simulations, we cannot compute the partition function, and hence, some thermodynamic behavior of these systems still needs proper investigation. For example, the free energy in the full BMN model and its variation with the deformation parameter still need numerical investigation [83].

Finally, we also plan to improve the numerical setup for the lower-dimensional theories; the current setup induces a slow convergence to equilibrium distributions of observables in Monte Carlo simulations. A recent setup using numerical bootstrap has been tested successfully, and it would be a viable option [122] to replace the traditional methods since it can give results really quickly compared to Monte Carlo lattice simulations. It has been tested for systems with two matrices [123, 124]. It would be interesting to use this method for systems with more matrices that are analytically unsolvable and have convergence issues in the lattice Monte Carlo simulations.

107107

108108